\documentclass[10pt, letterpaper]{article}
\usepackage[margin=1in]{geometry}
\usepackage{natbib}
\usepackage{algorithmic}
\usepackage{graphicx}
\usepackage{textcomp}
\usepackage{xcolor}
\usepackage{booktabs}

\usepackage{multirow}
\usepackage{epsfig}
\usepackage{paralist}
\usepackage[center]{subfigure}
\usepackage{enumitem}
\usepackage{color}
\usepackage{xspace}
\usepackage{bbm}

\usepackage{pifont}
\usepackage{bbold}
\usepackage{mathtools}
\usepackage{comment}
\usepackage{microtype}
\usepackage{algorithmic}
\usepackage{makecell}
\usepackage[ruled,vlined,linesnumbered,resetcount]{algorithm2e}
\usepackage[hidelinks]{hyperref}
\usepackage{amsmath,amsfonts, bm, amsthm, amssymb}

\usepackage[hidelinks]{hyperref}
\usepackage{mathrsfs}
\usepackage{amsmath,amsfonts, bm, amsthm, amssymb}

\newcommand\CommentNT[1]{\footnotesize\blue}
\let\oldnl\nl% Store \nl in \oldnl
\newcommand{\noLineNumber}{\renewcommand{\nl}{\let\nl\oldnl}}% Remove line number for one line
\newcommand{\Axioms}{\textbf{\textsc{Axioms}}\xspace}
\newcommand{\Axiom}{\textbf{\textsc{Axiom}}\xspace}
\newcommand{\axioms}{\Axioms}
\newcommand{\axiom}{\Axiom}
\newcommand{\change}{\color{blue}}
\newcommand{\returncolor}{\color{black}}

\newcommand\red[1]{\textcolor{red}{#1}}
\newcommand\blue[1]{\textcolor{blue}{#1}}

\newcommand\modify[1]{\textcolor{teal}{#1}}
\newcommand\kijung[1]{\textcolor{black}{#1}}

\definecolor{obs3green}{RGB}{44, 160, 44}
\definecolor{obs3blue}{RGB}{0, 130, 202}
\definecolor{boxplotblue}{RGB}{24, 123, 205}

\newcommand{\measure}{\textsc{HyperRec}\xspace}
\newcommand{\algo}{\textsc{FastHyperRec}\xspace}
\newcommand{\generator}{\textsc{ReDi}\xspace}
\newcommand{\dhgpa}{\generator}
\newcommand{\model}{\dhgpa}
\newcommand{\hyperpa}{\textsc{HyperPA}\xspace}
\newcommand{\supplelink}{\url{https://github.com/kswoo97/hyprec}}
\newcommand{\smallsection}[1]{{\vspace{0.02in} \noindent {{\underline{\smash{\bf #1:}}}}}}

\newcommand{\axiomnum}[1]{\textbf{\textsc{Axiom #1\xspace}}}

\let\oldnl\nl% Store \nl in \oldnl
\newcommand{\nonl}{\renewcommand{\nl}{\let\nl\oldnl}}% Remove line number for one line

\newtheorem{axiombox}{\textbf{Axiom}}
\newtheorem{newaxiombox}{\textbf{Generalized Axiom}}
\newtheorem{thm}{\textbf{Theorem}}
\newtheorem{corr}{\textbf{Corollary}}
\newtheorem{obs}{\textbf{Observation}}
\newtheorem{lemma}{\textbf{Lemma}}
\newtheorem{prop}{\textbf{Proposition}}

\newtheorem{remark}{Remark}

\newcommand{\itembound}{\normalfont{\textbf{A-I}}\xspace}
\newcommand{\itemzero}{\normalfont{\textbf{A-II}}\xspace}
\newcommand{\itemnonzero}{\normalfont{\textbf{A-III}}\xspace}
\newcommand{\itemzeroone}{\normalfont{\textbf{A-IV}}\xspace}
\newcommand{\itemzerotwo}{\normalfont{\textbf{A-V}}\xspace}
\newcommand{\itemoverlap}{\normalfont{\textbf{A-VI}}\xspace}

\newcommand{\newaxiom}[1]{%
  \theoremstyle{theorem} % upright type for examples
  \newtheorem{subaxiom#1}{Generalized Axiom}%
  \expandafter\renewcommand\csname thesubaxiom#1\endcsname{#1\Alph{subaxiom#1}}%
}

\newaxiom{1}
\newaxiom{2}
\newaxiom{3}

\newtheorem{newaxiomfull}{Generalized Axiom}
\newtheorem{newpropfull}{Proposition}

\newcommand{\newprop}[1]{%
  \theoremstyle{theorem} % upright type for examples
  \newtheorem{subprop#1}{Proposition}%
  \expandafter\renewcommand\csname thesubprop#1\endcsname{#1\Alph{subprop#1}}%
}

\newprop{1}
\newprop{2}
\newprop{3}

\newcommand\paire[1]{e_{#1}=\langle H_{#1},T_{#1} \rangle}
\newcommand\pairei{e_i=\langle H_i,T_i \rangle}

\newcommand{\satisfy}{\ding{52}}
\newcommand{\wrong}{\ding{55}}

\DeclareMathOperator*{\argmax}{argmax}
\DeclareMathOperator*{\argmin}{argmin}
\setlist[itemize]{leftmargin=*}

\DeclarePairedDelimiterX{\infdivx}[2]{(}{)}{%
  #1\;\delimsize\|\;#2%
}
\newcommand{\jsdpq}{\mathrm{JSD} \infdivx}

\newcommand{\contribfootnote}{\footnote{{This work is an extended version of \citep{kim2022reciprocity}, which was presented at the 22nd IEEE International Conference on Data Mining (ICDM 2022). 
In the extended version, we introduce several theoretical extensions: (a) generalized versions of the axioms in Section~\ref{sec:measure:axioms} and a proof of Theorem~\ref{thm:axiom} for the generalized versions (Appendix~\ref{section:theoremproof}), (b) seven baseline hypergraph reciprocity measures (Section~\ref{sec:baselinemeasure}), (c) a proof that none of the baseline measures satisfies all the axioms (Appendix~\ref{section:baselinefail}), and (d) proofs of Theorem~\ref{thm:algo:basic} and Corollary~\ref{corr:algo} (Appendix~\ref{section:theoremproof}).
In addition, we conduct additional experiments regarding (a) the efficiency of \algo (Figure~\ref{fig:ferret} and Table~\ref{tab:runtime} in Section~\ref{sec:ferret}), 
(b) the statistical significance of Observation~\ref{obs:observation1} (Table~\ref{tab:p-value} in  Section~\ref{section:subsecobservation}), 
(c) the robustness of \measure with respect to the choice of $\alpha$ (Tables~\ref{tab:hypergraphrobustness} and \ref{tab:arcrobustness} in Section~\ref{section:subsecobservation}), and (d) the verification of Observation~\ref{obs:observation2} in 12 more real and synthetic hypergraphs (Figure~\ref{fig:obs2} in \ref{section:subsecobservation} and Figure~\ref{fig:obs2gen} in Section~\ref{section:generation:eval}).
At last, we provide one additional reciprocal pattern in real-world hypergraph (Observation~\ref{obs:observation3}: Figure~\ref{fig:obs3} in Section~\ref{section:subsecobservation}) and verify whether \dhgpa can reproduce this pattern. 
}}
\returncolor}

\usepackage{authblk}
\date{} %% Removing Date Information

\begin{document}
%\include{pythonlisting}
%\title[Reciprocity in Directed Hypergraphs: Measures, Findings, and Generators]{Reciprocity in Directed Hypergraphs: Measures, Findings, and Generators}

\title{Reciprocity in Directed Hypergraphs: Measures, Findings, and Generators}

\begin{comment}
\author[1]{\fnm{Sunwoo} \sur{Kim}}\email{kswoo97@kaist.ac.kr}

\author[1]{\fnm{Minyoung} \sur{Choe}}\email{minyoung.choe@kaist.ac.kr}

\author[3]{\fnm{Jaemin} \sur{Yoo}}\email{jaeminyoo@cmu.edu}

\author*[1,2]{\fnm{Kijung} \sur{Shin}}\email{kijungs@kaist.ac.kr}

\affil[1]{\orgdiv{Kim Jaechul Graduate School of AI}, \orgname{KAIST}, \orgaddress{\city{Seoul}, \country{South Korea}}}

\affil[2]{\orgdiv{School of Electrical Engineering}, \orgname{KAIST}, \orgaddress{\city{Daejeon}, \country{South Korea}}}

\affil[3]{\orgdiv{Heinz College of Information Systems and Public Policy}, \orgname{Carnegie Mellon University}, \orgaddress{\city{Pittsburgh}, \country{USA}}}
\end{comment}

\author[1]{Sunwoo Kim\thanks{kswoo97@kaist.ac.kr}}
\author[1]{Minyoung Choe\thanks{minyoung.choe@kaist.ac.kr}}
\author[3]{Jaemin Yoo\thanks{jaeminyoo@cmu.edu}}
\author[1,2]{Kijung Shin\thanks{kijungs@kaist.ac.kr}}
\affil[1]{Kim Jaechul Graduate School of AI, KAIST}
\affil[2]{School of Electrical Engineering, KAIST}
\affil[3]{Heinz College of Information Systems and Public Policy, Carnegie Mellon University}

\maketitle

\abstract{Group interactions are prevalent in a variety of areas. Many of them, including email exchanges, chemical reactions, and bitcoin transactions, are directional, and thus they are naturally modeled as directed hypergraphs, where each hyperarc consists of the set of source nodes and the set of destination nodes.
For directed graphs, which are a special case of directed hypergraphs, reciprocity has played a key role as a fundamental graph statistic in revealing organizing principles of graphs and in solving graph learning tasks.
%understanding mutual connectivity pattern, and enhancing the edge prediction performance. 
For general directed hypergraphs, however, even no systematic measure of reciprocity has been developed.

In this work, we investigate the reciprocity of $11$ real-world hypergraphs. To this end, we first introduce eight axioms that any reasonable measure of reciprocity should satisfy. %, and show that they cannot be satisfied. by straightforward measures. 
Second, we propose \measure, \kijung{a family of principled measures} of hypergraph reciprocity that satisfy all the axioms. 
Third, we develop \algo, a fast and exact algorithm for computing the measures. %whose search space is up to $10^{147}\times$ smaller than that of naive computation. 
Fourth, using them, we examine 11 real-world hypergraphs and discover patterns that distinguish them from random hypergraphs. Lastly, we propose \model, an intuitive generative model for directed hypergraphs exhibiting the patterns.

}

\section{Introduction}{
    \label{section:intro}
    Beyond pairwise interactions, understanding and modeling group-wise interactions in complex systems have recently received considerable attention \citep{benson2018simplicial,comrie2021hypergraph,do2020structural,kook2020evolution,lee2021hyperedges}.
A \emph{hypergraph}, which is a generalization of a graph, has been used widely as an appropriate abstraction for such group-wise interactions.
Each hyperedge in a hypergraph is a set of any number of nodes, and thus it naturally represents a group-wise interaction. 

Many group-wise interactions are directional, and they are modeled as a \emph{directed hypergraph}, where each hyperarc consists of the set of source nodes and the set of destination nodes.
Examples of directional group-wise interactions include email exchanges (from senders to receivers), chemical reactions \citep{yadati2020nhp}, road networks \citep{luo2022directed}, and bitcoin transactions \citep{ranshous2017exchange}; and they are modeled as directed hypergraphs for various applications, including metabolic-behavior prediction \citep{yadati2020nhp} and traffic prediction \citep{luo2022directed}.
See Figure~\ref{fig:overview} for an example of hypergraph modeling.

\emph{Reciprocity} \citep{newman2002email,garlaschelli2004fitness}, which quantifies how mutually nodes are linked, has been used widely as a basic statistic of directed graphs, which are a special case of directed hypergraphs where every arc has exactly one source node and one destination node.
Reciprocity {increase} understanding of a graph, especially potential organizing principles of it, and has proved useful for various tasks, including trust prediction \citep{nguyen2010you}, persistence prediction \citep{hidalgo2008dynamics}, anomaly detection \citep{akoglu2012quantifying}, and analysis of the spread of a computer virus through emails \citep{newman2002email}.

\begin{figure}[t]
    \centering
    \subfigure[Example Citation Dataset]{\includegraphics[width=0.50\textwidth]{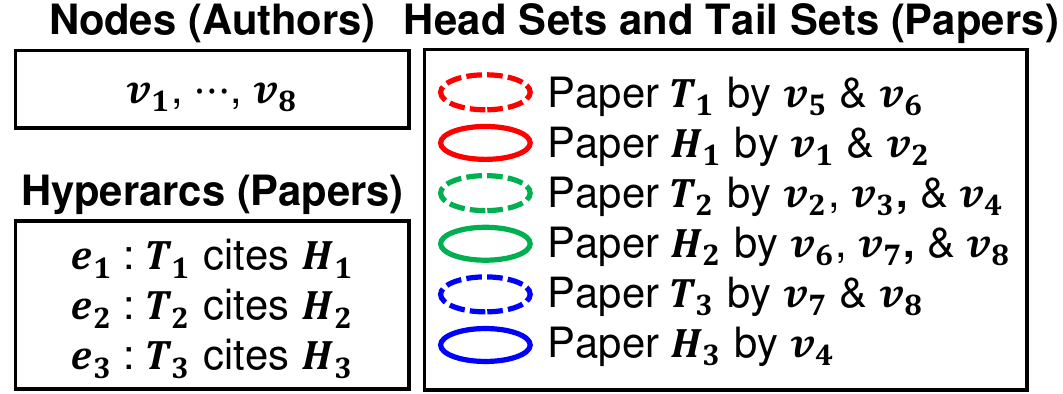}}
    \subfigure[Model]{\includegraphics[width=0.20\textwidth]{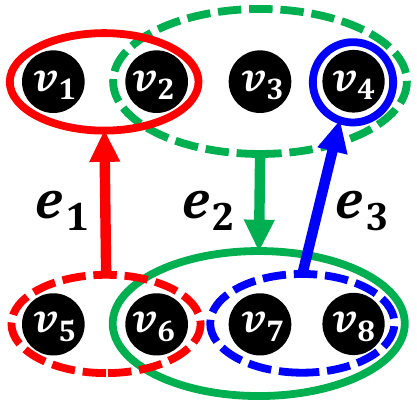}}
    \caption{\label{fig:overview} A citation dataset modeled as a directed hypergraph with 8 nodes and 3 hyperarcs. Nodes correspond to authors. Hyperarcs correspond to citations. The head set and tail set of each hyperarc correspond to sets of papers.}
\end{figure}

However, reciprocity has remained unexplored for directed hypergraphs, and to the best of our knowledge, no principled measure of reciprocity has been defined for directed hypergraphs. 
%One straightforward approach is to compute the reciprocity after replacing a given directed hypergraph into a directed graph by clique expansion (i.e., replacing each hyperarc with the directed bi-clique from its source nodes to its destination nodes), as suggested in \citep{pearcy2014hypergraph}.
\kijung{One straightforward approach is to first convert a directed hypergraph into an ordinary directed graph via clique expansion and then calculate standard reciprocity on the ordinary graph, as suggested in~\citep{pearcy2014hypergraph}.}
However, clique expansion may incur considerable information loss \citep{yadati2020nhp,dong2020hnhn,yoon2020much}. Thus, multiple directed hypergraphs whose reciprocity should differ, if they are determined by a proper measure, may become indistinguishable after being clique-expanded.

In this work, we investigate the reciprocity of real-world hypergraphs based on the first principled notion of reciprocity for directed hypergraphs.
Our contributions toward this goal are summarized as follows:\contribfootnote

\begin{itemize}
    \item \textbf{Principled Reciprocity Measure:} We design \measure, \kijung{a family of} probabilistic measures of hypergraph reciprocity. We prove that \measure satisfy eight axioms that any reasonable measure of 
    hypergraph reciprocity should satisfy, while baseline measures do not.
    
    \item \textbf{Fast and Exact Search Algorithm:} 
    The size of search space for computing \measure is exponential in the number of hyperarcs.
    We develop \algo, a fast and exact algorithm for computing \measure. %whose search space is up to $10^{147}\times$ smaller than that of naive computation.
    
    \item \textbf{Observations in Real-world Hypergraphs:} Using \measure and \algo, we investigate $11$ real-world directed hypergraphs, and discover three reciprocal patterns pervasive in them, which are verified using a null hypergraph model.
    \item \textbf{Realistic Generative Model:} To confirm our understanding of the patterns, we develop \model, a directed-hypergraph generator based on simple mechanisms on individual nodes. Our experiments demonstrate that \model yields directed hypergraphs with realistic reciprocal patterns.
\end{itemize}
For \textbf{reproducibility}, the code and data are available at \supplelink.

In Section~\ref{section:prelim}, we discuss preliminaries and related work. In Section~\ref{section:reciprocity}, we propose \kijung{a family of measures} of hypergraph reciprocity with a computation algorithm.
In Section~\ref{section:obs}, we discuss reciprocal patterns of real-world directed hypergraphs. %' which are verified by that of null hypergraphs. 
In Section~\ref{section:generation}, we propose a generative model for directed hypergraphs. Lastly, we offer a conclusion in Section~\ref{section:conclusion}.
}

\section{Basic Concepts and Related Work}{
    \label{section:prelim}
    In this section, we introduce some basic concepts and review related studies.
See Table~\ref{tab:symbols} for frequently-used symbols.

\subsection{Basic Concepts}
\label{section:prelim:concept}

A \textbf{\emph{directed hypergraph}} $G=(V,E)$ consists of a set of nodes $V=\{v_1,\cdots, v_{\lvert V \vert}\}$ and a set of \textbf{\emph{hyperarcs}} $E=\{e_1,\cdots, e_{\lvert E\vert}\} \subseteq \{\langle H,T \rangle  : H \subseteq V , T \subseteq  V  \}$.
%which is an ordered set of nodes 
For each hyperarc $\pairei \in E$, $H_i$ indicates the \textbf{\emph{head set}} %a set of nodes where a hyperarc is pointing at, 
and $T_i$ indicates the \textbf{\emph{tail set}}. %, a set of nodes which acts as a source of hyperarc. 
In Figure \ref{fig:overview}, the hyperarc $\paire{1} \in E $ is represented as an arrow that heads to $H_{1} = \{v_{1}, v_{2} \}$ from $T_{1} = \{v_{5}, v_{6} \}$.
It is assumed typically and also in this work that, in every hyperarc, the head set and the tail set are disjoint (i.e., $H_{i} \cap T_{i} = \emptyset, \forall i = 1, \cdots, \lvert E \vert$).
The \textbf{\emph{in-degree}} $d_{in}(v)=\lvert\{ e_i \in E: v\in H_i \}\vert$ of a node $v\in V$ is the number of hyperarcs that include $v$ as a head.
Similarly, the \textbf{\emph{out-degree}}  $d_{out}(v)=\lvert\{ e_{i} \in E: v\in T_{i}\}\vert$ of $v\in V$ is the number of hyperarcs that include $v$ as a tail.

% \smallsection{Incidence Matrix:} An \emph{incidence matrix} is an adjacency matrix of a hypergraph $C(G) \in \mathbf{R} ^{|V| \times \lvert E \vert}$ \citep{pearcy2014hypergraph}. Each head and tail set node is encoded as

% $$
%     c_{ij}=\begin{cases}
% 	1 & \text{if } v_{i} \in T_j \\
%     -1 & \text{if } v_{i} \in S_j \\
%     0 & \text{O.W.} \\
%     \end{cases}
% $$

%The \textbf{\emph{head size}} and \textit{tail size} of each hyperarc $\pairei\in E$ are defined as $|H_i|$ and $|T_i|$ respectively. 

% There are several special types of hypergraph: \emph{Forward Hypergraph} where head size of every hyperarc is equal to 1. (i.e., $|S_{i}| = 1 \quad \forall i = 1,...,\lvert E \vert$). E-mail dataset is an example of such case. \emph{Backward Hypergraph} where tail size of every hyperarc is equal to 1. (i.e., $|T_{i}| = 1 \quad \forall i = 1,...,\lvert E \vert$). Question and Answering datasets belong to this class of hypergraph.

From now on, we will use the term \textbf{\emph{hypergraph}} to indicate a \emph{directed hypergraph} and use the term \emph{undirected hypergraph} to indicate an undirected one.
We will also use the term \textbf{\emph{arc}} to indicate a hyperarc when there is no ambiguity.

\begin{table}[t]
    \caption{Frequently-used symbols.}\label{tab:symbols} % title of Table
    \vspace{2mm}
    \centering % used for centering table
    \scalebox{0.9}{
    \renewcommand{\arraystretch}{1.2}{
    \begin{tabular}{l  l} % centered columns (4 columns)
        \toprule
        \textbf{Notation} & \textbf{Definition} \\
        \midrule
        $G = (V, E)$ & hypergraph with nodes $V$ and hyperarcs $E$ \\
        $e_{i} = \langle H_{i}, T_{i} \rangle$ & hyperarc (or a target arc) \\
        $H_{i}, T_{i}$ & head set and tail set of a hyperarc $e_{i}$ \\
        $R_{i} = \{e'_{1}, \cdots , e'_{\lvert R_{i} \vert }\} $ & reciprocal set of a hyperarc $e_{i}$ \kijung{(see Section~\ref{sec:measure:axioms} for details)}\\
        $r(e_{i}, R_{i})$ & reciprocity of a target arc $e_{i}$ with a reciprocal set $R_{i}$\\
        $r(G) $ & reciprocity of a hypergraph $G$ \\
        $d_{in}(v), d_{out}(v)$ & in-degree and out-degree of a node $v$ \\
        $\vert A \vert$ & cardinality of a set $A$ (i.e., number of elements in $A$) \\
        \bottomrule
    \end{tabular}
    }
    } % is used to refer this table in the text
\end{table} 

\subsection{Related Work}\label{sec:relatedwork}

\smallsection{Reciprocity of Directed Graphs}
Reciprocity of directed graphs (i.e., a special case of directed hypergraphs where all head sets and tail sets are of size one) is a tendency of two nodes to be mutually linked \citep{newman2002email,garlaschelli2004fitness}.
This is formally defined as $\lvert E^{\leftrightarrow}\vert/\lvert E \vert$, where $\lvert E \vert$ is the number of edges in a graph, and $\lvert E^{\leftrightarrow}\vert$ is the number of edges whose opposite directional arc exists, i.e., $e = \langle \{v_{i}\}, \{v_{j}\} \rangle\in \lvert E^{\leftrightarrow}\vert$ if and only if $\langle \{v_{j}\} , \{v_{i}\}\rangle \in E$.
%Specifically, in \citep{newman2002email}, it is defined as the number of arcs pointing in both directions divided by the total number of arcs.
%and its upper and lower bounds from a given degree distribution of graph was studied \citep{jiang2015reciprocity}.
The notion was extended to weighted graphs \citep{squartini2013reciprocity,akoglu2012quantifying}, and using them, the relationship between degree and reciprocity was investigated \citep{akoglu2012quantifying}.
Moreover, the preferential attachment model \citep{albert2002statistical} was extended by adding a parameter that controls the probability of creating a reciprocal edge for generating  reciprocal graphs \citep{cirkovic2022preferential,wang2022asymptotic}.
Refer to Section~\ref{section:intro} for more applications of reciprocity.
% proposed a network generative model which can preserves reciprocity of a pairwise graph

%Squartini et al. \citep{squartini2013reciprocity} extended the it to weighted graphs. 
%Using them, 
%Akoglu et al. \citep{akoglu2012quantifying} proposed a probabilistic distribution of weighted graph's reciprocity pattern and analyzed . Recently, Cirkovic et al.\citep{cirkovic2022preferential} proposed a network generative model which can preserves reciprocity of a pairwise graph. Despite all these works enriched our understanding of reciprocal behavior in a network, mutual communication between groups still under-studied yet.

%Various analyses have been conducted regarding pairwise digraph's reciprocity. 

\smallsection{Patterns and Generative Models of Hypergraphs}
Hypergraphs have been used widely for modeling group-wise interactions in complex systems, and considerable attention has been paid to the structural properties of real-world hypergraphs, with focuses on node degrees \citep{do2020structural,kook2020evolution}, singular values \citep{do2020structural,kook2020evolution}, diameter \citep{do2020structural,kook2020evolution}, density \citep{kook2020evolution}, core structures \citep{bu2023hypercore}, the occurrences of motifs  \citep{lee2020hypergraph,lee2021thyme+}, simplicial closure \citep{benson2018simplicial}, ego-networks \citep{comrie2021hypergraph}, the repetition of hyperedges \citep{benson2018sequences,choo2022persistence}, and the overlap of hyperedges \citep{lee2021hyperedges}.
Many of these patterns can be reproduced by hypergraph generative models that are based on intuitive mechanisms \citep{benson2018sequences,do2020structural,kook2020evolution,lee2021hyperedges}. Such  models can be used for anonymization and graph upscaling in addition to testing our understanding of the patterns \citep{leskovec2008dynamics}.
All the above studies are limited to undirected hypergraphs, while this paper focuses on directed hypergraphs.

% As group interaction among entities are being highlighted, hypergraph has played a key role in modeling such characteristic. Lee et al. \citep{lee2020hypergraph} suggested 26 hypergraph motifs that frequently co-occured among same domain. These patterns were extended in a temporal way in \citep{lee2021thyme+}. Do et al. \citep{do2020structural} discovered the structural patterns of a hypergraph and proposed a preferential attachment generative model that successfully reproduced real-world datasets' structural characteristics. Lee and Choe et al. \citep{lee2021hyperedges} analyzed the overlapness property of a real-world hypergraphs and provided a generative model that captures real-world hypergraphs' pattern comprehensively.

% There have been many attempts where directed hypergraphs were used for certain tasks. Ranshous et al. \citep{ranshous2017exchange} investigated transaction patterns of bitcoin by expressing transactions as directed hypergraph. Luo et al. \citep{luo2022directed} used directed hypergraph attention network for accurate traffic forecasting. But these works focused on \textbf{using}  directed hypergraphs, rather than \textbf{analyzing} pervasive patterns which directed hypergraph have.

\smallsection{Directed Hypergraphs and Reciprocity}
Directed hypergraphs have been used for modeling chemical reactions \citep{yadati2020nhp}, knowledge bases \citep{yadati2021graph}, road networks \citep{luo2022directed}, bitcoin transactions \citep{ranshous2017exchange}, etc.
To the best of our knowledge, there has been only one attempt to measure the reciprocity of directed hypergraphs \citep{pearcy2014hypergraph}, where
%Pearcy et al. \citep{pearcy2014hypergraph} proposed a measure of hypergraph reciprocity.
(a) a hypergraph $G$ is transformed into a weighted digraph $\bar{G}$ by \textit{clique expansion},  {i.e., by replacing each arc $e_{i}=\langle H_{i},T_{i} \rangle$ with the bi-clique from $T_{i}$ to $H_{i}$, (b) a weighted digraph $\bar{G'}$ is obtained in the same way from a hypergraph $G'$ where, the perfectly reciprocal arc $\langle T_{i},H_{i} \rangle$ of each arc $e_{i}=\langle H_{i},T_{i} \rangle\in E$ is added if it is not already in $E$, and (c) ${tr(\bar{A}^{2})}/{tr(\bar{A'}^{2})}$, where $\bar{A}$ and $\bar{A'}$ are the weighted adjacency matrices of $\bar{G}$ and $\bar{G'}$, respectively, is computed as the reciprocity of $G$.
Note that $tr(\bar{A}^{2})$ corresponds to the weighted count of paths of length two in $\bar{G}$ that start and end at the same node, which is the same as the weighted count of mutually linked pairs of nodes in $\bar{G}$, and $tr(\bar{A'}^{2})$ is the count in the perfectly reciprocal counterpart.} \returncolor
%The square of adjacency matrix indicates number of paths of length 2, and trace computation provides number of paths that comes back to its original position within length 2. Thus, this measures how many mutually connected di-arcs exist compared to the perfectly reciprocal case.
However, as discussed in Section~\ref{section:intro}, clique expansion may cause substantial information loss \citep{yadati2020nhp,dong2020hnhn,yoon2020much}, and thus multiple directed hypergraphs whose reciprocities should differ, if they are determined by a proper measure, can be transformed into the same directed graphs by clique expansion. We further analyze the limitations of this approach based on axioms in the following section.
%This method cannot tell how much an individual hyperarc is reciprocal. Furthermore it carries problems of clique-expansion that high-order information of hyperarc is being destroyed and becomes indistinguishable. \citep{dong2020hnhn, yadati2020nhp}. Limitations of naive clique-expansion are covered in section~\ref{section:obs} by \axioms.

% which is called  clique-expansion. After clique-expansion, a specific hyperarc $e_{i} = \langle\{v_{1}, v_{2}\}, \{v_{3}, v_{4}\}\rangle$ becomes 4 arcs, $(e_{1},e_{3}), (e_{1},e_{4}), (e_{2},e_{3}), (e_{2},e_{4})$. 
% Beside, they assume there is perfectly reciprocal hypergraph $C'(G)$, where every hyperarc has its perfect opponent. (i.e., $e_{i}=\langle S_{i},T_{i} \rangle \in E'$, then $e'_{i}=\langle T_{i},S_{i}  \rangle \in E'$). They do same job with this new hypergraph ; $C'(G) \leftarrow A'(G)$. Their reciprocity is defined as 

}

\section{Directed Hypergraph Reciprocity}{
    \label{section:reciprocity}

%% Prof Comment: Section Summarize
In this section, we present eight necessary properties of an appropriate hypergraph reciprocity measure.
Then, we present \kijung{a family of reciprocity measures}, namely \measure, and an algorithm for fast computation of them. 
Lastly, we compare \measure with baseline measures to support its soundness.
%% Professor Comment: Reducing Size

\subsection{Framework and Axioms} \label{sec:measure:axioms}

We present our framework for measuring hypergraph reciprocity. Then, we suggest eight axioms that 
%explain conditions
any reasonable reciprocity measure must satisfy.

\smallsection{Framework for Hypergraph Reciprocity} Given a hypergraph $G$, we measure its reciprocity at two levels:

\begin{itemize}
    \item How much each arc (i.e., group interaction) is reciprocal.
    \item How much the entire hypergraph $G$ is reciprocal.
\end{itemize}

%We first discuss the arc-level reciprocity. 
For a \textbf{\emph{target arc}}, which we measure reciprocity for, multiple arcs should be involved in measuring its reciprocity inevitably. For example, in Figure \ref{fig:overview}, arc $e_{2}$'s head set and tail set overlap with $e_{1}$ and $e_{3}$'s tail set and head set, respectively, and thus we should consider both $e_{1}$ and $e_{3}$ in measuring $e_{2}$'s reciprocity. In graphs, however, only the arc with the opposite direction is involved in the reciprocity of an arc.
This unique characteristic of hypergraphs poses challenges in measuring reciprocity.
%from the digraph's reciprocity measure, where only its opposite direction is considered.
The \textbf{\emph{reciprocal set}} $R_{i}$ of a target arc $e_i$ is the set of \textbf{\emph{reciprocal arcs}} that we use to compute the reciprocity of $e_i$, %In other words, no hyperedge outside $R_{i}$ affects the reciprocity of  $e_i$. 
%Note that a reciprocal set may have an arbitrary cardinality, including $0$. 
%Unlike a pairwise graph, multiple arcs can be involved in measuring a single hyperarc's reciprocity. 
We use $r(e_{i},R_{i})$ to denote the \textbf{\emph{reciprocity of an arc}} $e_i$, %, where $0 \leq r(e_{i}|R_{i}) \leq 1$.
% \begin{equation}\label{eq:1}
%     \in [0,1]    
% \end{equation}
where the domain is $E \times 2^{E}$.\footnote{Note that all arcs in $R_{i}$ are used in computing the reciprocity of $e_{i}$, and thus it does not correspond to a search space.}
In graphs, a traditional reciprocity measure \citep{newman2002email} is defined as the proportion of arcs between nodes that point both ways, and if we assign $1$ to such an arc and $0$ to the others as reciprocity, the proportion is equivalent to the average reciprocity of arcs.
%the number of arcs pointing in both directions divided by the total number of arcs, i.e., 
Similarly, we regard, as the \textbf{\emph{reciprocity of a hypergraph $G$}}, the average reciprocity of arcs, i.e., %$r(G)=f( r(e_{1}|C_{1}), r(e_{2}|C_{2}), ..., r(e_{\lvert E \rvert }|C_{\lvert E \rvert }) )$, where $f(.)$ can be any aggregation function. Here, we used $f$ as a simple average.
\begin{equation} \label{eq:hypergraphreciprocity}
    r(G) := \frac{1}{\lvert E \vert } \sum\nolimits_{i=1}^{\lvert E \vert}r(e_{i}, R_{i}).
\end{equation}

\smallsection{Motivations of axioms}
What are the characteristics required for $r(e_{i} , R_{i})$ and $r(G)$?
We introduce eight axioms that any reasonable measure of $r(e_{i}, R_{i})$ (\axioms 1-5) and $r(G)$ (\axioms 6-8) should satisfy. {blue}
\kijung{We first provide the motivation and necessity of the proposed axioms.}
\begin{itemize}
    \item \textbf{Incremental changes}: Understanding when a value of a measure increases (decreases) helps users to understand how the measure works and have faith in the values the measure returns. Without this understanding, one cannot trust the measure, and this unreliability towards a measure may lead to misinterpretation of the measured value. 
    Thus, we propose \textsc{\textbf{Axioms}~\ref{ax:axiom1}-\ref{ax:axiom4}} (and their generalized axioms) to describe the cases where the value of a reciprocity measure increases (decreases).
    \item \textbf{Boundness}: Establishing a finite range for a measure helps an intuitive comprehension of the numerical extent of a characteristic. 
    For instance, if a measure does not lie in a fixed
    range, it becomes challenging to readily ascertain whether a particular hypergraph is reciprocal or not.  
    Furthermore, a measure with a defined finite range enhances the ability to make meaningful comparisons across diverse hypergraphs.
    Motivated by this intuition, we propose \textsc{\textbf{Axiom}}~\ref{axiom5} and \textsc{\textbf{Axiom}}~\ref{axiom7}, which suggest the bound of
    hyperarc and hypergraph reciprocity measures, respectively.
    \item \textbf{Reducibility}: Reciprocity in an ordinary directed graph is a well-known statistic that is widely used in various fields of study~\citep{nguyen2010you, hidalgo2008dynamics, newman2002email} (see Section~\ref{section:intro} for details). 
Since a directed hypergraph is a generalization of an ordinary directed graph, one would expect that a directed hypergraph reciprocity measure should be equivalent to the common directed graph reciprocity when applied to any hypergraph containing only hypercars with head sets and tail sets of size 1 (i.e., directed graph. where $\vert H_{i} \vert = \vert T_{i} \vert = 1, \forall \langle H_{i}, T_{i}\rangle \in E$). 
Thus, we propose \textsc{\textbf{Axiom}~\ref{axiom6}}, which suggests this characteristic
    \item \textbf{Reachability}: Identifying whether the upper bound of a measure is truly achievable or not plays a crucial role in ensuring the reliability of the measure's range and the accurate interpretability of its returned value. 
For example, let's consider a reciprocity measure with a known upper bound of 1, but it can actually only reach the value of 0.2.
In this scenario, if a particular hypergraph achieves the reciprocity value of 0.2, which is actually the maximum possible value for a hypergraph, one may think the corresponding hypergraph is not highly reciprocal, since the known upper bound of 1.
%Consequently, a user may distrust the suggested range of the measure, and misinterpret the measure's value.
Thus, we propose \textsc{\textbf{Axiom}~\ref{axiom8}} to formalize the reachability of the maximum reciprocity value.
\end{itemize}

\smallsection{Details of axioms} In \axioms 1-4, we compare the reciprocity of two target arcs $e_{i}$ and $e_{j}$ whose reciprocal sets are $R_{i}$ and $R_{j}$, respectively.
% intersection of target arc's head set and candiate set arcs' tail set (i.e., $H_i \cap T'_k \ e_k \in R_i$), and target arc's tail set and reciprocal set arcs' head set(i.e., $T_i \cap H'_k \ e_k \in R_i$).
Moreover, in \axiom 2-4, we commonly assume two target arcs $e_{i}$ and $e_{j}$ are of equal size (i.e., $\lvert H_{i}\vert  = \lvert H_{j}\vert $ and $\lvert T_{i} \vert = \lvert T_{j} \vert $). 
Here, we say two arcs $e_i$ and $e_k\in R_i$ \textbf{\emph{inversely overlap}} if and only if $H_i \cap T_k \neq \emptyset$ and $T_i \cap H_k \neq \emptyset$. 
Below, the statements in \axioms 1-4 are limited to the  examples in Figure~\ref{fig:AXIOMs} for simplicity.
%\footnote{\modify{Although the statements in \axioms 1-4: ``In Figure~\ref{fig:AXIOMs}(x), $r(e_{i}, R_{i}) < r(e_{j},R_{j})$ should hold'' are redundant, they are crucial for the \textit{formality} of each axiom. Removing these statements would leave axioms with only high-level verbal descriptions of certain characteristics, and they are insufficient for the axioms to maintain their formal nature.}} 
\kijung{Each statement of \axioms \textbf{1-4} is generalized in \textbf{\textsc{Generalized Axioms 1-4}}.}
%The statements in \axioms 1-4, however, are generalized and formalized in Appendix \ref{subsec:thm1}.
%they should satisfy. 

% From \textbf{\textsc{Axiom1}} to \textbf{\textsc{Axiom6}} are regarding the hyperarc reciprocity (i.e., $r(e_{i}|R_{i})$).
% They explain which hyperarc is more reciprocal than the other among a given pair of cases.
% Here, we consider that reciprocal set($R_{i}$) for each target arc($e_{i}$) is given. 
% After, \textbf{\textsc{Axiom7}} and \textbf{\textsc{Axiom8}} are defined in hypergraph level, necessary natures of hypergraph reciprocity ($r(G)$). 
% General idea of following \textbf{\textsc{Axioms}} is described in figure \ref{fig:AXIOMs}. 
%, while \textbf{\textsc{Axiom1}} does not have any size constraint.

%% Professor Comments: Summarize axiom in one sentence

\begin{figure*}
    \vspace{-2mm}
    \centering
    \includegraphics[width=1\textwidth]{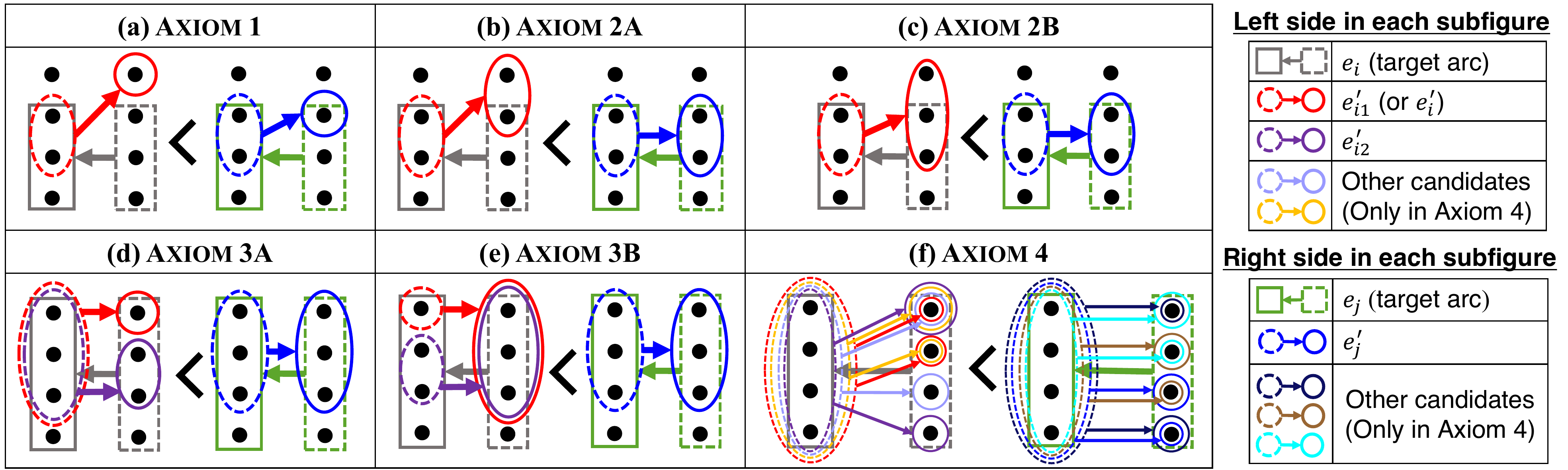}
    \caption{Examples for \Axioms 1-4.  
    In each subfigure, the reciprocity of the arc $e_i$ on the left side should be smaller than that of the arc $e_j$ on the right side. 
    This inequality holds by \measure (see Section~\ref{sec:measure:measure}) in all subfigures. Specifically, if $\alpha=1$, 
    {$r(e_i)$} \& {$r(e_j)$} are {$0.0000$} \& {$0.3605$} in (a), {$0.2697$} \& {$0.5394$} in (b), {$0.4444$} \& {$0.5394$} in (c), {$0.3167$} \& {$0.6466$} in (d), {$0.3233$} \& {$0.6466$} in (e), and {$0.2347$} \& {$0.2500$} in (f). 
    }
    \label{fig:AXIOMs}
\end{figure*}

\begin{axiombox}[Existence of Inverse Overlap]\label{ax:axiom1}%[No Inverse-Overlap]
In Figure~\ref{fig:AXIOMs}(a), $r(e_i , R_i) < r(e_j , R_j)$ should hold.
Roughly, an arc with at least one inverse-overlapping reciprocal arc is more reciprocal than an arc with no inverse-overlapping reciprocal arcs. 
% Original Foot
%\textcolor{white}{\footnote{\scriptsize \label{footnote:axiom1}Assume $R_i = \{e'_i\}$ and $R_j = \{e'_j\}$. Then, $\bm{\min(|H_i \cap T'_i|,|T_i\cap H'_i|)}$ $\bm{= 0}$ and $\bm{\min(|H_j \cap T'_j|, |T_j \cap H'_j|) \geq 1}$ imply $r(e_i | R_i) < r(e_j | R_j)$.}} 

%% Below is the initial expression
% \textcolor{white}{\footnote{\scriptsize \label{footnote:axiom1}Assume $R_i = \{e'_i\}$ and $R_j = \{e'_j\}$. Then, $\mathbf{\min(|H_i \cap T'_i|,|T_i\cap H'_i|)}$ $\mathbf{= 0}$ and $\mathbf{\min(|H_j \cap T'_j|, |T_j \cap H'_j|) \geq 1}$ imply $r(e_i | R_i) < r(e_j | R_j)$.}} 

%For simplicity,
% $$R_i = \{e'_i\} \text{ and } $\min(|H_i \cap T'_i|, |T_i \cap H'_i|) = 0 and (b) $R_j = \{e'_j\}$ where $\min(|H_j \cap T'_j|, |T_j \cap H'_j|) \geq 1$, then $r(e_i | R_i) < r(e_j | R_j)$$
\end{axiombox}
\vspace{-2mm}
{
\begin{newaxiombox}[Existence of Inverse Overlap]{
\label{newaxiom:inverse:overlap}
Consider two arcs $e_{i}$ and $e_{j}$.
If $R_{i}$ and $R_{j}$ satisfy
\begin{align*}
(i) \ \forall& e'_{i} \in R_{i}: \min(\lvert H_i \cap T'_i \vert , \lvert T_i \cap H'_i \vert ) = 0,  \\ 
(ii) \ \exists& e'_{j} \in R_{j}: \min(\lvert H_j \cap T'_j \vert , \lvert T_j \cap H'_j \vert ) \geq 1,
\end{align*}
then the following inequality holds:
$$
r(e_i , R_i) < r(e_j , R_j).
$$
}
\end{newaxiombox}
}

\begin{axiombox}[Degree of Inverse Overlap]%[More Inverse-Overlap (Intersection)]
In Figures~\ref{fig:AXIOMs}(b-c), $r(e_i , R_i) < r(e_j , R_j)$ should hold.
Roughly, an arc that inversely overlaps with reciprocal arcs to a greater extent (with a larger intersection and/or with a smaller difference, {which are considered separately in the generalized axioms}) is more reciprocal.
%\textcolor{white}{\tiny \footnote{\label{footnote:axiom2} \scriptsize Assume $R_i = \{e'_i\}$, $R_j = \{e'_j\}$, $|H'_i| = |H'_j|$, and $|T'_i| = |T'_j|$. Then, $\bm{0< |H'_i \cap T_i| < |H'_j \cap T_j|}$ and $\bm{0<|T'_i\cap H_i| < |T'_j \cap H_j|}$ imply $r(e_i | R_i) < r(e_j | R_j)$.}} \textcolor{white}{\tiny \footnote{\label{footnote:axiom2prime} \scriptsize Assume $R_i = \{e'_i\}$, $R_j = \{e'_j\}$, $|H'_i \cap T_i| = |H'_j \cap T_j|>0$, $|T'_i \cap H_i| = |T'_j \cap H_j| >0$, and $|T'_i| = |T'_j|$. Then, $\bm{|H'_i| > |H'_j|}$ implies $r(e_i | R_i) < r(e_j | R_j)$.}} % than an arc with less inversely overlapping reciprocal arcs. 
%% Below is the initial expression
% \textcolor{white}{\tiny \footnote{\label{footnote:axiom2} \scriptsize Assume $R_i = \{e'_i\}$, $R_j = \{e'_j\}$, $|H'_i| = |H'_j|$, and $|T'_i| = |T'_j|$. Then, $\mathbf{0< |H'_i \cap T_i| < |H'_j \cap T_j|}$ and $\mathbf{0< |T'_i \cap H_i| < |T'_j \cap H_j|}$ imply $r(e_i | R_i) < r(e_j | R_j)$.}}
\end{axiombox}
% \modify{
% Below, we generalize
% \axiomnum{2} by dividing it into two cases:
% \begin{itemize}
%     \item \textbf{\textsc{Axiom 2A:}}
%     An arc that inversely overlaps with reciprocal arcs with a \textbf{larger intersection} is more reciprocal,
%     \item \textbf{\textsc{Axiom 2B:}} An arc that inversely overlaps with reciprocal arcs with a \textbf{smaller difference} is more reciprocal. 
%     %$e'_{j}$ inverse overlap with $e_{j}$ more accurately than $e'_{i}$ inverse overlap with $e_{i}$.
% \end{itemize}
%Below, we separately generalize and prove \blue{the} two sub-axioms.

\kijung{
\begin{subaxiom2}[Degree of Inverse Overlap: More Overlap]
\label{newaxiom2a}
Consider two arcs  $e_{i}$ and $e_{j}$ of the same size $($i.e., ${(\lvert H_{i} \vert  = \lvert H_{j} \vert )} \wedge {(\lvert T_{i} \vert = \lvert T_{j} \vert )})$.
If $R_i = \{e'_i\}$ and $R_j = \{e'_j\}$ satisfy
\begin{gather*}
\lvert H'_i \vert = \lvert H'_j \vert, \ \lvert T'_i \vert = \lvert T'_j \vert, \ and \\
((i) \quad 0 < \lvert H'_i \cap T_i \vert < \lvert H'_j \cap T_j \vert \text{ and } 0 < \lvert T'_i \cap H_i \vert \leq \lvert T'_j \cap H_j \vert \quad \text{or} \\
(ii) \quad 0 < \lvert H'_i \cap T_i \vert \leq \lvert H'_j \cap T_j \vert \text{ and } 0 < \lvert T'_i \cap H_i \vert < \lvert T'_j \cap H_j \vert),
\end{gather*}
then the following inequality holds:
$$
r(e_i , R_i) < r(e_j , R_j).
$$
\end{subaxiom2}
\begin{subaxiom2}[Degree of Inverse Overlap: Small Difference]
\label{subaxiom2b}{
Consider two arcs  $e_{i}$ and $e_{j}$ of the same size $($i.e., ${(\lvert H_{i} \vert  = \lvert H_{j} \vert )} \wedge {(\lvert T_{i} \vert = \lvert T_{j} \vert)})$.
If $R_i = \{e'_i\}$ and $R_j = \{e'_j\}$ satisfy
\begin{gather*}
\lvert H'_i \vert > \lvert H'_j \vert, \quad \lvert T'_i \vert = \lvert T'_j \vert, \quad 0 < \lvert H'_i \cap T_i \vert = \lvert H'_j \cap T_j \vert, \ \text{and} \ 0 < \lvert T'_i \cap H_i \vert = \lvert T'_j \cap H_j \vert,
\end{gather*}
then the following inequality should hold:
$$
r(e_i , R_i) < r(e_j , R_j).
$$
}
\end{subaxiom2}
}
%% Below is the initial expression
% \textcolor{white}{\tiny \footnote{\label{footnote:axiom2prime} \scriptsize Assume $R_i = \{e'_i\}$, $R_j = \{e'_j\}$, $|H'_i \cap T_i| = |H'_j \cap T_j|>0$, $|T'_i \cap H_i| = |T'_j \cap H_j| >0$, and $|T'_i| = |T'_j|$. Then, $\mathbf{|H'_i| > |H'_j|}$ %, and (d) $|T'_i \cap H_i| < |T'_j \cap H_j|$. 
% implies $r(e_i | R_i) < r(e_j | R_j)$.}}
% \end{axiombox}

% \begin{axiombox}[Small Difference]%[More Inversely Overlap]
% An arc with more (with small difference) inversely overlapping reciprocal arcs is more reciprocal than an arc with less inversely overlapping reciprocal arcs.
% Simply, assume $R_i = \{e'_i\}$, $R_j = \{e'_j\}$, $|H'_i \cap T_i| = |H'_j \cap T_j|>0$, $|T'_i \cap H_i| = |T'_j \cap H_j| >0$, and $|T'_i| = |T'_j|$. If $\mathbf{|H'_i| > |H'_j|}$, as in Figure~\ref{fig:AXIOMs}(c), %, and (d) $|T'_i \cap H_i| < |T'_j \cap H_j|$. 
% then $r(e_i | R_i) < r(e_j | R_j)$ should hold.
% \end{axiombox}

\begin{axiombox}[Number of Reciprocal Arcs]
In Figures~\ref{fig:AXIOMs}(d-e), $r(e_i , R_i) < r(e_j , R_j)$ should hold.
%Roughly, %when two arcs inversely overlap equally with their reciprocal sets, an arc with fewer reciprocals is more reciprocal than the other. 
%an arc requiring fewer reciprocal arcs to inversely overlap to the same extent is more reciprocal.
\kijung{Roughly, when two arcs inversely overlap equally with their reciprocal sets, an arc with a single reciprocal arc is more reciprocal than one with multiple reciprocal arcs.}
\end{axiombox}

Below, we generalize \axiomnum{3} by dividing it into two cases.
\kijung{Although the below two statements compare an arc with a single reciprocal arc and an arc with exactly two reciprocal arcs, these statements can be further extended to encompass a comparison of the former and an arc with two or more reciprocal arcs. These extended statements remain valid for our proposed measure (refer to \textsc{\textbf{Remark~\ref{remark:canbegeneral}}} in Appendix~\ref{section:theoremproof}).}

\begin{subaxiom3}[Number of Reciprocal Arcs Differs: Identical Tail Sets]\label{subaxiom3a}
Let $e'_{k} \subseteq_{(R)} e_{k}$ indicate $H'_{k} \subseteq T_{k}$ and $T'_{k} \subseteq H_{k}$.
Consider two arcs  $e_{i}$ and $e_{j}$ of the same size $($i.e., ${(\lvert H_{i} \vert  = \lvert H_{j} \vert )} \wedge {(\lvert T_{i} \vert = \lvert T_{j} \vert)})$.
If $R_i = \{e'_{i1}, e'_{i2}\}$ and $R_j = \{e'_j\}$ satisfy 
\begin{gather*}
e'_{i1} \subseteq_{(R)} e_{i}, \quad e'_{i2} \subseteq_{(R)} e_{i}, \quad e'_{j} \subseteq_{(R)} e_{j}, \quad 
T'_{i1} = T'_{i2}, \quad \lvert T'_{i1} \vert = \lvert T_{j} \vert, \\ 
H'_{i1} \cap H'_{i2} = \emptyset, \quad \text{and} \quad 
\lvert (H'_{i1} \cup H'_{i2}) \cap T_{i} \vert = \lvert H'_{j}\cap T_{j} \vert,
\end{gather*}
then the following inequality should hold: 
$$
r(e_i , R_i) < r(e_j , R_j).
$$
\end{subaxiom3}

\begin{subaxiom3}[Number of Reciprocal Arcs: Identical Head Sets]\label{subaxiom3b}{
Let $e'_{k} \subseteq_{(R)} e_{k}$ indicate $H'_{k} \subseteq T_{k}$ and $T'_{k} \subseteq H_{k}$.
Consider two arcs  $e_{i}$ and $e_{j}$ of the same size $($i.e., ${(\lvert H_{i} \vert  = \lvert H_{j} \vert )} \wedge {(\lvert T_{i} \vert = \lvert T_{j} \vert)})$.
If $R_i = \{e'_{i1}, e'_{i2}\}$ and $R_j = \{e'_j\}$ satisfy
\begin{gather*}
e'_{i1} \subseteq_{(R)} e_{i}, \quad e'_{i2} \subseteq_{(R)} e_{i}, \quad e'_{j} \subseteq_{(R)} e_{j}, \quad 
H'_{i1} = H'_{i2}, \quad \lvert H'_{i1} \vert = \lvert H_{j} \vert , \\ 
T'_{i1} \cap T'_{i2} = \emptyset, \quad and \quad \lvert (T'_{i1} \cup T'_{i2}) \cap H_{i} \vert = \lvert T'_{j}\cap H_{j} \vert,
\end{gather*}
then the following inequality should hold: 
$$
r(e_i , R_i) < r(e_j , R_j).
$$
}
\end{subaxiom3}
\color{black}

\begin{axiombox}[Bias]\label{ax:axiom4} In Figure~\ref{fig:AXIOMs}(f), $r(e_i , R_i) < r(e_j , R_j)$ should hold.
Roughly, when two arcs inversely overlap \textbf{equally} with their reciprocal sets, an arc whose reciprocal arcs are equally reciprocal to all nodes in the arc is more reciprocal than one with reciprocal arcs \textbf{biased} towards a subset of nodes in the arc.
\end{axiombox}

\setcounter{newaxiomfull}{3}
\setcounter{newpropfull}{3}
\begin{newaxiomfull}[Bias]
\label{subaxiom4}{
Consider two arcs  $e_{i}$ and $e_{j}$ of the same size $($i.e., ${(\lvert H_{i} \vert  = \lvert H_{j} \vert )} \wedge {(\lvert T_{i} \vert = \lvert T_{j} \vert)})$.
If $R_{i}$ and $R_{j}$ satisfy
\begin{align}
      (i) & \quad   \lvert R_{i} \vert  = \lvert R_{j} \vert  = \lvert T_{i} \vert = \lvert T_{j} \vert, \quad \text{and} \quad \lvert T'_{i} \vert = \lvert T'_{j} \vert,
    \quad \forall e'_{i} \in R_{i}, \quad \forall e'_{j} \in R_{j}, \nonumber \\
      (ii) & \quad  T'_{i} = H_{i}, \quad H'_{i} \subset T_{i}, \quad \text{and} \quad \lvert H'_{i} \vert = 2, \quad \forall e'_{i} \in R_{i}, \nonumber  \\
      (iii) & \quad  T'_{j} = H_{j}, \quad H'_{j} \subset T_{j}, \quad \text{and} \quad \lvert H'_{j} \vert = 2, \quad \forall e'_{j} \in R_{j}, \nonumber \\
 (iv) & \quad \exists u,v \in T_{i} \quad \text{where} \quad \lvert  \{u\in H'_{i}  \ \mid \ e'_{i}\in R_{i} \} \vert \neq \lvert \{v\in H'_{i} \ \mid \ e'_{i}\in R_{i} \} \vert \label{eq:axiom4a}\\
 (v) & \quad  \forall u,v \in T_{j} \quad \text{where} \quad \lvert \{u\in H'_{j} \ \mid \ e'_{j}\in R_{j} \} \vert = \lvert \{v \in H'_{j} \ \mid \ e'_{j}\in R_{j}\} \vert, \label{eq:axiom4b}
\end{align}
then the following inequality should hold: 
$$
r(e_i , R_i) < r(e_j , R_j).
$$
}
\end{newaxiomfull}

% The reciprocity of any hypergraph should be within a fixed finite range.
% \kijung{Without loss of generality, we assume that the reciprocity values are within the range of $[0, 1]$. Therefore, for any hypergraph $G$, the reciprocity function $r : G \mapsto [0, 1]$ should hold.}

\begin{axiombox}[Upper and Lower Bounds] 
\label{axiom5}The reciprocity of any arc should be within within a fixed finite range.
\kijung{Without loss of generality, we assume that the reciprocity values are within the range of $[0, 1]$, 
i.e., for every $e_{i} \in E$ and $R_{i} \in 2^{E}$,  $r : E \times 2^{E} \mapsto [0, 1]$ should hold.
Note that any fixed finite range can be re-scaled to $[0,1]$.
}

\end{axiombox}
 
Now, we present the axioms defined at the hypergraph level.

\begin{axiombox}[Inclusion of Graph Reciprocity] 
\label{axiom6}
The graph reciprocity \citep{newman2002email} should be included as a special case.
That is, if $G$ is a graph (i.e., $\lvert H_i \vert = \lvert T_i \vert = 1, \forall i\in\{1, \cdots, 
\lvert E \vert \}$), then the following equality should hold:
\begin{equation}
    r(G)={\lvert E^{\leftrightarrow}\vert} / {\lvert E \vert }, \label{eq:graph_reciprociry}
\end{equation}
where $E^{\leftrightarrow}$ is the set of arcs between nodes that point each other in both directions.
%equal to the number of arcs in $G$ divded by the and $L^{\leftrightarrow}$ is the number of mutually linked arcs.
\end{axiombox}

% \frac{L^{\leftrightarrow}}{L}$ where $L$ is 
\begin{axiombox}[Upper and Lower Bounds] 
\label{axiom7}
The reciprocity of any hypergraph should be within a fixed finite range.
\kijung{Without loss of generality, we assume that the reciprocity values are within the range of $[0, 1]$, i.e., for any hypergraph $G$, the reciprocity function $r : G \mapsto [0, 1]$ should hold.
Note that any fixed finite range can be re-scaled to $[0,1]$.
}
%Specifically,
%for any hypergraph $G$, $r: G \mapsto [0, 1]$ should hold.
\end{axiombox}

\begin{axiombox}[\kijung{Reachability of Bounds}] 
\label{axiom8}
The maximum reciprocity, which is $1$ by \textbf{\Axiom~\textsc{\ref{axiom7}}}, should be reachable from any hypergraph by adding specific arcs. That is, for
every $G=(V,E)$, there exist $G^{+}=(V,E^{+})$ with $E^{+}\supseteq E$ such that $r(G^{+}) = 1$.
\kijung{Similarly, the minimum reciprocity, which is $0$ by \textbf{\Axiom~\textsc{\ref{axiom7}}}, should be reachable from any hypergraph by removing specific arcs. That is, for
every $G=(V,E)$, there exist $G^{-}=(V,E^{-})$ with $E^{-}\subseteq E$ such that $r(G^{-}) = 0$.}

\end{axiombox}

Note that \kijung{\textbf{\Axiom~\textsc{\ref{axiom8}}} is about whether the maximum and minimum values of a reciprocity measure are attainable or not (i.e., whether its bounds are tight or not), and it does not specify the situation when the value of a measure is maximized or minimized.}

\subsection{Proposed Measure of Hypergraph Reciprocity: \measure}
 \label{sec:measure:measure}

% In this subsection, we introduce a proposed \measure.

We propose \measure, \kijung{a family of principled hypergraph-reciprocity measures} based on transition probability. %, which we define first.

\smallsection{Transition Probability}
%\textbf{\emph{Reaching probability}} is a probability distribution which plays a key role in our proposal. 
For a target arc $e_i = \langle H_i, T_i \rangle$ and its reciprocal arcs $R_i$, the \textbf{\emph{transition probability}} $p_{h}(v)$ from a head set node $v_{h} \in H_{i}$ to each node $v$ is the probability of a random walker transiting from $v_{h}$ to $v$ when she moves to a uniform random tail-set node of a uniform random arc among the reciprocal arcs incident to $v_h$.
{For example, consider a target arc $e_1$ and a reciprocal set $R_{1} = \{e_{2}, e_{3} \}$ in Figure~\ref{fig:realidealfig}(a).
For a head set node $v_2$ of the target arc, the reciprocal arcs incident to it are $\{e_{2}, e_{3}\}$. 
The node $v_7$ is only in the head set of $e_2$, and thus the transition probability $p_2(v_7)=P(e_2 \mid \{e_{2}, e_{3}\})\times P(v_7 \mid H_2)=0.5\times0.5 =0.25$.
Similarly, since the node $v_6$ is in the head set of both $e_2$ and $e_3$, $p_2(v_6)=P(e_2 \mid \{e_{2}, e_{3}\})\times P(v_6 \mid H_2)+P(e_3 \mid \{e_{2}, e_{3}\})\times P(v_6 \mid H_3)=0.5\times0.5+0.5\times0.5 =0.5$.
Since, $v_8$ is not in tail set of any reciprocal arc, $p_2(v_8)=0$.
For a head set node $v_3$ of the target arc, the only reciprocal arc incident to it is $e_{2}$. 
Since the node $v_6$ is in the head set of $e_2$, $p_3(v_6)=P(e_2 \mid \{e_{2}\})\times P(v_6 \mid T_2)=1.0\times0.5 =0.5$.}
\returncolor

\begin{figure}[t]
    \centering
    \subfigure[Non-optimal case]{\includegraphics[width=0.22\textwidth]{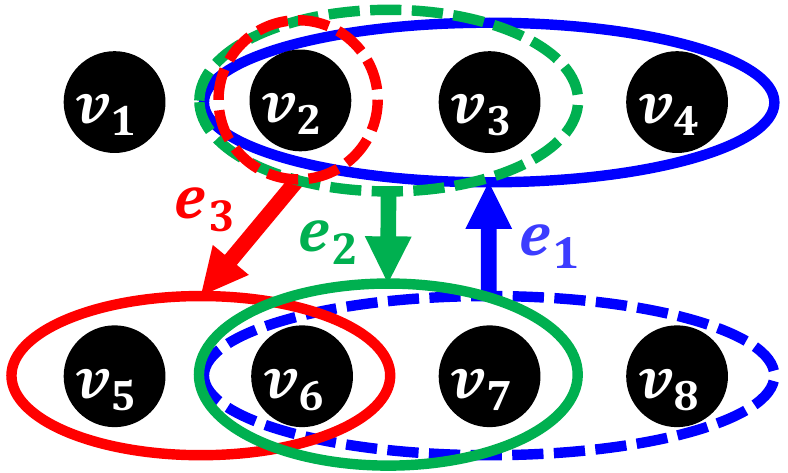}}
    \hspace{3mm}
    \subfigure[Optimal]{\includegraphics[width=0.22\textwidth]{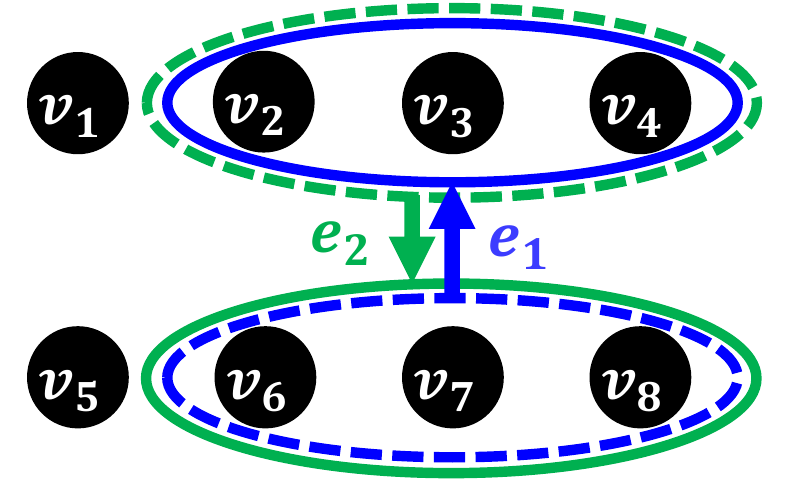}}
    \caption{\label{fig:realidealfig}  {Examples for describing the computation of transition probabilities.} \returncolor}
\end{figure}

% As we only consider paths that departure from target arc's head set nodes, 
There might be some head set nodes that are not incident to any reciprocal arc. For example, $v_{4}$ in Figure \ref{fig:realidealfig}(a) is such a node when the target arc is $e_1$ and the reciprocal set is $R_{1} = \{e_{2}, e_{3} \}$.
We assume that, from such a node, the random walker always transits to the virtual \textbf{\emph{sink node} $v_{sink}\notin V$.}

% Some of the head-set nodes may not be covered by reciprocal arcs (e.g., ), and their reaching probability distributions are not properly defined.
% For such a node $v$, we introduce a \emph{sunken node} $v_{sink}$, an arbitrary node that does not belong to the target arc's tail set $T_{i}$, and assume a virtual arc from $v$ to $v_{sink}$. 
% % This is the shown as the edge from $v_4$ to $v_1$ in the figure.
% This makes $v$ have zero reaching probabilities to the tail-set nodes with a valid probability distribution.

Then, for each head set node $v_h\in H_i$ of a target arc $e_i$, a \textbf{\emph{transition probability distribution}} over $V\cup\{v_{sink}\}$ is defined, and we use $p_h$ to denote it.
We also denote an \textbf{\emph{optimal transition probability distribution}} by $p^*_h$, which is a transition probability distribution when the perfectly reciprocal arc $e^{*}_{i} = \langle H^{*}_{i} = T_{i} , T^{*}_{i} = H_{i} \rangle$ is assumed as the reciprocal arc of $e_i$, i.e., $R_i=\{e^*_i\}$. For example, in Figure \ref{fig:realidealfig}(b), $e_2$ is the perfectly reciprocal arc of $e_1$. The following equality always holds:
\begin{equation*}
    p^{*} _{h}(v) = \begin{cases}
    	\frac{1}{\lvert T_{i} \vert } & \textrm{if } v \in T_{i}, \\
        0 & \textrm{otherwise.}
        \end{cases}
\end{equation*}

\smallsection{Proposed Measures}
Based on the above concepts, we propose \measure (\underline{\smash{Hyper}}graph \underline{\smash{Rec}}iprocity) as a \kijung{family of} principled measures of hypergraph reciprocity.
We notice that reciprocal arcs in a graph lead to paths of length two that start and end at the same node. Thus, intuitively, in a hypergraph, a target arc should become more reciprocal if its reciprocal arcs allow for heading back to the head-set nodes of the target arc more ``accurately''.
In order to measure numerically the accuracy for a target arc $e_i$, we compare the transition probability distribution $p_h$ from each head-set node $v_h\in H_i$ with the optimal distribution $p^*_h$.
% Let the general forms of probability distributions $p_{h}(v)$ and $p^{*}_{h}(v)$ for all target nodes be $p _{h}$ and $p^{*} _{h}$, respectively. 
% In order to decide how much a target arc is reciprocal with a given reciprocal set, we measure the \textbf{\emph{probabilistic distance between the observed reaching probability distribution $\bm{p_{h}}$ and the optimal reaching probability distribution $\bm{p^{*}_{h}}$}}.
% Our main idea is that a node is more reciprocal if its reaching probability distribution is closer to the optimal distribution.
% By using appropriate distance function $\mathcal{L}(p, q)$, we can measure probabilistic distance between observed and optimal reaching probabilities.

While any distance function $\mathcal{L}$ can be used to quantify the difference between $p_h$ and $p^*_h$, we use the \emph{Jensen-Shannon Divergence} (JSD) \citep{lin1991divergence} since it is a symmetric measure that can handle zero mass in both distributions.
The JSD between distributions $p$ and $q$ with domain $D$ is defined as

$$
    \mathcal{L}(p,q) := \sum_{i\in D}
        \left({\frac{p(i)}{2} \log{\frac{2p(i)}{p(i) + q(i)}}
        + \frac{q(i)}{2} \log{\frac{2q(i)}{p(i) + q(i)}}}\right).
$$
% \begin{multline*}
%     \mathcal{L}(p_h,p^*_h)= \sum_{v_i\in V\cup\{v_{sink}\}}
%         \left({\frac{p_h(v_i)}{2} \log{\frac{2p_h(v_i)}{p_h(v_i) + p^*_h(v_i)}} \\
%         + \frac{p^*_h(v_i)}{2} \log{\frac{2p^*_h(v_i)}{p_h(v_i) + p^*_h(v_i)}}}\right).
% \end{multline*}
%% Professor comment: Reason of JSD
% \begin{figure}
%     \centering
%     \includegraphics[width=0.45\textwidth]{FIG/COMPOSECi/FIG4_Ci.pdf}
%     \caption{Comparison between different approaches to compose the reciprocal set $C_4$ of an arc $e_{4}$: (a) a hypergraph and (b) how baselines \textbf{B6 (All hyperarcs) and B7 (Inverse-overlap hyperarcs)} and \measure form $C_4$.  \red{TODO}}
%     \label{fig:composereciprocal}
% \end{figure}
%\textbf{\textsc{Probabilistic Distance Reciprocity:}}
Based on $\mathcal{L}$, we define \measure  of an arc $e_{i}$ whose reciprocal set is $R_{i}$ as 
\begin{equation} \label{eq:proposal}
    r(e_i,R_i) :=  \left( \frac{1}{\lvert R_i \vert} \right)^\alpha \left(1 - \frac{\sum_{v_{h} \in H_i}
    \mathcal{L}(p_{h}, p^{*}_{h})}{\lvert H_i \vert \cdot \mathcal{L}_\mathrm{max} }\right),
\end{equation}
where $\alpha \in (0, 1]$ is a constant controlling the degree of penalization of a large reciprocal set, and $\mathcal{L}_\mathrm{max}$ is the maximum value of the distance measure $\mathcal{L}$, which is $\log 2$ for the JSD.
Note that the value of $r(e_i, R_i)$ becomes larger if $\mathcal{L}(p_{h}, p^{*}_{h})$ becomes small, implying that an arc is more reciprocal if its transition distribution becomes closer to the optimal distribution.

\kijung{Note that as the value of $\alpha$ increases, the penalty that occurs from the size of the reciprocal set (i.e., $\vert R_{i} \vert $) increases. 
This allows \measure to effectively capture the user's preferences regarding the impact of reciprocal set size on reciprocity.
For instance, in certain domains, users may believe that, under the same conditions, an arc with a larger number of reciprocal arcs is considerably less reciprocal compared to an arc with fewer reciprocal arcs. To reflect this belief, a higher value of $\alpha$ can be set.
However, in domains where the size of the reciprocal set is relatively less significant in relation to reciprocity, a smaller value of $\alpha$ can be employed. The flexibility of choosing $\alpha$ allows \measure to adapt to different scenarios and user preferences.}

\kijung{Since the value of $r(e_{i}, R_{i})$ varies depending on the specific choice of $\alpha$, \measure can be regarded as a family of hypergraph-reciprocity measures.
Throughout the remainder of the paper, we use \measure to denote any measure that falls within the \measure family, unless explicitly mentioned otherwise.}

% $\mathcal{L}(p_{h}, p^{*}_{h})$ indicates a probabilistic distance between observed and optimal reaching probability. As distance between two cases are getting lower, overall value of measure increases, which implies higher reciprocity of target arc. Note that \measure lies in $r(e_{i} | R_{i}) \in [0,1]$.

\smallsection{Composing Reciprocal Sets} 
The value of $r(e_i, R_i)$ is dependent on how we select the reciprocal set $R_i$ from the set $E$ of all arcs.
For each target arc $e_i$, we propose to choose non-empty $R_{i} \subseteq E$ that maximizes  the reciprocity $r(e_i, R_i)$ of $e_i$, i.e.,
\begin{equation} \label{eq:reciprocal-obj}
    R_i := \argmax_{R_i' \subseteq E, R_i'\neq \emptyset} {r(e_i , R_i')}.
\end{equation}
% That is, $R_i$ contains all necessary hyperarcs that are inversely overlapped with the target hyperarc $e_i$ to maximize $r(e_i | R_i)$.
In summary, according to \measure, the reciprocity of an arc $e_i\in E$ is
\begin{equation} \label{eq:final}
r(e_i):=\max_{R_i \subseteq E, R_i\neq \emptyset} r(e_i, R_i),
\end{equation} and by Eq.~\eqref{eq:hypergraphreciprocity}, the reciprocity of $G$ is $r(G) := \frac{1}{\lvert E \rvert} \sum_{i=1}^{\lvert E \rvert }r(e_{i})$.

\smallsection{Strengths of \measure}
\measure satisfies all proposed \axioms regardless of the value of $\alpha > 0$, as stated in Theorem~\ref{thm:axiom}, while none of the considered baseline measures, which are described below, does.

\begin{thm}[Soundness of \measure]  \label{thm:axiom}
\emph{\measure always satisfies} \Axioms \textsc{\textbf{\ref{ax:axiom1}-\ref{axiom8}}} and {\textsc{\textbf{Generalized Axioms}}
\textsc{\textbf{\ref{newaxiom:inverse:overlap}}},
\textsc{\textbf{\ref{newaxiom2a}}},
\textsc{\textbf{\ref{subaxiom2b}}},
\textsc{\textbf{\ref{subaxiom3a}}},
\textsc{\textbf{\ref{subaxiom3b}}}, and
\textsc{\textbf{\ref{subaxiom4}}}}
\end{thm}

\begin{proof}
See the caption of Figure~\ref{fig:AXIOMs} for the numerical values for the examples.
{Full proofs can be found in Appendix~\ref{subsec:thm1}.} \returncolor
\end{proof}
\kijung{Note that while any distance function can be utilized as $\mathcal{L}$ in Eq~\eqref{eq:proposal}, the selected function should enable \measure to satisfy all the axioms to ensure the soundness of \measure.}

\smallsection{Limitations of \measure} 
\kijung{One limitation of \measure is its high computational cost, as it involves finding maximum reciprocity value over $O(2^{\lvert E \rvert})$ potential reciprocal sets. This search, however, is necessary to satisfy the proposed axioms, as discussed in Section~\ref{sec:baselinemeasure}
In Section~\ref{sec:ferret}, we discuss methods to reduce the search space and thus mitigate this computational burden, without affecting the reciprocity value.
}

\subsection{\textbf{Baseline Approaches and Axiomatic Analysis}}\label{sec:baselinemeasure}

%% Professor Comment: Summarize subsection in one sentence.

Below, we present the baseline measures considered in our work.

% Professor Comment: High-level idea of each baseline -> Detail equations can be sent to appendix.

%\smallsection{B1. Pearcy et al. \cite{pearcy2014hypergraph}} %This approach was proposed in \cite{pearcy2014hypergraph}.

\smallsection{B1. \citet{pearcy2014hypergraph}} %This approach was proposed in
{Refer to the last paragraph of Section~\ref{section:prelim} for details of this measure.
This measure is defined only for the entire hypergraph. 
For the arc-level axioms (i.e., \axioms 1-5), we use the hypergraphs that consist only of arcs that are mentioned in each axiom. 
That is, we compare the reciprocity of $G_i = (V, \{e_i\} \cup R_i)$ and $G_j = (V, \{e_j\} \cup R_j)$.
} \returncolor

% \smallsection{B2. Weighted Clique Expansion Reciprocity} This method also uses clique-expanded adjacency matrix. But rather than using trace computation as did in \cite{pearcy2014hypergraph}, it directly applies weighted graph reciprocity measure proposed in \cite{squartini2013reciprocity}. 
% \vspace{-1mm}
% \begin{equation*}
%     r(G) = \frac{\sum_{i,j} ^{|V|} {min(A(G)_{ij} , A(G)_{ji})}}{\sum_{i,j} ^{|V|} {A(G)_{ij}}} \quad 
% \end{equation*}

\smallsection{B2. Ratio of Covered Pairs} 
{This measures, roughly, how accurately pair interactions within a target arc are matched with those within reciprocal arcs.
We define the pair interactions within $e_i$ as:
\begin{equation*}
K(e_{i}) = \{ \langle v_{h}, v_t \rangle \mid \ v_{h} \in H_i, v_t \in T_i \}.
\end{equation*}
We also define the set of inverse pair-interactions for $e_k$ as:
\begin{equation*}
K^{-1}(e_{k}) = \{ \langle v_t, v_{h} \rangle \mid \ v_{h} \in H_k, v_t \in T_k \}.
\end{equation*}
Then, inspired by the Jaccard Index, we define the measure as the ratio of identical members between $K(e_{i})$ and $K^{-1}(e_{k})$ for all $e_{k} \in R_{i}$ as follows:
\begin{equation*}
r(e_i, R_i) = \frac{\lvert K(e_{i}) \cap  \bigcup_{e_k \in R_{i}} K^{-1}(e_{k}) \vert }
{\lvert K(e_{i}) \cup  \bigcup_{e_k \in R_{i}} K^{-1}(e_{k}) \vert }.
\end{equation*}
} \returncolor

\smallsection{B3. Penalized Ratio of Covered Pairs} 
{This measure is an extension of \textbf{B2}, where large reciprocal sets are penalized as in \measure:
\begin{equation*}
r(e_i, R_i) = \left(\frac{1}{\lvert R_i \vert}\right)^{\alpha} \times \frac{\lvert K(e_{i}) \cap \bigcup_{e_k \in R_{i}} K^{-1}(e_{k}) \vert}
{\lvert K(e_{i}) \cup \bigcup_{e_k \in R_{i}} K^{-1}(e_{k}) \vert }.
\end{equation*}}\returncolor

{To demonstrate the necessity of the reciprocal-set penalty term ($(1/\lvert R_{i} \vert )^{\alpha}$) and the normalizing term $\lvert H_{i} \vert $ in Eq.~\eqref{eq:proposal}, we consider two variants of \measure where these two terms are removed respectively from Eq.~\eqref{eq:proposal}.} \returncolor

\smallsection{B4. \measure w/o Normalization} {This measure is a variant of \measure where the normalization by $\lvert H_{i} \vert$ is removed from Eq.~\eqref{eq:proposal} as follows:
\begin{equation*}
% \resizebox{0.48\textwidth}{!}{$ 
r(e_i , R_i) = \left(\frac{1}{\lvert R_i \vert}\right)^{\alpha} \left(\lvert H_{i} \vert -
        {\frac{\sum_{v_{h} \in H_i}\mathcal{L}(p_{h}, p^{*}_{h})}{\mathcal{L}_\mathrm{max}}} \right)
        % $}
\end{equation*}} \returncolor

\smallsection{B5. \measure w/o Size Penalty}
{This measure is a variant of \measure where the reciprocal-size penalty term $(1/\lvert R_{i} \vert)^{\alpha}$ is removed from Eq.~\eqref{eq:proposal} as follows:
\begin{equation*}
    r(e_i , R_i) = 1 - \frac{\sum_{v_{h} \in H_i}\mathcal{L}(p_{h}, p^{*}_{h})}{\lvert H_i \vert \cdot \mathcal{L}_\mathrm{max}}
\end{equation*}} \returncolor

\begin{table}% [ht]
    \caption{\label{tab:AXIOMsatisfy1} {\measure satisfies all axioms, while all the others do not.} \returncolor} % title of Table
    \centering
    \subfigure[Arc-level Axioms (B6 and B7 are exactly the same with \measure regarding arc-level reciprocity)]{
    \centering
    \scalebox{1}{
    \renewcommand{\arraystretch}{1.2}{
    \begin{tabular}{l c c c c c} % centered columns (4 columns)
        \toprule
         & \multicolumn{5}{c}{\axioms} \\
        \textbf{\textsc{Measure}} & \textbf{\textsc{1}} & \textbf{\textsc{2}} & \textbf{\textsc{3}} 
        & \textbf{\textsc{4}} & \textbf{\textsc{5}}\\
        \midrule
        B1 (\citet{pearcy2014hypergraph})  & \satisfy & \satisfy & \wrong & \satisfy & \satisfy \\ 
        B2 (Ratio of Covered Pairs) & \satisfy & \satisfy & \wrong & \wrong & \satisfy \\ 
        B3 (Penalized Ratio of Covered Pairs) & \satisfy & \satisfy & \satisfy & \wrong & \satisfy \\ 
        B4 (\measure w/o Normalization) & \satisfy & \satisfy & \satisfy & \satisfy & \wrong \\ 
        B5 (\measure w/o Size Penalty) & \satisfy & \satisfy & \wrong & \satisfy & \satisfy \\ 
        \midrule
        \measure (proposed) & \satisfy & \satisfy & \satisfy & \satisfy & \satisfy\\ 
        \bottomrule
    \end{tabular}
    }
    }
    }
    \subfigure[Hypergraph-level Axioms (B2-B5 are not applicable)]{
    \centering
    \scalebox{1}{
    \hspace{-3.5mm}
    \renewcommand{\arraystretch}{1.2}{
    \begin{tabular}{l c c c} % centered columns (4 columns)
        %\toprule
        \toprule
         & \multicolumn{3}{c}{\axioms} \\
        \textbf{\textsc{Measure}} & \textbf{\textsc{6}} & \textbf{\textsc{7}} & \textbf{\textsc{8}}\\
        \midrule
        B1 (\citet{pearcy2014hypergraph}) & \wrong  & \satisfy & \satisfy \\ 
        B6 (\measure w/ All Arcs as Reciprocal Set) & \wrong & \satisfy & \wrong\\ 
        B7 (\measure w/ Inversely Overlapping Arcs as Reciprocal Set) & \satisfy & \satisfy & \wrong\\ 
        \midrule
        \measure (proposed) & \satisfy & \satisfy & \satisfy\\
        \bottomrule
        %\bottomrule
    \end{tabular}
    }
    }
    }
    % is used to refer this table in the text
\end{table} 

{The baseline measures \textbf{B1} - \textbf{B5} are other forms of arc-level reciprocity $r(e_i , R_i)$ given $R_i$.
Below, we suggest two more baseline measures that are variants of \measure with different ways of forming $R_i$.} \returncolor

% To clarify why $R_i$ which can maximize reciprocity measure is an appropriate form of reciprocal set, 

\smallsection{B6. \measure w/ All Arcs as Reciprocal Set}
{This measure is a variant of \measure where the reciprocal set is always defined as $R_i = E$.} \returncolor
% instead of  Eq.~\eqref{eq:reciprocal-obj}.

\smallsection{B7. \measure w/ Inversely Overlapping Arcs as Reciprocal Set} 
{This measure is a variant of \measure where the reciprocal set is always defined as
$$
R_i = \{e_k \in E: \min(\lvert H_i \cap T_k \vert , \lvert T_i \cap H_k \vert ) \geq 1 \}.
$$
That is, \textbf{all} inversely overlapping arcs (see Section~\ref{sec:measure:axioms} for the definition) are used as the reciprocal set.} \returncolor
%instead of Eq.~\eqref{eq:reciprocal-obj}

{As summarized in Table~\ref{tab:AXIOMsatisfy1}, none of the considered baseline measures satisfies all the axioms, while \measure satisfies all (Theorem~\ref{thm:axiom}).
%\red{Refer to Figure~\ref{fig:AXIOMs} for counter examples regarding \Axioms 1-5 for the baseline measures, and counter examples regarding \Axioms 6-8 are given in \onlineappendix \cite{appendix}.}
Refer to Appendix~\ref{section:baselinefail} for specific counter-examples for the baseline measures.} \returncolor
\kijung{Especially, the failure of $\textbf{B6}$ and $\textbf{B7}$ highlights the necessity of finding the maximum reciprocity value over all potential reciprocal arcs, as described in Eq~\eqref{eq:reciprocal-obj}, in order for \measure to satisfy all the axioms.}

% An example regarding composing reciprocal sets is illustrated in Figure \ref{fig:composereciprocal}.
% Table \ref{table:reciprocal-sets} shows that \measure satisfies all the \textbf{\textsc{Axioms}}, while the two baselines fail.

% To demonstrate the soundness of our proposal, we show that proposed argument max reciprocal set can only satisfy every \axiom by the proof of the \textsc{Theorem 2}, while other baselines cannot satisfy at least one \axiom (see Table~\ref{tab:AXIOM_satisfy2}). 
% Details regarding Table~\ref{tab:AXIOM_satisfy2} are also analyzed in \textbf{Online Appendix}. 

% \textbf{\textsc{Theorem 2.}} $max_{R_{i} \in 2^{E}}r(h_{i}|R_{i})$ satisfies \textbf{\textsc{Axiom}} 6-8. 

% \vspace{1mm}
% \textbf{\textsc{Proof:}} See \textbf{Online Appendix} \\
% %% Professor comment: Why other measures fail

% \begin{table} \label{table:reciprocal-sets}
%     \caption{Reciprocity with argument max reciprocal set satisfies \textbf{\textsc{Axiom}}6-8, while other baselines do not.}\label{tab:AXIOM_satisfy2}
%     \centering % used for centering table

%     }
% \end{table}

%%%%%%%%%%%%%%%%%%%%%%%%%%%%%%%%%%%%
% \begin{figure}[t]
%     \centering
%     \includegraphics[width=0.4\textwidth]{FIG/SEARCH/FIG5_SEARCH.pdf}
%     \caption{Describing proposed efficient argument max reciprocal set searching method (a) Original hypergraph (b) How interaction sets($\Phi_{i}$) are composed and arcs are classified into them. \red{TODO}}
%     \label{fig:searchspace}
% \end{figure}

\begin{algorithm}[t]
    \small
    \caption{{\algo for Exact and Rapid Computation of \measure\label{alg:recipmeasure}} \returncolor}
    \KwIn{Hypergraph $G = (V, E)$}
    \KwOut{The reciprocity $\{r(e_1), \cdots r(e_{\lvert E \rvert })\}$ of arcs in $E$}
    %\begin{algorithmic}[1]
    \ForEach{$e_i \in E$}{
        $\Phi_{i} \leftarrow$ a mapping table whose default value is $\emptyset$
        
        $\Psi_{i} \leftarrow \{\}$
        
        $\Omega_i \leftarrow \{e_j : \min(\lvert H_{i} \cap T_{j}\vert , \lvert T_{i} \cap H_{i} \vert ) \geq 1 \}$
        
        \If{$\Omega_{i} = \emptyset$ \label{alg:ferret:emptycond}}{
        
            $r(e_{i}) \leftarrow 0$\label{alg:ferret:empty}
        }
        \uElseIf{$\langle T_{i}, H_{i}\rangle  \in \Omega_{i}$\label{alg:ferret:perfectcond}}{
            
            $r(e_{i}) \leftarrow 1$\label{alg:ferret:perfect}
        }
        \Else{\label{alg:ferret:partialcond}
            %Create $\Phi_{i} = \{\langle H'_{i}, T'_{i}\rangle  \mid H'_{i} \subseteq S_{i} , T'_{i} \subseteq T_{i} \}$  
        
            \ForEach{$\paire{k} \in \Omega_{i}$}{
            
               % \ForEach{$\langle H'_{i}, T'_{i}\rangle  \in \Phi_{i}$}{
                
                    %\If{$((S_{i} \cap T_{k} = H'_{i}) \wedge (T_{i} \cap S_{k} = T'_{i}))$}{
                        $H'_{i} \leftarrow H_{i} \cap T_{k}$;
                        
                        $T'_{i} \leftarrow T_{i} \cap H_{k}$
                        
                        $\Phi_{i}(\langle H'_{i},T'_{i}\rangle ) \leftarrow \Phi_{i}(\langle H'_{i},T'_{i}\rangle ) \cup \{e_{k}\}$ %\hfill \  \tcp{\footnotesize $T_{\langle A,B\rangle }$ is initially an empty set for any $A$ and $B$} 
                    %}
               % }
            }
            
            \ForEach{$\langle H'_{i},T'_{i}\rangle$ where $\Phi_{i}(\langle H'_{i},T'_{i}\rangle)\neq \emptyset$}{
              %  \If{$SG_{\langle H'_{i}, T'_{i}\rangle } \neq \emptyset$}{
                    
                    $e'_{i} \leftarrow \argmin_{e_{j} \in \Phi_{i}(\langle H'_{i},T'_{i}\rangle)}  \lvert H_{j} \vert $
                    
                    $\Psi_{i} \leftarrow \Psi_{i} \cup \{e'_{i}\}$ 
             %   }
            }
            $r(e_i)\leftarrow \max_{R_{i} \subseteq \Psi_{i}} r(e_{i}, R_{i})$
            \label{alg:ferret:partial}
        }
        \textbf{return} $\{r(e_1), \cdots r(e_{\lvert E \rvert })\}$
    } 
    %\end{algorithmic}
\end{algorithm}

\subsection{\textbf{Exact and Rapid Search for Reciprocal Sets}}\label{sec:ferret}
%% Professor Comment: Summary of subsection

We propose \algo (\underline{\textbf{F}}ast and \underline{\textbf{E}}xact Algo\underline{\textbf{r}}ithm for Hypergraph R\underline{\textbf{e}}ciproci\underline{\textbf{t}}y), an approach for rapidly searching for the reciprocal set $R_i$ of Eq.~\eqref{eq:reciprocal-obj}. %, which takes exponential time with a naive approach.
We prove the exactness of \algo and demonstrate its efficiency in real-world hypergraphs. %.that its search space is up to $10^{147} \times$ smaller than that of naive computation.
% The size of search space for computing \measure is exponential in the number of hyperarcs.
% $R_i = \argmax_{R_i \in 2^{E}} {r(e_i | R_i)}$ is not always tractable, since it has exponential complexity $O(2^{\lvert E \rvert -1})$.
% For even a relatively small hypergraph where its number of edge is $\lvert E \rvert  = 100$, we need $O(2^{99}) \approx 10^{30}$ computations with a naive approach.
% To address this issue, we propose \algo, which enables obtaining adequate reciprocal set within a reasonable time. The procedure is described in algorithm \ref{alg:recipmeasure}. Note that this algorithm is about finding hyperarc reciprocity rapidly.

\smallsection{High-level ideas} 
\kijung{The computational overhead of \measure lies in finding the maximum reciprocity value over all possible subsets of the hyperarc set $E$ (i.e., $\max_{R_{i} \subseteq E, R_{i} \neq \emptyset } r(e_{i},R_{i})$). 
Conducting an exhaustive search over the entire search space results in the time complexity of $O(2^{\vert E\vert})$, which becomes infeasible for the considered real-world hypergraphs (refer to Table~\ref{tab:datastat} for the sizes of the real-world hypergraphs).
To address this issue, \algo explores $2^{\Psi_{i}}$ instead of $2^{E}$, where $\Psi_{i} \subseteq E$ holds, without affecting the computed reciprocity value.
Specifically, \algo first creates disjoint groups of hyperarcs, which will be further explained in the following paragraph.
Subsequently, it constructs $\Psi_{i}$ by selecting solely the hyperarcs with the smallest head set size from each group.
Then, \algo computes $\max_{R_{i} \subseteq \Psi_{i}, R_{i} \neq \emptyset} r(e_{i},R_{i})$.
Importantly, $\vert \Psi_{i} \vert  \ll \vert E \vert$ for most, if not all, real-world hypergraphs.
That is, by employing \algo, the computation of $r(e_{i}, R_{i})$ in \measure is performed for a significantly smaller number of reciprocal sets $R_{i}$, leading to a substantial reduction in the overall computational time. 
Again, it is important to highlight that \algo is an exact algorithm that gives the precise value of Eq~\eqref{eq:final}.
}

\smallsection{Detailed Procedure}
\algo is described in Algorithm \ref{alg:recipmeasure}.
For each arc $e_i$, we first retrieve the set $\Omega_{i}$ of inverse-overlapped arcs (see Section~\ref{sec:measure:axioms} for the definition) and check whether $e_i$ is (1) non-reciprocal, (2) perfectly reciprocal, or (3) partially reciprocal. 
Reciprocity for the first two cases is $0$ (lines \ref{alg:ferret:emptycond}-\ref{alg:ferret:empty}) and $1$ (lines \ref{alg:ferret:perfectcond}-\ref{alg:ferret:perfect}), respectively. 
For a partially reciprocal case (lines \ref{alg:ferret:partialcond}-\ref{alg:ferret:partial}), we group the arcs in $\Omega_{i}$ using a mapping table $\Phi_{i}$ where the key of each arc $e_k\in \Omega_{i}$ is the head-set and tail-set nodes of $e_i$ that it covers (i.e., $\langle H'_{i}, T'_{i}\rangle$ where
 $H'_{i} \leftarrow H_{i} \cap T_{k}$ and $T'_{i} \leftarrow T_{i} \cap H_{k}$).
%, which uses the covered interactions  as its key and the arcs with the same interactions as its value.
For each group with the same key $\langle H'_{i}, T'_{i}\rangle$, we choose an arc with the minimum number of head set nodes. Then, we create a new search space $\Psi_{i}$ containing only the chosen arcs. After that, every subset $R_i$ of $\Psi_{i}$ is considered to maximize Eq.~\eqref{eq:proposal}, and we return the maximum value as the reciprocity $r(e_i)$ of $e_i$.

\smallsection{Theoretical Properties}
As stated in Theorem~\ref{thm:algo:basic}, \algo finds the best reciprocal set, as in Eq.~\eqref{eq:reciprocal-obj}, and thus it computes the reciprocity of each arc exactly, as in Eq.~\eqref{eq:final}.
%we theoretically guarantee that the search result from $\Psi_{i}$ yields the same reciprocity as from $E$, while the search cost is dramatically reduced. \\

\begin{thm}[Exactness of \algo] \label{thm:algo:basic}
For every $e_{i}\in E$,
$\max_{R_{i} \subseteq E} r(e_{i}, R_{i})$ is identical to the $\max_{R_{i} \subseteq \Psi_{i}} r(e_{i}, R_{i})$.
\end{thm}
\begin{proof}
% See Appendix~\ref{section:proof}.
See Appendix~\ref{subsec:thm2}.
\end{proof}
%\textbf{\textsc{Proof:}} See \onlineappendix. \\
{For a special type of hypergraphs,
we further reduce the search space based on Corollary~\ref{corr:algo}. 
Recall that $\Psi_{i}$ is a final search space produced by \algo.}
\returncolor 

{\begin{corr} \label{corr:algo}
%In a hypergraph $G$ where $\lvert T_{i} \vert = 1, \forall i \in \{1 ,\cdots ,\lvert E \rvert \}$, if $\alpha\rightarrow 0$, then $\argmax_{R_{i} \subseteq E} r(e_{i}, R_{i}) = \Psi_{i}$.
 Consider a hypergraph $G$ where every arc's tail set size is $1$ $($i.e., $\lvert T_{i} \vert = 1, \forall i \in \{1 ,\cdots ,\lvert E \rvert \})$, and let $\Gamma_{i, k}$ be a subset of $\Psi_{i}$ that satisfies $ \lvert \Gamma_{i,k} \vert = k \text{ and }\lvert H_{s} \vert \leq \lvert H_{t} \vert , \forall e_{s} \in \Gamma_{i, k}, \forall e_{t} \in \{\Psi_{i} \setminus \Gamma_{i, k}\}$.
 Then, $\argmax_{(R_{i} \subseteq E \ s.t. \lvert R_{i} \vert  = k)} r(e_{i}, R_{i})$ is identical to $\Gamma_{i, k}, \forall k\in \{1,\cdots, \lvert \Psi_{i} \vert \}$, which implies that $\argmax_{R_{i} \subseteq \Psi_{i}, R_{i}\neq \emptyset} r(e_{i}, R_{i})$ is identical to $\argmax_{R_{i} \in \{\Gamma_{i, 1}, \cdots , \Gamma_{i, \lvert \Psi_{i}\vert }\}} r(e_{i}, R_{i})$. That is, the size of the search space for $R_i$ is reduced to $O(\lvert \Psi_{i} \vert )$.
\end{corr}
\begin{proof}
% See Appendix~\ref{section:proof}.
See Appendix~\ref{subsec:col1}.
\end{proof}
} \returncolor
\begin{figure}[t]
 %   \vspace{-2mm}
    \centering
    \includegraphics[width=0.8\textwidth]{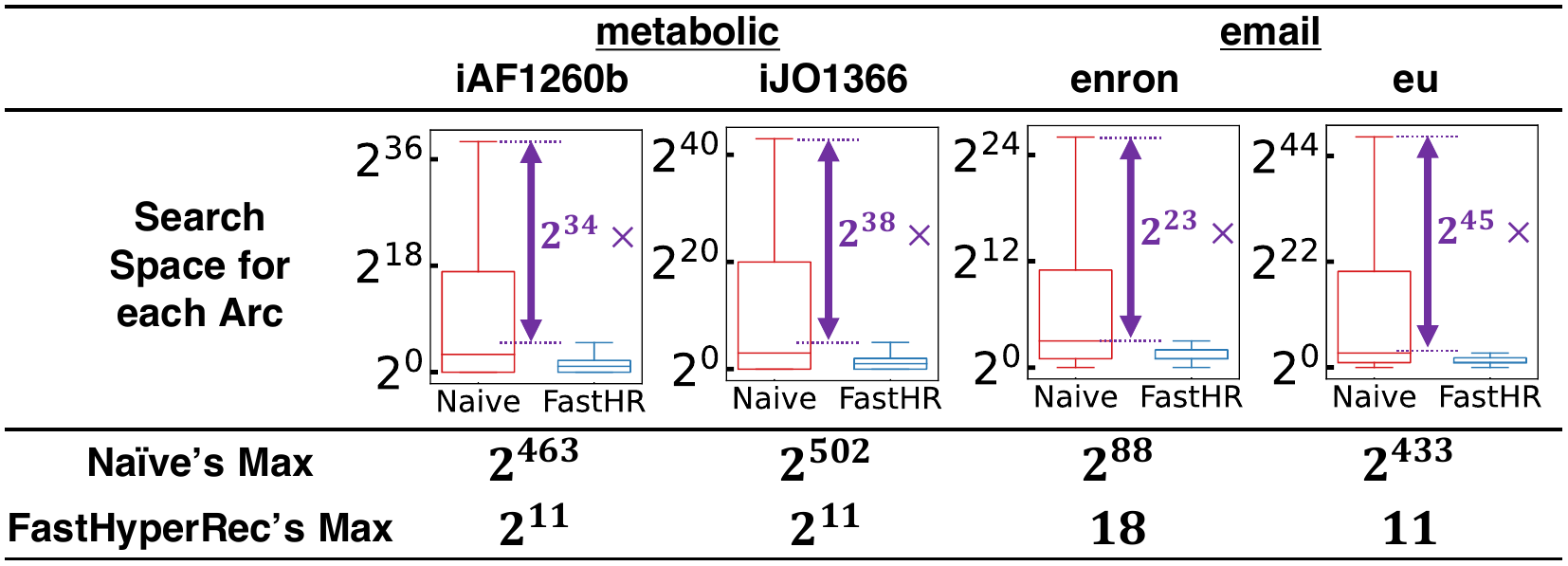}
    % \vspace{-3mm}
    \caption{
    \algo (FastHR in short, left in each box plot) reduces the search space by up to $2^{491}(\approx10^{147})\times$, compared to that of the naive method (right in each box plot)  \kijung{in the iJO1366 dataset, which has 1,805 nodes and 2,251 arcs}. %We do not assume $\alpha \rightarrow 0$. 
    \kijung{To improve legibility, we remove several outliers that lie outside the interquartile range (refer to Footnote~\ref{footnote:iqr}).}
    \label{fig:ferret}
    }
\end{figure}
\returncolor
\begin{table*}[t]
\centering
\caption{{Running time in seconds of \algo and searching $\Omega_{i}$. O.O.T.: out of time ($\geq 12$ hours).} \returncolor} \label{tab:runtime}
\vspace{2mm}
    %\scalebox{0.55}{
    \resizebox{\columnwidth}{!}{
    \renewcommand{\arraystretch}{1.0}
        \centering
        \begin{tabular}{c c c c c c c c c c c c}
            \toprule
            & \multicolumn{2}{c}{metabolic} & \multicolumn{2}{c}{email} & \multicolumn{2}{c}{citation} & \multicolumn{2}{c}{qna} 
            & \multicolumn{3}{c}{bitcoin}  \\
            & iAF1260b & iJO1366 & enron & eu & data mining & software & math & server & 2014 & 2015 & 2016  \\
            \midrule
            \midrule
            
            \algo & 0.382 & 0.567 & 0.220 & 8.221 & 20.766 & 422.764 &
            2.820 & 2.981 & 12297.318 & 762.645 & 428.283  \\
            \midrule 
            
            Searching $\Omega_{i}$ & O.O.T. & O.O.T. & O.O.T. & O.O.T. & O.O.T. & O.O.T. &
            O.O.T. & O.O.T. & O.O.T. & O.O.T. & O.O.T.  \\
            
            \bottomrule
        \end{tabular}
        }
\end{table*}

\smallsection{Complexity Analysis and Evaluation in Real-world Hypergraphs}
After the reduction above, the size of the search space for $R_{i}$ becomes $O(2^{\lvert \Psi_{i} \vert })$ in general and $O(\lvert \Psi_{i} \vert )$ for the case where every arc's tail set size is $1$ (i.e., $\lvert T_{i} \vert = 1, \forall i \in \{1,\cdots,\lvert E \rvert \}$). 
Although the complexity is still exponential, we demonstrate that the search space is reasonably small, and thus a search can be performed within a reasonable time period for real-world hypergraphs.
{To evaluate the effectiveness of \algo, we compare the size of search space of \algo and a naive method that finds the best reciprocal set in $\Omega_{i}$ in four real-world hypergraphs, which are described in the following section.
As shown in Figure~\ref{fig:ferret},\footnote{\kijung{Note that, to improve legibility, we remove data points that lie outside the interquartile range (i.e., $[Q_{1} - 1.5(Q_{3} - Q_{1}), Q_{3} + 1.5(Q_{3} - Q_{1})]$, where $Q_{3}$ and $Q_{1}$ denote the third and first quantile of the corresponding distribution) from the box plots.}\label{footnote:iqr}} 
%the search space of \algo is up to $2^{491}(\approx10^{147}) \times$ smaller than that of the naive method.
in the iJO1366 dataset, which has 1,805 nodes and 2,251 arcs, the naive method searches at most $2^{502} (\approx 10^{148})$ sets for the reciprocity of an arc, while \algo searches at most $2^{11}(=2048)$ sets.}
\returncolor
In addition, we report the running time taken to compute the reciprocity of each of the 11 real-world hypergraphs, which are described in detail in the following section. As shown in Table~\ref{tab:runtime}, \algo terminates within 3.5 hours for every hypergraph, while naively searching $\Omega_{i}$ does not terminate within 12 hours for any of the hypergraphs.

%that \algo can reduce search space at most $10^{147}\times$ than naive search ($\Omega_{i}$ ; only uses inverse-overlap arcs).

%in terms of 

%the size of its search space, which is $O(2^{|\Psi_{i}|})$ and that of naive computation, 
%After narrowing down, we have a complexity of $O(2^{|\Psi_{i}|})$ for each arc $e_{i}$. 
%Although this has exponential complexity, we have found that real-world hypergraphs do not exceed reciprocity computation of $2^{27}$ with our proposal.

}

\section{Datasets and Observations}{
    \label{section:obs}
    \begin{table}[t!]
    \vspace{2mm}
    \caption{{Summary of 11 real-world hypergraphs from 5 domains: the number of nodes $\lvert V \vert $, the number of arcs $\lvert E \vert $, the average size of head set $\overline{\lvert H_{i} \vert }$, the average size of tail set $\overline{\lvert T_{i} \vert }$, the average node in-degree $\overline{\lvert d_{in}(v) \vert }$ and the average node out-degree $\overline{\lvert d_{out}(v) \vert }$.} \returncolor} % title of Table
    %\vspace{-3mm}
    \centering % used for centering table
    \scalebox{0.9}{
    \renewcommand{\arraystretch}{1.5}{
    \begin{tabular}{l  r  r  r  r  r  r }
        \hline 
        \textbf{Dataset} & $\vert V \vert$ & $\vert E \vert$ & $\overline{\vert H_{i} \vert}$
        & $\overline{ \vert T_{i} \vert}$ & $\overline{ \vert d_{in}(v) \vert }$ & $ \overline{ \vert d_{out}(v) \vert}$\\
        \hline % inserts single horizontal line
        \hline 
        metabolic-iaf1260b & 1,668 & 2,083 & 2.267 & 1.998 & 2.831 & 2.495 \\ 
        metabolic-iJO1366 & 1,805 & 2,251 & 2.272 & 2.026 & 2.833 & 2.527 \\ 
        \hline
        email-enron & 110 & 1,484 & 2.354 & 1.000 & 31.764 & 13.491 \\ 
        email-eu & 986 & 35,772 & 2.368 & 1.000 & 85.900 & 36.280 \\ 
        \hline
        citation-dm & 27,164 & 73,113 & 3.038 & 3.253 & 8.177 & 8.755 \\ 
        citation-software & 16,555 & 53,177 & 2.717 & 2.927 & 8.729 & 9.401 \\ 
        \hline
        q\&a-math & 34,812 & 93,731 & 1.000 & 1.779 & 2.692 & 4.789 \\ 
        q\&a-server & 17,2330 & 27,2116 & 1.000 & 1.747 & 1.579 & 2.759 \\ 
        \hline
        bitcoin-2014 & 1,697,625 & 1,437,082 & 1.697 & 1.478 & 1.437 & 1.251 \\ 
        bitcoin-2015 & 1,961,886 & 1,449,827 & 1.744 & 1.568 & 1.288 & 1.159 \\ 
        bitcoin-2016 & 2,009,978 & 1,451,135 & 1.715 & 1.495 & 1.238 & 1.079 \\ 
        \hline
    \end{tabular}
    }
    }
\label{tab:datastat} % is used to refer this table in the text
\end{table} 

In this section, we investigate the reciprocal patterns of real-world hypergraphs using \measure and \algo.
After introducing used real-world hypergraph datasets and null hypergraphs, we discuss our observations at three different levels: \emph{hypergraph, arc, and node}.
The significance of the patterns are verified by a comparison with the null hypergraphs.

\begin{table*}[t!]
\centering
\caption{{Hypergraph reciprocity $r(G)$ of 11 datasets when $\alpha \approx$ $0$, $\alpha=0.5$, and $\alpha=1$.} \returncolor
} \label{tab:absolutereciprocity}
\vspace{2mm}
    %\scalebox{0.6}{
    \resizebox{\columnwidth}{!}{
    \renewcommand{\arraystretch}{1.2}
        \centering
        \begin{tabular}{c  c  c c c c c c c c c c c}
            \toprule
            \multicolumn{2}{c}{} & \multicolumn{2}{c}{metabolic} & \multicolumn{2}{c}{email} & \multicolumn{2}{c}{citation} & \multicolumn{2}{c}{q\&a} 
            & \multicolumn{3}{c}{bitcoin}  \\
            \multicolumn{2}{c}{} & iAF1260b & iJO1366 & enron & eu & data mining & software & math & server & 2014 & 2015 & 2016  \\
            \midrule
            \midrule
            
            \multirow{3}{*}{$r(G)$} & $\alpha \approx 0$ & 21.455 & 22.533 & 59.161 & 79.489 & 12.078 & 15.316 & 9.608 & 13.219 & 10.829 & 6.923 & 3.045  \\
            & $\alpha = 0.5$ & 17.756 & 18.497 & 49.480 & 65.477 & 10.840 & 13.984 & 9.283 & 13.196 & 10.654 & 6.845 & 2.988  \\
            & $\alpha = 1.0$ & 16.654 & 17.385 & 44.459 & 58.139 & 10.585 & 13.704 & 9.236 & 13.193 & 10.606 & 6.828 & 2.977  \\
            \bottomrule
        \end{tabular}
        }
\end{table*}

\begin{table}[t!]
\centering
\caption{{Hypergraph reciprocity $r(G)$ is robust to the choice of $\alpha$.
Although their absolute value may differ (see Table~\ref{tab:absolutereciprocity}), their relative values are not sensitive to the to the choice of $\alpha$, as supported by the fact that all the measured Pearson correlation coefficients and Spearman rank correlation coefficients are greater than 0.99.} \returncolor
} \label{tab:hypergraphrobustness}
\vspace{2mm}
    %\scalebox{0.9}{
    \resizebox{\columnwidth}{!}{
    \renewcommand{\arraystretch}{1.0}
        \centering
        \begin{tabular}{c  c  c  c}
            \toprule
            \multicolumn{2}{c}{} & \textbf{Pearson Correlation} & \textbf{Spearman Rank Correlation} \\
            \midrule
            \midrule
            \multirow{3}{*}{$r(G)$} & $\alpha \approx 0 \leftrightarrow \alpha = 0.5$ & 0.999 & 1.0  \\
            & $\alpha \approx 0 \leftrightarrow \alpha = 1.0$ & 0.999 & 0.991 \\
            & $\alpha = 0.5 \leftrightarrow \alpha = 1.0$ & 0.999 & 0.991 \\
            \bottomrule
        \end{tabular}
        }
\end{table}

\subsection{\textbf{Datasets and Null Hypergraphs}}

\smallsection{Datasets}
We use 11 real-world hypergraphs from five different domains. Refer to Table~\ref{tab:datastat} for basic statistics of them.
All duplicated edges are removed, and detailed pre-processing steps are described in the Appendix~\ref{section:datadescription}.

\begin{itemize}
    \item \textbf{Metabolic} (iAF1260b and iJO1366 \citet{yadati2020nhp}): Each network models chemical reactions among various genes. Nodes correspond to genes, and arcs indicate reactions.
    \item \textbf{Emails} (email-enron \citet{chodrow2020annotated} and email-eu \citet{snapnets}): Each node is an email account, and each arc consists of two ordered sets of senders and receivers of an email.
    \item \textbf{Citations} (DBLP-data mining and DBLP-software \citet{sinha2015overview}). Each node is a researcher, and each head set and tail set indicates a paper. Arcs represent citations, as in Figure~\ref{fig:overview}.
    % That is, if paper A cites paper B where each paper is written by $\{v_{1}, v_{2}\}$ and $\{v_{3}, v_{4}, v_{5} \}$, this arc becomes $e=\langle H = \{v_{3}, v_{4}, v_{5}\},T = \{v_{1}, v_{2}\} \rangle$, as done in \cite{yadati2021graph}.
    \item \textbf{Question and Answering} (math-overflow and stack-exchange server fault \citet{qnadataset}). Each node is a user, and each arc corresponds to a post. The questioner of a post becomes the head of an arc and the answerers compose its tail set.
    \item \textbf{Bitcoin Transactions} (bitcoin-2014, 2015, 2016 \citet{wu2020detecting}). Each node is an address in bitcoin transactions, and each arc is a transaction among users. The three datasets contain the first 1,500,000 transactions of Nov 2014, June 2015, and Jan 2016, respectively. We filtered out all transactions where the head set and the tail set are the same. %which do not satisfy general condition of directed hypergraph.
\end{itemize}

\smallsection{Null Hypergraphs}
Patterns observed in real-world graphs can be caused by chance. In order to demonstrate discovered reciprocal characteristics are distinguishable from random behavior, we measure the same statistics and patterns in randomized hypergraphs, which we call \textbf{\emph{null hypergraphs}}.
Given a real hypergrpah, we create a null hypergraph with the same number of nodes and the same distribution of arc sizes (i.e., the size of the head set and tail set of each arc).
To create each arc, we draw nodes uniformly at random and compose a head set and a tail set with the chosen nodes. To minimize the randomness of experiments, we create 30 null hypergraphs from each dataset and report the statistics averaged over them. %$r(G_{null, j}) : j=1\cdots 30$ 

\subsection{\textbf{Observations}}\label{section:subsecobservation}

We investigate the reciprocal patterns of real-world hypergraphs at three different 
levels: Hypergraph, Arc, and Node. %, and our observations are summarized as follows.

\begin{table*}[t!]
\caption{Observations~\ref{obs:observation1} and the superiority of \model. Reciprocity in (a) real-world hypergraphs, (b) null hypergraphs, (c) those generated by \model (Section~\ref{section:generation}), and (d) those generated by a baseline generator is reported.
\kijung{Specifically, we generate five synthetic hypergraphs using each generator (Null, \dhgpa, and Baseline) and the statistics from each dataset, and report the average ($\overline{r(G)}$) and standard deviation ($sd(r(G))$) of hypergraph-level reciprocity values of the generated hypergraphs.}
As the arc-level difference, we report the D-statistic (the lower the better) between each distribution of arc-level reciprocity  and that in the corresponding real-world hypergraph. 
Values below $10^{-6}$ are all marked with ${}^{*}$.
In each column, the hypergraph reciprocity closest to that in the real-world hypergraph and the minimum D-statistic are \underline{underlined}. %$\alpha$ is fixed to a value close to $0$.
Note that real-world hypergraphs are more reciprocal than null hypergraphs, and our proposed generator, \model, successfully reproduces the reciprocity in real-world hypergraphs.
} \label{tab:observation1}
\vspace{2mm}
    \centering
    %\scalebox{0.55}{
    \resizebox{\columnwidth}{!}{
    \renewcommand{\arraystretch}{1.0}
        \centering
        \begin{tabular}{c  c  c c c c c c c c c c c}
            \toprule
            \multicolumn{2}{c}{} & \multicolumn{2}{c}{metabolic} & \multicolumn{2}{c}{email} & \multicolumn{2}{c}{citation} & \multicolumn{2}{c}{q\&a} 
            & \multicolumn{3}{c}{bitcoin}  \\
            \multicolumn{2}{c}{} & iAF1260b & iJO1366 & enron & eu & data mining & software & math & server & 2014 & 2015 & 2016  \\
            \midrule
            \midrule
            
            \textbf{Real World} & \textbf{$r(G)$} & 21.455 & 22.533 & 59.001 & 79.416 & 12.078 & 15.316 &
            9.608 & 13.219 & 10.829 & 6.923 & 3.045  \\
            \midrule
            
            \multirow{3}{*}{\textbf{Null}} & $\overline{r(G)}$ & 0.306 & 0.270 & 14.862 & 4.633 & 0.094 & 0.147 &
            0.018 & 0.002 & 0.0001 & $0.000^{*}$ & $0.000^{*}$  \\
            
            & \kijung{\textbf{sd($r(G)$)}} & \kijung{0.054} & \kijung{0.091} & \kijung{0.296} & \kijung{0.110} & \kijung{0.005} & \kijung{0.006} & \kijung{0.001} & \kijung{0.005} & \kijung{$0.000^{*}$} & \kijung{$0.000^{*}$} & \kijung{$0.000^{*}$} \\
            
            & \textbf{\textit{D-Stat}} & 0.625 & 0.642 & 0.539 & 0.807 & 0.355 & 0.377 &
            0.124 & 0.160 & 0.147 & 0.100 & 0.050 \\
            \midrule
            
            \multirow{3}{*}{\makecell{\textbf{\dhgpa} \\ (Section~\ref{section:generation})}} & $\overline{r(G)}$ & \underline{21.727} & \underline{22.185} & \underline{59.161} & \underline{79.489} & \underline{12.601} & \underline{14.279} & \underline{9.427} & \underline{13.229} & \underline{10.267} & \underline{6.587} & \underline{3.497}  \\
            
            & \kijung{\textbf{sd($r(G)$)}} & \kijung{1.811} & \kijung{0.562} & \kijung{2.895} & \kijung{1.013} & \kijung{0.586} & \kijung{0.448} & \kijung{0.004} & \kijung{0.083} & \kijung{0.451} & \kijung{0.121} & \kijung{0.796} \\
            
            & \textbf{\textit{D-Stat}} & \underline{0.098} & \underline{0.104} & \underline{0.053} & \underline{0.043} & \underline{0.212} & \underline{0.151} &
            \underline{0.011} & \underline{0.005} & \underline{0.045} & \underline{0.033} & \underline{0.017}  \\
            \midrule
            
            \multirow{3}{*}{\makecell{\textbf{Baseline} \\ (Section~\ref{section:generation})}} & $\overline{r(G)}$ & 0.412 & 0.851 & 23.846 & 31.190 & 0.048 & 0.004 & 1.622 & 0.002 & 0.002 & 0.002 & 0.001  \\
            
            & \kijung{\textbf{sd($r(G)$)}} & \kijung{0.117} & \kijung{0.882} & \kijung{1.437} & \kijung{0.273} & \kijung{0.642} & \kijung{0.543} & \kijung{0.004} & \kijung{0.009} & \kijung{0.002} & \kijung{0.002} & \kijung{0.002} \\
            
            & \textbf{\textit{D-Stat}} & 0.625 & 0.623 & 0.403 & 0.535 & 0.328 & 0.367 &
            0.103 & 0.160 & 0.147 & 0.099 & 0.050  \\
            \bottomrule
        \end{tabular}
        }
\end{table*}

\begin{table}[t!]
\centering
\caption{{P-value testing results on the 11 considered datasets. The null hypotheses are all rejected, which implies that real-world hypergraphs are significantly more reciprocal than null hypergraphs. A p-value smaller than 0.00001 is denoted by $0.0000^{*}$.} \returncolor} \label{tab:p-value}
\vspace{3mm}
    %\scalebox{0.58}{
    \resizebox{\columnwidth}{!}{
    \renewcommand{\arraystretch}{1.1}
        \centering
        \begin{tabular}{cccccccccccc}
            \toprule
            & \multicolumn{2}{c}{metabolic} & \multicolumn{2}{c}{email} & \multicolumn{2}{c}{citation} & \multicolumn{2}{c}{q\&a} 
            & \multicolumn{3}{c}{bitcoin}  \\
            & iAF1260b & iJO1366 & enron & eu & data mining & software & math & server & 2014 & 2015 & 2016  \\
            \midrule
            \midrule
            \textbf{Z-stat} & -1502.52 & -1789.79 & -241.13 & -3835.98 & -17548.20 & -9605.19 & -8884.98 & -88965.12 & -691316.77 & -555709.95 & -325308.06  \\
            \textbf{P-value} & $0.0000^{*}$ & $0.0000^{*}$ & $0.0000^{*}$ & $0.0000^{*}$ & $0.0000^{*}$ & $0.0000^{*}$ & $0.0000^{*}$ & $0.0000^{*}$ & $0.0000^{*}$ & $0.0000^{*}$ & $0.0000^{*}$  \\
             \textbf{Null hypothesis} & Reject & Reject & Reject & Reject & Reject & Reject & Reject & Reject & Reject & Reject & Reject  \\
            \bottomrule
        \end{tabular}
        }
\end{table}

\smallsection{L1. Hypergraph Level}
We first demonstrate that hypergraph reciprocity $r(G)=\frac{1}{\lvert E \vert }\sum_{e_{i} \in E}r(e_{i})$ is robust to the choice of $\alpha$, i.e. the size penalty term for reciprocal sets. 
{As shown in Table \ref{tab:hypergraphrobustness}, although absolute reciprocity values vary depending on $\alpha$, their ranks in real-world hypergraphs remain almost the same, as supported by the fact that both the Pearson and rank correlation coefficients are near $1$.} 
Based on this result, we fix $\alpha$ to a value near zero for the investigation below.

\begin{table}[t!]
\centering
\caption{{Arc-level reciprocity $r(e, R)$ is robust to the choice of $\alpha$. Although their absolute values may differ, their relative values are not sensitive to the choice of $\alpha$, as supported by the fact that the measured Pearson correlation coefficients and Spearman rank correlation coefficients are at least $0.678$ and in many cases even close to $1$.} \returncolor
} \label{tab:arcrobustness}
\vspace{2mm}
    %\scalebox{0.58}{
    \resizebox{\columnwidth}{!}{
    \renewcommand{\arraystretch}{1.1}
        \centering
        \begin{tabular}{c  c c c c c c c c c c c c}
            \toprule
             \multicolumn{2}{c}{} & \multicolumn{2}{c}{metabolic} & \multicolumn{2}{c}{email} & \multicolumn{2}{c}{citation} & \multicolumn{2}{c}{q\&a} 
            & \multicolumn{3}{c}{bitcoin}  \\
            \multicolumn{2}{c}{} & iAF1260b & iJO1366 & enron & eu & data mining & software & math & server & 2014 & 2015 & 2016  \\
            \midrule
            \midrule
            \multirow{3}{*}{\textbf{Pearson}} & $\alpha \approx 0 \leftrightarrow \alpha = 0.5$ & 0.961 & 0.957 & 0.928 & 0.836 & 0.984 & 0.986 & 0.994 & 0.999 & 0.997 & 0.998 & 0.997  \\
             & $\alpha \approx 0 \leftrightarrow \alpha = 1.0$ & 0.916 & 0.913 & 0.828 & 0.678 & 0.973 & 0.977 & 0.992 & 0.999 & 0.995 & 0.997 & 0.996  \\
            & $ \alpha = 0.5 \leftrightarrow \alpha = 1.0$ & 0.985 & 0.986 & 0.975 & 0.967 & 0.998 & 0.998 & 0.999 & 0.999 & 0.999 & 0.999 & 0.999  \\
            \midrule
            \multirow{3}{*}{\textbf{Spearman Rank}} & $\alpha \approx 0 \leftrightarrow \alpha = 0.5$ & 0.973 & 0.969 & 0.947 & 0.817 & 0.998 & 0.998 & 0.999 & 0.999 & 0.999 & 0.999 & 0.999  \\
             & $\alpha \approx 0 \leftrightarrow \alpha = 1.0$ & 0.925 & 0.918 & 0.869 & 0.721 & 0.996 & 0.996 & 0.999 & 0.999 & 0.999 & 0.999 & 0.999  \\
            & $ \alpha = 0.5 \leftrightarrow \alpha = 1.0$ & 0.975 & 0.973 & 0.978 & 0.983 & 0.999 & 0.999 & 0.999 & 0.999 & 0.999 & 0.999 & 0.999  \\
            \bottomrule
        \end{tabular}
        }
\end{table}

%These results imply the robustness of hypergraph reciprocity to the change of $\alpha$ (see Table \ref{tab:hypergraphrobustness}).

% In order to measure the significance of difference between real-world hypergraph and randomized hypergraph's reciprocity, we measure Z-statistic ($Z$-stat) as follows. 
% \vspace{-3mm}

% \begin{gather*}
%     Z = \frac{r(G) - \overline{r(G_{null})}}{\frac{sd(r(G_{null}))}{\sqrt{n}}} ~ \text{where } \overline{r(G_{null})} : \text{Average of } r(G_{null, j}) \\ sd(r(G_{null})) : \text{Standard Deviation of } r(G_{null, j})
% \end{gather*}

% By approximating computed $Z$-stat to the standard normal distribution, we report p-value of it to see whether average of randomized hypergraph reciprocity is significantly different from that of real-world hypergraph. \\

As shown in Table \ref{tab:observation1}, the hypergraph reciprocity is several orders of magnitude greater in real-world hypergraphs than in corresponding null hypergraphs. 
%Moreover, their differences are all statistically significant under the significance level of $0.01$, as the p-values in all dataset are below $0.001$ (see Table \ref{tab:p-value}). 
{To statistically verify this, we conduct statistical tests for all the datasets where the alternative hypothesis is that $\overline{r(G)}$ is statistically-significantly greater than $\overline{r(G_{null})}$.
The detailed numerical results of the tests are provided in Table~\ref{tab:p-value}.} \returncolor
In summary, we demonstrate that the hypergraph reciprocity is statistically-significantly greater in real-world hypergraphs than in corresponding null hypergraphs.
% This implies that real-world $r(G)$ is significantly different from $r(G_{null})$.

\begin{obs}~\label{obs:observation1}
Real-world hypergraphs are more reciprocal than randomized hypergraphs.
\end{obs}

%\textbf{\textsc{Observation 1.}} \textit{} \\

% To statistically verify the significance of the difference between real-world hypergraphs' reciprocity and that of randomized hypergraphs, we measure p-value of the null-hypothesis of \emph{randomized hypergraph's average $r(G_{null})$ is identical to the real-world hypergraph's $r(G)$}. \\ 

\smallsection{L2. Arc Level} We first show the robustness of arc-level reciprocity to the choice of $\alpha$, i.e. the size penalty term for reciprocal sets. 
{We measure the Pearson and Rank correlation coefficients between arc-level reciprocity values in each pair of settings with different $\alpha$ values (spec., $0.0 \leftrightarrow 0.5$, $0.0 \leftrightarrow 1.0$, and $0.5 \leftrightarrow 1.0$). 
As shown in Table~\ref{tab:arcrobustness}, the correlation coefficients are at least 0.721 and in most cases close to $1$, implying that relative values of arc-level  reciprocity remain almost the same regardless of $\alpha$ values.} \returncolor
Due to this robustness, \textbf{we fix $\alpha$ to a value close to $0$ for all the following experiments}.

At the arc level, we examine the relations between the degree of arcs and their reciprocity. 
We define degrees at the arc level as follows:
\begin{equation}\label{eq:dout}
    \text{Head set out-degree: } d_{H,out}(e_{i}) = \frac{1}{\lvert H_{i} \vert }\sum_{v \in H_{i}} d_{out}(v)
\end{equation}
\begin{equation}\label{eq:din}
    \text{Tail set in-degree: } d_{T,in}(e_{i}) = \frac{1}{ \lvert T_{i} \vert }\sum_{v \in T_{i}} d_{in}(v)
\end{equation}
Refer to Section~\ref{section:prelim:concept} for the definitions of $d_{out}(v)$ and $d_{in}(v)$.
Then, we compare the distributional difference of these statistics (i.e., Eqs.~\eqref{eq:dout} and \eqref{eq:din}) between the arcs of zero reciprocity and those of non-zero reciprocity.

As shown in Figure~\ref{fig:obs2},  the degrees at arcs with non-zero reciprocity tend to be greater than those at arcs with zero reciprocity.
%degree distributions are having higher value than that of zero reciprocity hyperarcs. 
This is intuitive since arcs where their head sets are frequently being pointed and tail sets are frequently pointing others tend to have higher chance to be reciprocal. Such tendency, however, is not clear in null hypergraphs.

\begin{obs}~\label{obs:observation2}
Arcs with non-zero reciprocity tend to have higher head set out-degree and tail set in-degree than arcs with zero reciprocity.
\end{obs}

%\textbf{\textsc{Observation 2.}} \textit{.} \\

\begin{figure*}[t!]
  \centering
  \includegraphics[width=1.0\textwidth]{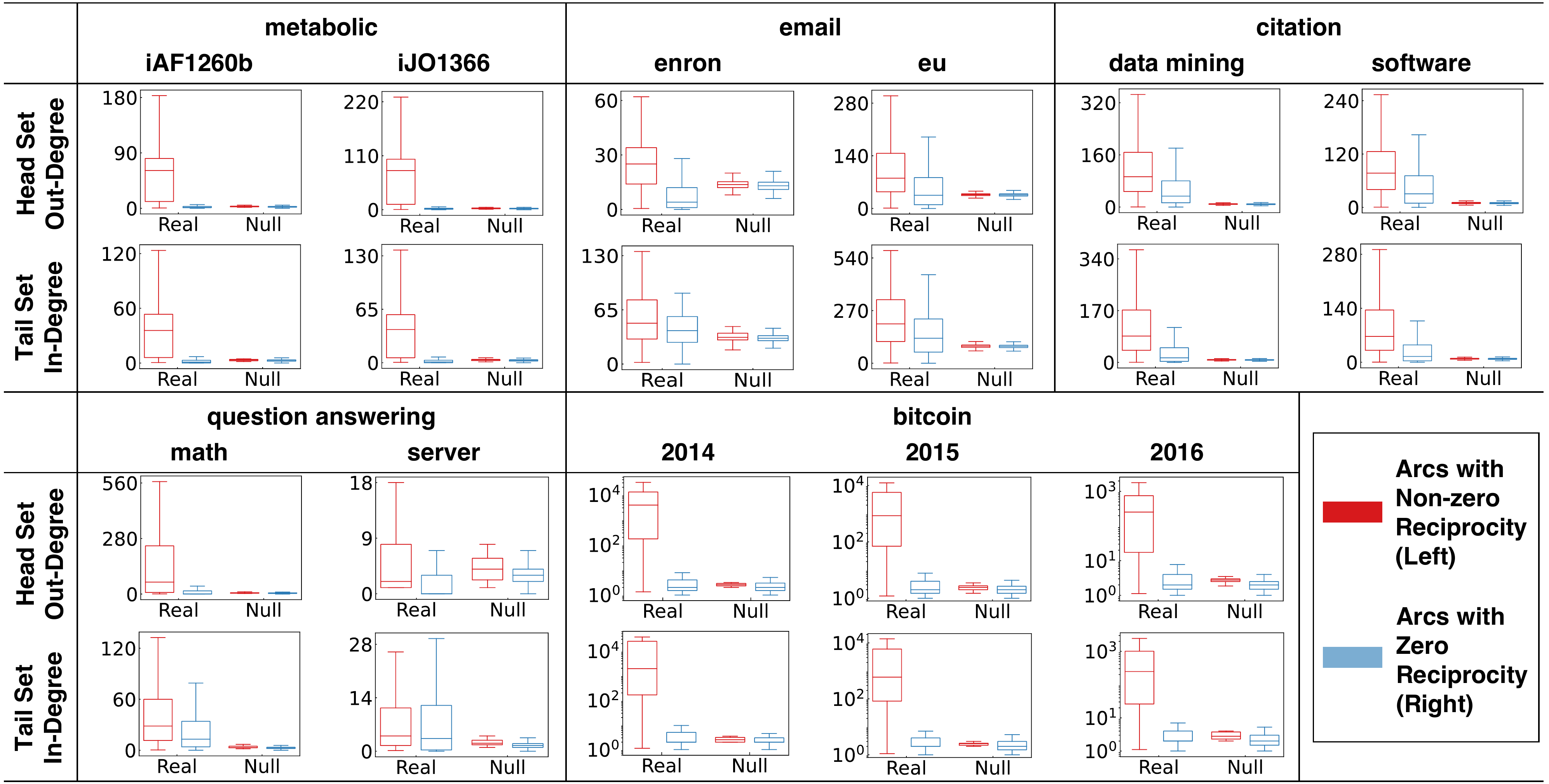}
  \caption{\label{fig:obs2}
  {Observation~\ref{obs:observation2}.
  In real-world hypergraphs,
  the (a) head set out-degree and the (b) tail set in-degree tend to be larger at arcs with non-zero reciprocity (left in each box plot) than at  arcs with zero reciprocity (right in each box plot),
  while there is no such trend in null hypergraphs.} \returncolor}
%  For each box-plot, two left side plots describe real-world hyperarcs' behavior, and right two show null hyperarcs'. }
%   By comparing  with \textbf{\textcolor{boxplotblue}{zero reciprocity arcs}}, we discover that real-world hyperarcs with non-zero reciprocity tend to show high head set out-degree and tail set in-degree than zero reciprocity arcs.\red{TODO} \label{fig:obs2}  }
\end{figure*}

% \begin{table}[t]
%     \caption{Relative values of arc-level reciprocity are robust to the choice of $\alpha$, as supported by the large correlation coefficients.} % title of Table
%     \centering % used for centering table
%     \scalebox{1.0}{
%     \renewcommand{\arraystretch}{1.2}{
%     \begin{tabular}{c | c  c  c | c  c  c } % centered columns (4 columns)
%         \toprule %inserts double horizontal lines
%         \multirow{2}{*}{Dataset} & \multicolumn{3}{c}{\textbf{\underline{Pearson Correlation}}} & \multicolumn{3}{|c}{\textbf{\underline{Rank Correlation}}} \\
%         & $0 \leftrightarrow 0.5 $ & $0 \leftrightarrow 1 $ & $0.5 \leftrightarrow 1 $ & 
%         $0 \leftrightarrow 0.5 $ & $0 \leftrightarrow 1 $ & $0.5 \leftrightarrow 1 $ \\
%         \midrule
%         iaf1260b & 0.961 & 0.916 & 0.985 & 0.973 & 0.925 & 0.975 \\ 
%         %\midrule 
%         email-enron & 0.928 & 0.828 & 0.975 & 0.947 & 0.869 & 0.978 \\
%         %\midrule
%         citation-dm & 0.984 & 0.973 & 0.998 & 0.998 & 0.996 & 0.999 \\ 
%         %\midrule 
%         q\&a-math & 0.994 & 0.992 & 0.999 & 0.999 & 0.999 & 0.999 \\
%         %\midrule
%         bitcoin-2014 & 0.997 & 0.995 & 0.999 & 0.999 & 0.999 & 0.999 \\ 
%         \bottomrule
%     \end{tabular}
%     }
%     }
% \label{tab:arcrobustness} % is used to refer this table in the text
% \end{table} 

\begin{figure*}[t!]
\vspace{-3mm}
  \centering
  \includegraphics[width=\textwidth]{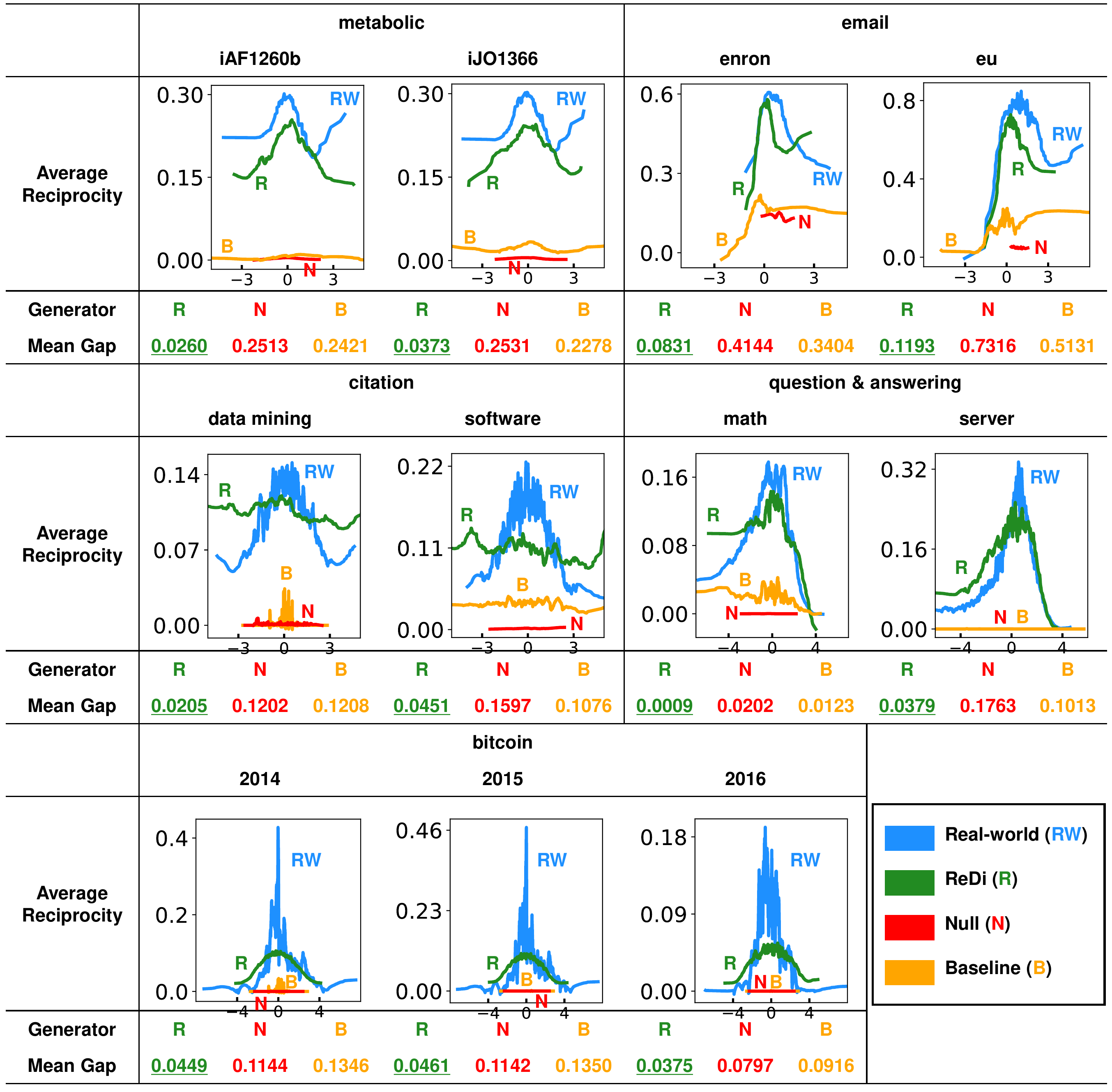}
%  \vspace{-1mm}
  \caption{\label{fig:obs3}\change \kijung{Observation~\ref{obs:observation3} and the superiority of \model.
  Relations between the degree balance and average reciprocity of nodes in
  real-world hypergraphs (RW), synthetic hypergraphs generated by \model (R) (Section~\ref{section:generation}), null hypergraphs (N), and those generated by a baseline generator (B)
  are reported. 
  For each dataset, we also report the mean gap values (Eq.~\eqref{eq:meangap}) from the real-world hypergraph.
  In real-world hypergraphs and those generated by \model,  node-level reciprocity tends to increase as nodes' in- and out-degrees become balanced. The curves from the hypergraphs generated by \model are most similar to those from real-world hypergraphs, as supported numerically by the smallest gaps from the real-world hypergraphs.} \returncolor
  }
\end{figure*}

\smallsection{L3. Node Level}
{Lastly, we investigate reciprocal patterns at the node level.
A node is called \textit{degree-balanced}, when its in-degree and out-degree are similar.
One may suspect that degree-balanced nodes tend to be involved in highly reciprocal arcs, as the number of incoming arcs and outgoing arcs are similar at them.
To verify this hypothesis, we measure the \textit{degree balance of each node $v$}, which we define as $x(v) := log(d_{in}(v) + 1) - log(d_{out}(v) + 1)$, and measure \textit{node-level reciprocity} $r(v)$, which we define as 
\begin{equation}\label{noderecip}
    r(v) = \frac{1}{ \lvert E_{v} \vert } \sum_{e_{k} \in E_{v}} r(e_{k}) \\
\end{equation}
where $E_{v}=\{e_{k} : v \in (H_{k} \cup T_{k})\}$ is the set of arcs where $v$ is included in its head set or its tail set.
} \returncolor
{Figure~\ref{fig:obs3} shows how the average node-level reciprocity depends on the degree balance of nodes after applying the Savitzky–Golay filter \citep{savitzky1964smoothing} for smoothing the curves.   
%Each coordinate of plot is $(x(v), y_{v})$ where $y_{v}$ is an average of $\{r(v') : v' = x(v)\}$. 
The curves from the real-world hypergraphs are bell-shaped with maximum values around zero, implying node-level reciprocity gets larger as nodes' in- and out-degrees become balanced. On the other hand, such a tendency is not clear in null hypergraphs.
} \returncolor
% From this pattern we discover that nodes of high degree balance tend to participate in high reciprocity arcs.
{\begin{obs}~\label{obs:observation3}
There is a tendency that
degree-balanced nodes participate more in arcs with high reciprocity.
\end{obs}
} \returncolor
% \red{KJ: I am here}
% \textbf{\textsc{Observation 3.}} \textit{Hyperarcs consist of degree-balanced nodes tend to have higher reciprocity.} \\

}

\section{Directed Hypergraph Generation: \model}{
    \label{section:generation}

In this section, we propose \model (\underline{\textbf{Re}}ciprocal and \underline{\textbf{Di}}rectional Hypergraph Generator), a realistic generative model of directed hypergraphs.
We first describe \model. Then, we demonstrate its successful reproduction of the reciprocal properties of real-world hypergraphs examined in Section~\ref{section:obs}. 
In addition to testing our understanding of the patterns, \model can also be used for anonymization, graph upscaling, etc \citep{leskovec2008dynamics}.

%.After introducing \generator, we validate whether observation 1, 2 and 3 are maintained in generated hypergraphs. 

\subsection{Model Description} 

% After providing a high-level idea behind \model, we describe each step in detail. % the explanation of each generation step.

\smallsection{High-level Introduction to \model}
Given some basic hypergraph statistics and three hyperparameter values, \model generates a directed hypergraph with realistic structural and reciprocal patterns.
\model is largely based on \hyperpa \citep{do2020structural}, an extension of the preferential attachment model \citep{albert2002statistical} to hypergraphs. %Among them, \hyperpa \cite{do2020structural} extends the preferential attachment model to hypergraphs. 
In \hyperpa, each new node forms hyperedges with groups of nodes that are drawn with probability proportional to the degree of groups (i.e., the number of hyperedges containing each group).
Introducing the degree of groups, instead of the degree of individual nodes, tends to lead to more realistic higher-order structures of generated graphs \citep{do2020structural}.
\model extends \hyperpa, which only can generate undirected hypergraphs, to generate directed hypergraphs and especially those with realistic reciprocal patterns. In a nutshell, \model stochastically creates reciprocal arcs while controlling the number of reciprocal arcs and their degree of reciprocity.

\begin{algorithm}[t!]
    \small
    \caption{{\dhgpa: Realistic Directed Hypergraph Generator} \returncolor}\label{alg:dhgpa}\footnotetext{$y_0$ denotes the initial value.}
    \KwIn{(1) Number of nodes $n$ and number of initial arcs $N$ \\
        \hspace{10mm} (2) Distribution of hyperedge head, tail set size $f_{HD}, f_{TD}$ \\
        \hspace{10mm}  (3) Distribution of number of new hyperedges $f_{NP}$ \\
        \hspace{10mm}  (4) Reciprocal  hyperparameter $\beta_{1}$ and $\beta_{2}$}
    \KwOut{Generated hypergraph $G=(V,E)$.}
    \setcounter{AlgoLine}{0}
    % $B^{*}(n,p)$ denotes binomial random sampling with parameter n and p. \\
    % $p.i.$ denotes proportional to in-degree. \\
    % $p.o.$ denotes proportional to out-degree. \\
    % $GS$ denotes group-wise sampling. 
    %\Statex aaa \\

    Initialize $G$ with $N$ arcs (w/ $1$ head \& $1$ tail) with $2N$ nodes 
    
    \ForEach{node $v_i \in \{v_1,\cdots, v_{n}\}$}{
        
        $k\leftarrow$  a number sampled from $f_{NP}$
        
        $V\leftarrow V \cup \{v_i\}$
        
        $E_i \leftarrow \{\}$
        
        \For{$j \leftarrow 1 \ \text{to} \ k$}{
           
            \While{True}{
                $recip. \leftarrow B(1, \beta_{1});$ \ $head \leftarrow B(1, 0.5)^a$
                
                $h, t \leftarrow$ arc sizes sampled from $f_{HD},f_{TD}$
                
                % $recip$ $\leftarrow$  \\
                \If{$recip. = 0$}{
                    \If{$head=1$}{
                        $H' \leftarrow \{v_i\}$; \ $T'\leftarrow \emptyset$
                    }
                    \Else{
                    
                        $T' \leftarrow \{v_i\}$; \ $H'\leftarrow \emptyset$
                    }
                    % \If{$h > |S_{ij}|$}{
                        $H' \leftarrow  H'$ $\cup$ (($h - \lvert H' \vert $) nodes sampled${}^{b}$)
                    %}
                    
                        $T' \leftarrow  T'$ $\cup$ (($t - \lvert T' \vert $) nodes sampled${}^{c}$)
                }
                \Else{
                    $e_o \leftarrow$ an arc sampled${}^{d}$ from $E_i$
                    
                    % \If{$E_i=\emptyset$}{
                    %     $e \leftarrow$ sample an arc from $E_i$
                    % }
                    % \Else{
                    %     $e \leftarrow$ sample an arc from $E_i$ 
                    % }
                    $n_{H} \leftarrow $ $B(\min(h, \lvert T_o \vert ), \beta_2)^a$
                    
                    $H'\leftarrow$  $\max(n_{H},1)$ nodes sampled${}^{e}$ from $T_o$
                    
                    $n_{T} \leftarrow $ $B(\min(t, \lvert H_o \vert ), \beta_2)^a$
                    
                    $T'\leftarrow$  $\max(n_{T},1)$ nodes sampled${}^{f}$ from $H_o$
                    
                    \If{$head=1$ \normalfont{and} $h > \lvert H' \vert $}{
                        $H' \leftarrow H' \cup \{v_i\}$
                    }
                    \uElseIf{$head=0$ \normalfont{and} $t > \lvert T' \vert $}{
                        $T' \leftarrow T' \cup \{v_i\}$
                    }
                    % \If{$h > |S_{ij}|$}{
                        $H' \leftarrow  H' \cup$ (($h - \lvert H'\vert $) nodes sampled${}^{b}$)
                        
                    % }
                        $T' \leftarrow  T' \cup$ (($t - \lvert T' \vert$) nodes sampled${}^{c}$)
                }
                \If{$H' \cap T' = \emptyset$}{
                    \textbf{break} the while loop 
                }  
            }
            $E \leftarrow E \cup \{\langle H',T'\rangle\}$; \ $E_i = E_i \cup \{\langle H',T'\rangle\}$
        }
    }
    \textbf{return} $G=(V,E)$ 
    
    %\nonl ${}^a$ $B(n,p)$ denotes binomial sampling with parameters $n$ and $p$ \\
    %\nonl ${}^{b}$  with probability proportional to group in-degree \citep{do2020structural} \\
    %\nonl ${}^{c}$ with probability proportional to group out-degree \citep{do2020structural} \\
    %\nonl ${}^{d}$ uniformly at random from $E_i$ (or from $E$ if $E_i=\emptyset$) \\
    %\nonl ${}^{e}$ with probability proportional to node in-degree  \\
    %\nonl ${}^{f}$ with probability proportional to node out-degree  \\
    %\nonl ${}^{f}$ with probability proportional to node out-degree  \\
    
    \nonl${}^{a}$ $B(n,p)$ denotes binomial sampling with parameters $n$ and $p$
    
    \nonl ${}^{b}$  with probability proportional to group in-degree \citep{do2020structural}
    
    \nonl ${}^{c}$ with probability proportional to group out-degree \citep{do2020structural}
    
    \nonl ${}^{d}$ uniformly at random from $E_i$ (or from $E$ if $E_i=\emptyset$) 
    
    \nonl ${}^{e}$ with probability proportional to node in-degree 
    
    \nonl ${}^{f}$ with probability proportional to node out-degree
    
\end{algorithm}

%  \cite{do}, which is an extension of 

% First concept of \dhgpa is a \emph{reciprocity preserving preferential attachment}. As preferential attachment  has its capacity in expressing real-world graph's degree distribution, our model is also based on it. 
% However, naive preferential attachment cannot reproduce the reciprocity of real-world hypergraph. To maintain the real reciprocal patterns, we stochastically create reciprocal arcs. 
% We control how many arcs to be reciprocal, and extent of reciprocity of reciprocal arcs by using certain hyperparameters.

% Second concept is a \emph{group-level interaction}. Idea and observations of group-level interaction was proposed in \cite{do2020structural}, which indicates a specific group interaction is frequently occurring together in a hypergraph. 
% For example, if an undirected hyperarc $e_{i}=\{v_{1}, v_{2}, v_{3}\}$ is found in a given undirected hypergraph, it is likely that interaction of $e_{j} = \{v_{1}, v_{2}, v_{3}, v_{4}\}$ or $e_{j} = \{v_{1}, v_{2}, v_{3}, v_{5}\}$ also exist in the same hypergraph. 
% To enable hypergraph generator reproduce such characteristic, they proposed \textsc{HyperPa} which samples nodes in a group-level. 
% Similarly, we also sample nodes in a group-wise manner with respect to the group-wise degree.

\smallsection{Details of \model} 
The pseudocode of \model is provided in Algorithm \ref{alg:dhgpa}. It requires three hyperparameters: (a) a proportion $\beta_{1} \in [0, 1]$ of reciprocal arc, (b) their extent $\beta_{2} \in [0, 1]$ of reciprocity, and (c) the number $N$ of initial nodes. 
In addition, \model requires the following statistics that it preserves in expectation: (a) the number $n$ of nodes, (b) the distributions $f_{HD}$ and $f_{TD}$ of the head-set and tail-set sizes, and (c) the distribution $f_{NP}$ of the number of new arcs per node.
\kijung{We adopt $NP$ distribution suggested in~\citep{do2020structural} as our $f_{NP}$.}

At each step, \model introduces a new node $v_i$ and creates $k$ arcs where $k$ is sampled from $f_{NP}$. %We use the same $f_{NP}$ as suggested in \cite{do2020structural}.
Before creating a new arc, we decide whether it to be reciprocal (with prob. $\beta_{1}$) or ordinary.
After deciding the size of a new arc according to the sizes sampled from $f_{HD}$ and $f_{TD}$, we decide whether to include $v$ into the head set (with prob. $0.5$) or the tail set. 

If a new arc is decided to be ordinary, we include \modify{$v_i$} in either the head set or the tail set according to the choice made beforehand. Subsequently, we fill the new arc with nodes sampled based on in- and out-degrees of groups (i.e., the number of arcs that include the group in their head set and tail set, respectively). Note that the head set and the tail set should be disjoint for both reciprocal and ordinary arcs.

If a new arc is decided to be reciprocal, we choose an opponent arc $e_{o}$ uniformly at random among those with $v_i$ (or among all existing arcs if no arc contains $v_i$).
Then, we decide how many nodes are brought from the opponent arc's head set and tail set by binomial sampling with  probability $\beta_{2} \in [0, 1]$. 
After sampling nodes from the opponent arc with probability proportional to their degree, we fill the new arc with $v_i$ and those sampled based on in- and out-degrees of groups.

%Together with \dhgpa generation result, we provide generated hypergraphs of this baseline models for comparison.

\subsection{Evaluation of \model} 
\label{section:generation:eval}
We evaluate how well \model can reproduces the reciprocal patterns of real-world hypergraphs discussed in Section~\ref{section:obs}. For each real-world hypergraph, we generate 5 hypergraphs using their statistics and report the average of generated statistics.\footnote{The search space of $\beta_{1}$ is (a) $[0.05, 0.1, \cdots , 0.6]$ for the small datasets where $\lvert V \vert \leq 10^{4}$, and (b) $[0.001, 0.0015, \cdots ,0.005]$ for the dense large datasets where $\lvert V \vert  > 10^{4}$ and ${\lvert E \vert}/{\lvert V \vert} \geq 3$, and (c) $[0.01, 0.02, \cdots 0.15]$ for the other sparse large datasets. The search space of $\beta_{2}$ is fixed to $\in [0.1, 0.1, \cdots , 0.5]$ for all datasets.} 
In addition, we introduce a naive preferential attachment model, as a \textbf{baseline model} for comparison, to clarify the necessity of the reciprocal edge generation step.
The baseline model is identical to \dhgpa, except only for that it always decides to create ordinary arcs, i.e., $\beta_{1}=\beta_{2}=0$.\footnote{As bitcoin transactions are made among randomly chosen accounts, the repetition of (partial) group interactions is rarely observed. Due to this intrinsic characteristic of the datasets, we use the degrees of individual nodes instead of the degrees of groups when \dhgpa and the baseline model are given the statistics from bitcoin datasets.
The same strategy is also used for the baseline model when the input statistics are from the q\&a server dataset. Without the strategy, the baseline model takes more than $12$ hours.}

\begin{figure}[t!]
  \centering
  \includegraphics[width=1.0\textwidth]{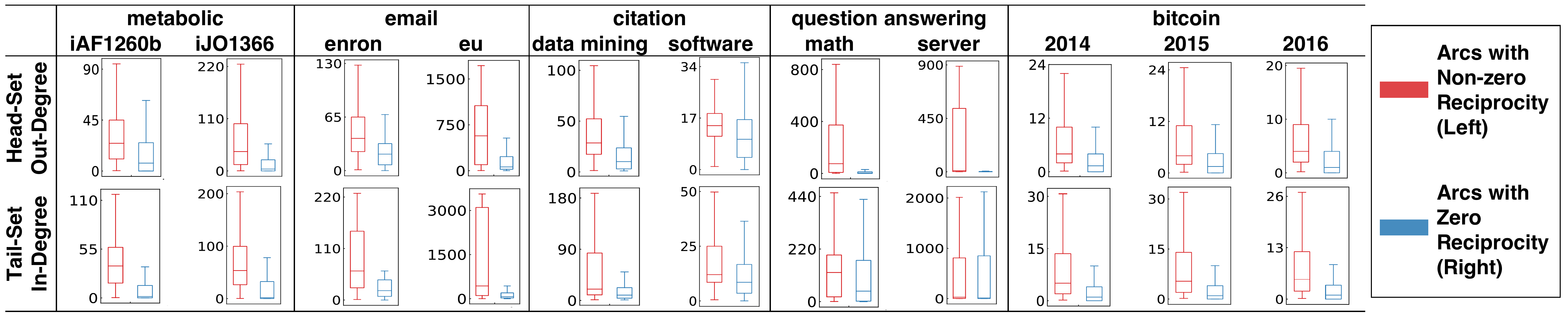}
  \caption{{Hypergraphs generated by \model exhibits Observation~\ref{obs:observation2}, which is a pervasive pattern in real-world hypergraphs, as shown in Figure~\ref{fig:obs2}} \returncolor}\label{fig:obs2gen}
\end{figure}

\smallsection{Reproducibility of Observation~\ref{obs:observation1}} 
We measure the reciprocity of generated hypergraphs at the hypergraph and arc levels and compare it with that of real-world hypergraphs. 
As shown in Table~\ref{tab:observation1}, \model generates hypergraphs whose reciprocity is very close to that in the corresponding real-world hypergraphs both at the hypergraph and arc levels. The baseline model fails to reproduce high enough reciprocity in most cases.

%Proposal's generation result's hypergraph reciprocity tend to show similar absolute value with the real-world hypergraph. In addition, computed D-statistics are much smaller in \dhgpa than baseline and null hypergraphs which implies proposed method reproduced real-world hyperarc reciprocity distribution more realistically than other approaches (see ).

% We compare absolute value of real-world hypergraph's reciprocity with that of randomized one in observation 1. Similarly, 

%In addition, to investigate how much generated arc-level reciprocity is far from real-world, we measure D-statistic \cite{hodges1958significance} between them. We do the same process to the baseline generators either.

\smallsection{Reproducibility of Observation~\ref{obs:observation2}} Moreover, as shown in Figure~\ref{fig:obs2gen}, in the hypergraphs generated by \model, arcs with non-zero reciprocity tend to have higher (a) head set out-degree and (b) tail set in-degree than arcs with zero reciprocity, just as in the real-world hypergraphs.

%In order to verify this pattern in generated hypergraphs, we plot the degree distribution difference of non-zero reciprocity arcs and zero reciprocity arcs. 
%As shown in Figure \ref{fig:obs2gen}, generated hypergraphs also show similar distributional difference between non-zero reciprocity arcs and zero reciprocity arcs.

\smallsection{Reproducibility of Observation~\ref{obs:observation3}}
{Furthermore, as shown in Figure~\ref{fig:obs3}, the bell-shaped relation between the degree balance and average reciprocity of nodes in hypergraphs generated by \model is close to the relation in the corresponding real-world hypergraphs, as supported numerically by the smallest mean gaps (i.e., mean of squared differences) from the real-world hypergraphs. Formally, the mean gap is defined as follows:
\begin{equation}\label{eq:meangap}
    \text{\textit{mean-gap}}(f,f') =  \frac{1}{\lvert D \vert } \sum_{x \in D} (f(x) - f'(x))^{2},
\end{equation}
where $D$ is the intersection of the domains of $f$ and $f'$.
Specifically, the mean gaps are up to $13\times$ smaller in hypergraphs generated by \model than in those generated by the baseline model.} \returncolor

% To quantitatively compare the three generation methods, we measure \emph{mean-gap}, which is calculated as the average of the squared-gap between two plots, for the points where $x(v)$ values are identical. Among three, \dhgpa shows the least difference from the real-world hypergraphs.

% In real-world dataset, nodes where their in-degree and out-degree is balanced tend to be involved in relatively high reciprocity arcs. 
% We draw same distribution of observation 3 about real-world, \dhgpa, null hypergraphs and baseline. 
% One can clearly notice that \dhgpa (green) successfully captures the general bell-shaped pattern of real-world hypergraph. (See Figure \ref{fig:obs3})
% Neither null hypergraphs (red) nor baseline's hypergraphs (orange) show the real-world hypergraph's behavior.

}

\section{Conclusion}{
    \label{section:conclusion}
    In this paper, we perform a systematic and extensive study of reciprocity in real-world hypergraphs. We propose \measure, a family of probabilistic measures of reciprocity that guarantee all eight desirable properties (Table~\ref{tab:AXIOMsatisfy1}).
Our algorithmic contribution is to develop \algo, which enables rapid yet exact computation of \measure (Theorem~\ref{thm:algo:basic},  Figure~\ref{fig:ferret}, and Table~\ref{tab:runtime}). Using both, we discover several unique reciprocal patterns (Table~\ref{tab:observation1} and Figures~\ref{fig:obs2}-\ref{fig:obs3}) that distinguish real-world hypergraphs from random hypergraphs.
Lastly, we design \model, a simple yet powerful generator that yields realistic directed hypergraphs (Table~\ref{tab:observation1} and Figures~\ref{fig:obs3}-\ref{fig:obs2gen}).
For \textbf{reproducibility}, we make the code and all datasets available at \supplelink.
}

\section*{Declarations}
\begin{comment}
\subsection*{Funding}
\red{TBD}
This work was supported by Samsung Electronics Co., Ltd., National Research Foundation of Korea (NRF) grant funded by the Korea government (MSIT) (No. NRF-2020R1C1C1008296), and Institute of Information \& Communications Technology Planning \& Evaluation (IITP) grant funded by the Korea government (MSIT) (No. 2022-0-00157,
Robust, Fair, Extensible Data-Centric Continual Learning) (No. 2019-0-00075, Artificial Intelligence Graduate School Program (KAIST)).
\end{comment}

\subsection*{Competing interests}
The authors have no relevant financial or non-financial interests to disclose.

\bibliography{ref.bib}

\appendix

%\section{Appendix: Motivation and Necessity of Axioms}{
%    \label{section:axiomintuition}
%    \input{207NewAxiomIntuition}
%}
\section{Appendix: Proof of Theorems}{
    \label{section:theoremproof}

In this section, we provide proofs of theorems in the main paper. 
We first introduce some preliminaries of proofs and then prove why \kijung{\measure, the proposed family of reciprocity measures,} satisfies all of \textsc{\textbf{Axioms 1-8}}.
Then, we prove the exactness of \textbf{\algo} and the related complexity reduction techniques.

\subsection{Preliminaries of Proofs}

In this subsection, we first give the general form of our proposed measure. 
Then, we introduce several important characteristics of \textit{Jensen-Shannon Divergence} (JSD) \citep{lin1991divergence}, which plays a key role in our proofs. 
After that, we examine how these concepts are applied to our measure.
Basic symbols used for hypergraphs and arcs are defined in Section \ref{section:prelim:concept}.

\measure, the proposed reciprocity measure for an arc $e_{i}$ and for a hypergraph $G$ is defined as
\begin{equation}
    r(e_{i}) := \max_{R_{i} \subseteq E , R_{i} \neq \emptyset}{\left(\frac{1}{\lvert R_{i} \vert }\right)^{\alpha} 
    \left( 1 - \frac{\sum_{v_{h} \in H_{i}} \mathcal{L}(p_{h} , p^{*}_{h})}
    {\lvert H_{i} \vert \cdot \mathcal{L}_{max}} \right)},
\end{equation}
\begin{equation}
    r(G) := \frac{1}{\lvert E \vert}\sum_{i=1}^{\lvert E \vert} r(e_{i}),
\end{equation}
where $\mathcal{L}(p_{h}, p^{*}_{h})$ denotes  Jensen-Shannon Divergence \citep{lin1991divergence} between a transition probability distribution $p_{h}$ and the optimal transition probability distribution $p^{*}_{h}$. 

%Here, the transition probability from a head set node $v_h \in H_i$ to each node $v$, when the reciprocal set consists of a single arc $e'_{i}$ (i.e., $R_{i} = \{e'_{i}\}$), is defined as
%\begin{equation*}
%    p_{h}(v) = \begin{cases}
%    \frac{1}{\lvert H'_{i} \vert } & \text{if $v \in H'_{i}$} \\
%    0 & \text{otherwise}. \\
%\end{cases}
%\end{equation*}
%This is the case when $v_{h} \in H_{i} \cap T'_{i}$. 
%For each $v_{h} \in H_{i} \setminus T'_{i}$, $p_{h}(v)=1$ if $v = v_{sink}$ and $0$ otherwise. 
For a target arc $e_{j}$ with an arbitrary non-empty reciprocal set $R_{j}$, the transition probability is defined as
\begin{comment}
\begin{equation*}
    p_{h}(v) = \begin{cases}
    p_{h, 1}(v) & \text{if $v_h \in \bigcup_{e_{s} \in R_{j}}T_{s}$}, \\
    p_{h, 2}(v) & \text{otherwise}, \\
\end{cases}
\end{equation*}
where
\begin{align*}
p_{h, 1}(v)& = \frac{\sum_{e_{s} \in R_{j}}
           \left( \frac{\mathbf{1}[v_{h} \in T_{s} , v \in H_{s}]}{\lvert H_{s} \vert }\right)}
                    {\sum_{e_{s} \in R_{j}} (\mathbf{1}[v_{h} \in T_{s}])}, \\
p_{h, 2}(v)& = \begin{cases}
    1 & \text{if $v = v_{sink}$} \\
    0 & \text{otherwise},
    \end{cases}, \ \ \mathbf{1}[\text{TRUE}]=1, \ \text{and} \ \mathbf{1}[\text{FALSE}]=0.
\end{align*}
% Throughout the proof, we use these formal expressions for both single-reciprocal-arc and multiple-reciprocal-arc cases.
\end{comment}
%Now we discuss several theoretical properties of $\jsdpq{P}{Q}$. 

\begin{equation*}
    p_{h}(v) = \begin{cases}
    \frac{\sum_{e_{k} \in R_{j}}
           \left( \frac{\mathbf{1}[v_{h} \in T_{k} , v \in H_{k}]}{\lvert H_{k} \vert }\right)}
                    {\sum_{e_{k} \in R_{j}} (\mathbf{1}[v_{h} \in T_{k}])} & \text{if $v_h \in \bigcup_{e_{k} \in R_{j}}T_{k}$}, \\
    1 & \text{if $v = v_{sink}$ and $v_h \notin \bigcup_{e_{k} \in R_{j}}T_{k}$}, \\
    0 & \text{otherwise,}
\end{cases}
\end{equation*}
where $\mathbf{1}[\text{TRUE}]=1$, and $\mathbf{1}[\text{FALSE}]=0$.

In Lemma~\ref{lemma:jsd},
we provide several theoretical properties of JSD that are used for our proofs.
Note that a general form of $\jsdpq{P}{Q}$ is defined as
\begin{equation}\label{eq:JSD}
    \mathcal{L}(P,Q)=\sum_{i=1}^{n}
        \ell(p_{i},q_{i}),
\end{equation}
where
\begin{equation}
    \ell(p_{i},q_{i}) = {\frac{p_i}{2} \log{\frac{2p_i}{p_i + q_i}} 
        + \frac{q_i}{2} \log{\frac{2q_i}{p_i + q_i}}}. \label{eq:ell}   
\end{equation}
\begin{lemma}[Basic Properties of Jensen-Shannon Divergence]
\label{lemma:jsd}
The Jensen-Shannon Divergence (JSD) has the following properties:
\begin{itemize}%[label=\textbf{{A.\Roman*.}},ref=\Roman*]
  \item \itembound. For any two discrete probability distributions $P$ and $Q$, $0 \leq \jsdpq{P}{Q} \leq \log 2$ holds. 
  \item \itemzero. For two discrete probability distributions $P$ and $Q$ where their non-zero-probability domains do not overlap (i.e., $p_{i}q_{i} = 0, \ \forall i = \{1, \cdots ,\lvert V \vert \} $), $\jsdpq{P}{Q}$ is maximized, and the maximum value is $\log{2}$. 
  \item \itemnonzero. Consider two discrete probability distributions $P$ and $Q$. If there exists a value where both $P$ and $Q$ have non-zero probability, $\jsdpq{P}{Q} < \log{2}$ holds. 
\end{itemize}
\end{lemma}
\begin{proof}
\begin{itemize}
    \item \textbf{(Proof of \itembound)} Refer to \citep{lin1991divergence} for a proof of \itembound.
    \item \textbf{(Proof of \itemzero)} Let $\mathcal{X}_{p}$ be the domain where $P$ has non-zero probability, and let $\mathcal{X}_{q}$ be the domain where $Q$ has non-zero probability. 
    Since $\mathcal{X}_{p}$ and $\mathcal{X}_{q}$ do not overlap, Eq.~\eqref{eq:JSD} is rewritten as 
    \begin{equation*}
        \mathcal{L}(P,Q) = \sum_{i\in\mathcal{X}_{p}}\frac{p_{i}}{2}\log 2 + \sum_{i\in\mathcal{X}_{q}}\frac{q_{i}}{2}\log 2 
        = \frac{\log 2}{2}\left(\sum_{i \in \mathcal{X}_{p}}p_{i} + \sum_{i \in \mathcal{X}_{q}}q_{i} \right)  
        = \log 2.
    \end{equation*}
  \item \textbf{(Proof of \itemnonzero)} Let $k$ be a point where $p_{k}q_{k} \neq 0$ holds. Then,
  Eq.~\eqref{eq:JSD} is rewritten as
    \begin{equation}\label{eq:nonoverlap}
        \mathcal{L}(P,Q) \  \leq
        \sum_{i\in\mathcal{X}_{p}\setminus k}\frac{p_{i}}{2}\log2 + \sum_{i\in\mathcal{X}_{q}\setminus k}\frac{q_{i}}{2}\log2 
        + \left({\frac{p_k}{2} \log{\frac{2p_k}{p_k + q_k}} 
        + \frac{q_k}{2} \log{\frac{2q_k}{p_k + q_k}}}\right).
    \end{equation}
    The below inequality implies that Eq.~\eqref{eq:nonoverlap} is smaller than $\log 2$.
    $$
    \left({\frac{p_k}{2} \log{2} 
    + \frac{q_k}{2} \log{2}}\right) - 
    \left({\frac{p_k}{2} \log{\frac{2p_k}{p_k + q_k}} 
    + \frac{q_k}{2} \log{\frac{2q_k}{p_k + q_k}}}\right) > 0
    $$
    $$
    \equiv {\frac{p_k}{2} \log{\left(1 + \frac{q_k}{p_k}\right)} 
    + \frac{q_k}{2} \log{\left(1 + \frac{p_k}{q_k}\right)}} > 0 \quad (\because p_{k},q_{k} > 0).
    $$
    Since $\log(x)>0$ holds for any $x>1$, the last inequality holds. Thus, we can conclude that $\jsdpq{P}{Q} < \log 2$ holds in this case.
\end{itemize}
\end{proof}

In Lemma~\ref{lemma:measure},
we provide several basic properties of \measure, our proposed measure of reciprocity in hypergraphs.

\begin{lemma}[Basic Properties of \measure]
\label{lemma:measure}
\measure \normalfont{(}i.e., defining $r(e_i , R_i)$ as in Eq.~\eqref{eq:proposal}\normalfont{)} has the following properties:
\begin{itemize}%[label=\textbf{{A.\Roman*.}},ref=\Roman*]
  \item \itemzeroone. If a target arc's head set and the tail sets of its reciprocal arcs do not overlap, the target arc's reciprocity becomes zero. Formally,
  $$\text{If } H_{i} \cap \bigcup_{e_{k} \in R_{i}} T_{k} = \emptyset \quad \text{ then } \quad r(e_{i} , R_{i}) = 0.$$ 
  \item \itemzerotwo. If a target arc's tail set and the head sets of its reciprocal arcs do not overlap, the target arc's reciprocity becomes zero. Formally,
  $$\text{If } T_{i} \cap \bigcup_{e_{k} \in R_{i}} H_{k} = \emptyset \quad \text{ then } \quad r(e_{i} , R_{i}) = 0.$$ 
  \item \itemoverlap. If (a) a target arc's head set and the tail sets of its reciprocal arcs overlap and (b) the target arc's tail set and the head sets of its reciprocal arcs overlap, then the target arc's reciprocity is greater than zero. Formally,
  $$\text{If }  \sum_{e_{k} \in R_{i}}\lvert H_{i} \cap T_{k}\vert  \cdot \lvert T_i \cap H_{k} \vert  \geq 1 \quad \text{ then } \quad r(e_{i} , R_{i}) > 0.$$ 
\end{itemize}
\end{lemma}
\begin{proof}
Below, we use $\mathcal{L}_{max}$ to denote the maximum value of JSD, which is $\log 2$.
\begin{itemize}
    \item \textbf{(Proof of \itemzeroone)}
    For this case, as mentioned in Section~\ref{sec:measure:measure}, the probability mass is non-zero only at $v_{sink}$. 
    On the other hand, the optimal transition probability $p^{*}$ is non-zero only at each $v \in T_{i}$.
    Since $v_{sink} \not\in T_{i}$, the non-zero-probability domains of the transition probability and the optimal transition probability do not overlap, and by \itemzero, the probabilistic distance between them is maximized. 
    This happens for all $v \in H_{i}$. Therefore,
    \begin{align*}{
    r(e_i , R_i) &=  \left( \frac{1}{\lvert R_{i} \vert } \right)^\alpha \left(1 - \frac{\sum_{v_h \in H_i}
        \mathcal{L}_{max}}{\lvert H_i \vert \cdot \mathcal{L}_{max} }\right) \\
    &=  \left( \frac{1}{\lvert R_{i} \vert } \right)^\alpha  \left(1 - \frac{\lvert H_{i} \vert  \cdot
        \mathcal{L}_{max}}{\lvert H_i \vert \cdot \mathcal{L}_{max} }\right) = \left( \frac{1}{\lvert R_{i} \vert } \right)^\alpha \times 0 = 0.
    }\end{align*}
    \item \textbf{(Proof of \itemzerotwo)} As in \itemzeroone, the non-zero probability domains of the transition probability and the optimal transition probability do not overlap since $T_{i} \cap \bigcup_{e_{k} \in R_{i}} H_{k} = \emptyset$. 
    Again, the probabilistic distance is maximized.
    This happens for all $H_{i}$, i.e.,
    \begin{align*}{
    r(e_i , R_i) &=  \left( \frac{1}{\lvert R_{i} \vert} \right)^\alpha \left(1 - \frac{\sum_{v_h \in H_i}
        \mathcal{L}_{max}}{\lvert H_i \vert \cdot \mathcal{L}_{max} }\right) \\
    &=  \left( \frac{1}{\lvert R_{i} \vert } \right)^\alpha  \left(1 - \frac{\lvert H_{i} \vert \cdot
        \mathcal{L}_{max}}{\lvert H_i \vert  \cdot \mathcal{L}_{max} }\right) = 0
    }\end{align*}
    \item \textbf{(Proof of \itemoverlap)} According to the statement, there exists at least one reciprocal arc $e_{k}$ whose (a) tail set overlaps with the target arc's head set (i.e., $\lvert T_{k} \cap H_{i} \vert \geq 1$) and (b) head set overlaps with the target arc's tail set (i.e., $ \lvert H_{k} \cap T_{i} \vert \geq 1$).
    Thus, for $v_{h} \in H_{i} \cap T_{k}$, $p_{h}$ and $p^{*}_{h}$ share non-zero probability domains, which implies {$\mathcal{L}(p_{h}, p^{*}_{h}) < \log{2}$} by \itemnonzero.
    Hence, we can derive the following inequality: % (\textbf{\underline{\smash{see the inequality sign between two middle terms}}}).
    \begin{align*}{
    r(e_i , R_i) 
    &=  \left( \frac{1}{\lvert R_{i} \vert } \right)^\alpha \left(1 - \frac{\sum_{v_h \in H_i}
        \mathcal{L}(p_{h} , p^{*}_{h})}{\lvert H_i \vert \cdot \mathcal{L}_{max} }\right)  \\
    &> \left( \frac{1}{\lvert R_{i} \vert} \right)^\alpha \left(1 - \frac{\lvert H_{i} \vert \cdot
        \mathcal{L}_{max}}{\lvert H_i \vert \cdot \mathcal{L}_{max} }\right) = 0. %\qedhere
    }\end{align*}
\end{itemize}
\end{proof}

\subsection{Proof of Theorem 1.}\label{subsec:thm1}
In this section, %we provide a generalized version of each axiom and 
we show that the proposed measure \measure satisfies all the generalized axioms.
For a proof of {\textbf{\textsc{Generalized Axiom 1}}}, we simply show that the former's reciprocity gets zero (i.e., $r(e_{i} , R_{i}) = 0$), while the latter's reciprocity gets a positive value (i.e., $r(e_{j} , R_{j}) > 0$). 
For proofs of {\textbf{\textsc{Generalized Axioms 2-4}}}, we first show how the formal statement of each axiom can be written in terms of the probabilistic distance. 
Then, we derive a less reciprocal case has a higher probabilistic distance between the transition probability and the optimal transition probability for every head set node of a target arc. 
%An overview of \textsc{\textbf{Axioms 1-4}} is illustrated in Figure~\ref{fig:AXIOMs}. 
%\textbf{In \textsc{Axioms} 2-4, we commonly assume that two target arcs $e_{i}$ and $e_{j}$ have the same size (i.e., ${(\lvert H_{i} \vert  = \lvert H_{j} \vert )} \wedge {(\lvert T_{i}\vert  = \lvert T_{j}\vert )}$)}.
%% Identical size is keep mentioned. Thus removed the sentences.

%-----------------------------    AXIOM1  --------------------------------

\subsubsection{Proof of the Fact that \measure Satisfies Axiom 1}\label{subsec:axiom1} 

% \red{\axiomnum{1} explains a arc which has at least one inversely-overlapping arc is more reciprocal than a arc without any inversely-overlapping arc (formal definition of inversely-overlapping arc is explained in axioms explanation of Section~\ref{sec:measure:axioms}).} 

Through an example, \axiomnum{1} states that an arc that has at least one inversely-overlapping arc should be more reciprocal than an arc without any inversely-overlapping arc.
\kijung{The generalized statement of \axiomnum{1} is formalized in \textbf{\textsc{Generalized Axiom~\ref{newaxiom:inverse:overlap}}}.}
%Below, we generalize \axiomnum{1} and prove that \measure satisfies the generalized version.

\begin{comment}
\begin{newaxiombox}[Existence of Inverse Overlap]{
\label{newaxiom:inverse:overlap}
Consider two arcs $e_{i}$ and $e_{j}$.
If $R_i = \{e'_i\}$ and $R_j = \{e'_j\}$ satisfy
\begin{gather*}
\min(\lvert H_i \cap T'_i \vert , \lvert T_i \cap H'_i \vert ) = 0 \ \textrm{and} \ \min(\lvert H_j \cap T'_j \vert , \lvert T_j \cap H'_j \vert ) \geq 1,
\end{gather*}
then the following inequality holds:
$$
r(e_i , R_i) < r(e_j , R_j).
$$
}
\end{newaxiombox}
\end{comment}

\begin{prop}
\measure \normalfont{(}i.e., defining $r(e_i , R_i)$ as in Eq.~\eqref{eq:proposal}\normalfont{)}
satisfies Generalized Axiom~\ref{newaxiom:inverse:overlap}.
\end{prop}
\begin{proof}
\kijung{
We first show that $r(e_{i}, R_{i}) = 0$. 
The suggested condition ($\forall e'_{i} \in R_{i}: \min(\vert H_{i} \cap T'_{i}\vert , \vert T_{i} \cap H'_{i}\vert) = 0$) implies that every reciprocal arc $e'_{i} \in R_{i}$ lies in one of the following cases (for simplicity, we refer to $e'_{i} \in R_{i}$ as a \textit{non-influential} arc if $e'_{i}$ {does not contribute to making} the non-zero probability domains of the transition probability that {overlap} with that of the optimal transition probability):
(1) When $(H_{i} \cap T'_{i} = \emptyset) \wedge (T_{i} \cap H'_{i} \neq \emptyset)$ holds, as shown in the proof of \itemzeroone, $e'_{i}$ is \textit{non-influential}.
(2) When $(H_{i} \cap T'_{i} \neq \emptyset) \wedge (T_{i} \cap H'_{i} = \emptyset)$ holds, as shown in the proof of \itemzerotwo, $e'_{i}$ is \textit{non-influential}.
(3) When $(H_{i} \cap T'_{i} = \emptyset) \wedge (T_{i} \cap H'_{i} = \emptyset)$ holds, as shown in the proof of \itemzero and \itemzerotwo, $e'_{i}$ is \textit{non-influential}.
That is, all reciprocal hyperarcs in $R_{i}$ are \textit{non-influential}, and by \itemzero, the distance between the transition probability and the optimal transition probability is maximized. 
This happens for all $v \in H_{i}$. Therefore, 
\begin{align*}{
    r(e_i , R_i) &=  \left( \frac{1}{\lvert R_{i} \vert } \right)^\alpha \left(1 - \frac{\sum_{v_h \in H_i}
        \mathcal{L}_{max}}{\lvert H_i \vert \cdot \mathcal{L}_{max} }\right) \\
    &=  \left( \frac{1}{\lvert R_{i} \vert } \right)^\alpha  \left(1 - \frac{\lvert H_{i} \vert  \cdot
        \mathcal{L}_{max}}{\lvert H_i \vert \cdot \mathcal{L}_{max} }\right) = \left( \frac{1}{\lvert R_{i} \vert } \right)^\alpha \times 0 = 0.
    }\end{align*}
}
\kijung{We now show that $r(e_{j}, R_{j}) > 0$ holds. 
The suggested condition ($\exists e'_{i} \in R_{i}: \min(\vert H_{i} \cap T'_{i}\vert , \vert T_{i} \cap H'_{i}\vert) \geq 1$) is equivalent to the condition of \itemoverlap. 
Thus, by \itemoverlap, the inequality $r(e_{j}, R_{j}) > 0$ holds.
Since $r(e_{i} , R_{i}) = 0$ and $r(e_{j} , R_{j}) > 0$, the following inequality holds: $r(e_{i} , R_{i}) < r(e_{j} , R_{j})$. 
}
\end{proof}
%\red{Sunwoo: About to prove this statement.}
\begin{comment} % Existing proof of the previous statement
\begin{proof} We first show that $r(e_{i} , R_{i}) = 0$. 
The condition $\min(\lvert H_{i} \cap T'_{i} \vert  , \lvert T_{i} \cap H'_{i} \vert ) = 0$ implies that at least one of $H_{i} \cap T'_{i}$ and $T_{i} \cap H'_{i}$ is an empty set. 
By \itemzeroone and \itemzerotwo, 
we can derive that reciprocity for such a case is zero, i.e., $r(e_{i} , R_{i}) = 0$.
Now we show that $r(e_{j} , R_{j}) > 0$. 
As $\min(\lvert H_j \cap T'_j \vert , \lvert T_j \cap H'_j \vert ) \geq 1$, we can guarantee that both $H_{j} \cap T'_{j}$ and $T_{j} \cap H'_{j}$ are non-empty sets. 
Thus, by \itemoverlap, the inequality $r(e_{j} , R_{j}) > 0$ holds. 
Since $r(e_{i} , R_{i}) = 0$ and $r(e_{j} , R_{j}) > 0$, the following inequality holds: $r(e_{i} , R_{i}) < r(e_{j} , R_{j})$. 
\end{proof}
\end{comment}

%-----------------------------    AXIOM2  --------------------------------
\subsubsection{Proof of the Fact that \measure Satisfies Axiom 2} \label{subsec:axiom2}

% \red{In this axiom, we discuss which arc should be more reciprocal betweeen two cases when a single reciprocal arc is given respectively.
% Here, let two target arcs $e_{i}$ and $e_{j}$, and their corresponding reciprocal arcs as $e'_{i}: R_{i} = \{e_{i}\}$ and $e'_{j}: R_{j} = \{e_{j}\}$.

Through an example, \axiomnum{2} states that
an arc that inversely overlaps with reciprocal arcs to a greater extent should be more reciprocal.
In addition, \axiomnum{2} is divided into two cases, {which are formalized in \textsc{\textbf{Generalized Axiom}~\ref{newaxiom2a}} and {\textsc{\ref{subaxiom2b}}} respectively.}

\begin{comment}
Below, we generalize
\axiomnum{2} and prove it by dividing it into two cases:
\begin{itemize}
    \item \textbf{\textsc{Axiom 2A:}}
    An arc that inversely overlaps with reciprocal arcs with a \textbf{larger intersection} is more reciprocal,
    \item \textbf{\textsc{Axiom 2B:}} An arc that inversely overlaps with reciprocal arcs with a \textbf{smaller difference} is more reciprocal. 
    %$e'_{j}$ inverse overlap with $e_{j}$ more accurately than $e'_{i}$ inverse overlap with $e_{i}$.
\end{itemize}
%Below, we separately generalize and prove \blue{the} two sub-axioms.

\begin{subaxiom2}[Degree of Inverse Overlap: More Overlap]
\label{newaxiom2a}

Consider two arcs  $e_{i}$ and $e_{j}$ of the same size $($i.e., ${(\lvert H_{i} \vert  = \lvert H_{j} \vert )} \wedge {(\lvert T_{i} \vert = \lvert T_{j} \vert )})$.
If $R_i = \{e'_i\}$ and $R_j = \{e'_j\}$ satisfy
\begin{gather*}
\lvert H'_i \vert = \lvert H'_j \vert, \ \lvert T'_i \vert = \lvert T'_j \vert, \ and \\
((i) \quad 0 < \lvert H'_i \cap T_i \vert < \lvert H'_j \cap T_j \vert \text{ and } 0 < \lvert T'_i \cap H_i \vert \leq \lvert T'_j \cap H_j \vert \quad \text{or} \\
(ii) \quad 0 < \lvert H'_i \cap T_i \vert \leq \lvert H'_j \cap T_j \vert \text{ and } 0 < \lvert T'_i \cap H_i \vert < \lvert T'_j \cap H_j \vert),
\end{gather*}
then the following inequality holds:
$$
r(e_i , R_i) < r(e_j , R_j).
$$
\end{subaxiom2}
\end{comment}
\begin{subprop2}{
\measure \normalfont{(}i.e., defining $r(e_i , R_i)$ as in Eq.~\eqref{eq:proposal}\normalfont{)}
satisfies  Generalized Axiom~\ref{newaxiom2a}.}
\end{subprop2}

\begin{proof}
Since $\lvert R_{i} \vert  = \lvert R_{j} \vert  = 1$, the cardinality penalty terms on both sides can be discarded, i.e.,
\begin{align*}
    r(e_{i} , R_{j}) = \left(\frac{1}{\lvert R_{i} \vert }\right)^{\alpha} 
    \left( 1 - \frac{\sum_{v_{i} \in H_{i}} \mathcal{L}(p_{h} , p^{*}_{h})}
    {\lvert H_{i} \vert \cdot \mathcal{L}_{max}} \right) 
    = 
    1 - \frac{\sum_{v_{i} \in H_{i}} \mathcal{L}(p_{h} , p^{*}_{h})}
    {\lvert H_{i} \vert \cdot \mathcal{L}_{max}}. %\text{Same for $r(e_{j} , R_{j})$ with a subscript $j$.}
\end{align*}
% Goal of a proof, , is equivalent to
% \begin{gather*} 
%     \frac{\sum_{v_{i} \in H_{i}} \mathcal{L}(p_{h} , p^{*}_{h})}
%     {\lvert H_{i} \vert \cdot \mathcal{L}_{max}} > \frac{\sum_{v_{j} \in H_{j}} \mathcal{L}(p_{h} , p^{*}_{h})}
%     {\lvert H_{j} \vert \cdot \mathcal{L}_{max}}
% \end{gather*}
Since $\lvert H_{i} \vert  = \lvert H_{j} \vert $, 
%the denominators are identical for both $i$ and $j$. }
%By removing them, 
the main inequality is rewritten as
\begin{align}
& r(e_i , R_i) <  r(e_j , R_j) \nonumber \\ 
&\equiv \sum_{v_h \in H_i}
    \mathcal{L}(p_{h} , p^{*}_{h}) > 
    \sum_{v_h \in H_j}
    \mathcal{L}(p_{h} , p^{*}_{h}). \label{eq:axiom2asummary}
\end{align}
Each head set can be divided into two parts: $H_{k} \setminus T'_{k} \ \text{and} \ H_{k} \cap T'_{k}, \ \forall k = i, j$.
For $H_{k} \setminus T'_{k}$, as described in \itemzeroone, the probabilistic distance is maximized to $\mathcal{L}_{max} = \log{2}$.
For $H_{k} \cap T'_{k}$, by using the fact that $T_{k} \cap H'_{k} \neq \emptyset$, $\forall k = i,j$, we can derive $\mathcal{L}(p_{h}, p^{*}_{h}) < \log{2}$ holds, $\forall v_{h} \in H_{k} \cap T'_{k}$, $\forall k=i,j$ by \itemoverlap. 
One more notable fact is that, since there is a single reciprocal arc for the target arc, $\mathcal{L}(p_{h}, p^{*}_{h})$ is the same for every $v_{h} \in H_{k} \cap T'_{k}$.
Here, let $\bar{p}_{k}, \forall k = i,j$ be the transition probability distribution regarding the target arc $e_{k}$ and its reciprocal set $R_{k}$. 
%\red{We first show that  the inequality~\eqref{eq:axiom2asummary} is equivalent to  $\mathcal{L}(p_{h, i} , p^{*}_{h, i}) > \mathcal{L}(p_{h, j} , p^{*}_{h, j})$ for both cases of (i) and (ii).}
We rewrite the inequality~\eqref{eq:axiom2asummary} 
\begin{align*}
&\sum_{v_h \in H_i} \mathcal{L}(p_{h} , p^{*}_{h}) > \sum_{v_h \in H_j} \mathcal{L}(p_{h} , p^{*}_{h}) \\
&
\equiv 
\lvert H_{i} \setminus T'_{i} \vert \times\log{2} + \lvert H_{i} \cap T'_{i}\vert \times\mathcal{L}(\bar{p}_{i} , \bar{p}^{*}_{i}) > \lvert H_{j} \setminus T'_{j} \vert \times\log{2} + \lvert H_{j} \cap T'_{j} \vert \times\mathcal{L}(\bar{p}_{j} , \bar{p}^{*}_{j}). 
\end{align*}
Below, we show that this inequality holds for Case $(i)$ and then Case $(ii)$.

\smallsection{Case $(i)$}
For Case $(i)$,
we first show that the inequality~\eqref{eq:axiom2asummary} is equivalent to $\mathcal{L}(\bar{p}_{i} , \bar{p}^{*}_{i}) > \mathcal{L}(\bar{p}_{j} , \bar{p}^{*}_{j})$.
The intersection of the target arc's head set and the reciprocal arc's tail set is larger for $e_{j}$ than for $e_{i}$.
Thus, the following inequality hold:
\begin{align*}
\lvert H_{i} \setminus T'_{i} \vert\times\log{2} + \lvert H_{i} \cap T'_{i} \vert\times\mathcal{L}(\bar{p}_{i} , \bar{p}^{*}_{i})  \geq \lvert H_{j} \setminus T'_{j} \vert\times\log{2} + \lvert H_{j} \cap T'_{j} \vert\times\mathcal{L}(\bar{p}_{i} , \bar{p}^{*}_{i}).
\end{align*}
Therefore,  Eq.~\eqref{eq:axiom2asummary} is implied by
\begin{align*}
&\lvert H_{j} \setminus T'_{j} \vert\times\log{2} + \lvert H_{j} \cap T'_{j} \vert\times\mathcal{L}(\bar{p}_{i} , \bar{p}^{*}_{i}) > \lvert H_{j} \setminus T'_{j} \vert\times\log{2} + \lvert H_{j} \cap T'_{j} \vert\times\mathcal{L}(\bar{p}_{j} , \bar{p}^{*}_{j}) \\
& \equiv \lvert H_{j} \cap T'_{j} \vert\times\mathcal{L}(\bar{p}_{i} , \bar{p}^{*}_{i}) > 
\lvert H_{j} \cap T'_{j} \vert\times\mathcal{L}(\bar{p}_{j} , \bar{p}^{*}_{j}) \\
&\equiv \mathcal{L}(\bar{p}_{i} , \bar{p}^{*}_{i}) > \mathcal{L}(\bar{p}_{j} , \bar{p}^{*}_{j}).
\end{align*}
Now, we show that $\mathcal{L}(\bar{p}_{i} , \bar{p}^{*}_{i}) > \mathcal{L}(\bar{p}_{j}, \bar{p}^{*}_{j})$ holds.
%Now, we only need to show that the probabilistic distance between the transition probability and the optimal transition probability is greater in $e_{i}$ than in $e_{j}$. 
To this end, denote the size of the intersection regions as $F_{1} = \lvert T_{i} \cap H'_{i} \vert \ < \ F_{2} = \lvert T_{j} \cap H'_{j} \vert$. 
We can decompose the domain of $v \in V$ into four parts as
$$
T_{k} \cap H'_{k}, \quad T_{k} \setminus H'_{k}, \quad 
H'_{k} \setminus T_{k}, \ \text{and} \ V \setminus \{H'_{k} \cup T_{k}\}, \quad \forall k = i,j
$$
For the last part, both the transition probability and the optimal transition probability of it have zero mass, i.e., $p_{h}(v) = p^{*}_{h}(v) = 0$, which results in no penalty. 
We only need to consider the first three parts for comparison. Here, the probabilistic distance can be explicitly written as
\begin{align*}
    \mathcal{L}(\bar{p}_{i}, \bar{p}^{*}_{i}) &= F_{1} \times\ell\left(\frac{1}{T}, \frac{1}{A}\right) + (A - F_{1} )\times\frac{1}{2A} \log{2} + (T - F_{1})\times\frac{1}{2T} \log{2}, \\ 
    \mathcal{L}(\bar{p}_{j}, \bar{p}^{*}_{j}) &= F_{2}\times\ell\left(\frac{1}{T}, \frac{1}{A}\right) + (A - F_{2} )\times\frac{1}{2A} \log{2} + (T - F_{2} )\times\frac{1}{2T} \log{2},
\end{align*}
where $\ell$ denotes a single-element comparison of $\jsdpq{P}{Q}$ in Eq.~\eqref{eq:ell}. Let $A = \lvert H'_{i} \vert = \lvert H'_{j} \vert$ and $T = \lvert T_{i} \vert = \lvert T_{j} \vert$. 
Then, we can rewrite $\mathcal{L}(\bar{p}_{i} , \bar{p}^{*}_{i}) - \mathcal{L}(\bar{p}_{j} , \bar{p}^{*}_{j}) > 0$ as
\begin{align*}
    &\mathcal{L}(\bar{p}_{i} , \bar{p}^{*}_{i}) - \mathcal{L}(\bar{p}_{j} , \bar{p}^{*}_{j}) > 0 \\ 
    &\equiv ( F_{1} - F_{2})\times\ell\left(\frac{1}{T}, \frac{1}{A}\right) + 
    ( F_{2} - F_{1} )\times\left(\frac{1}{2A} + \frac{1}{2T}\right)\log{2} > 0 \\
    &\equiv ( F_{2} - F_{1} ) \times \left(\frac{1}{2A}\log2 + \frac{1}{2T}\log2 - \ell\left(\frac{1}{T}, \frac{1}{A}\right) \right) > 0 \\
    &\equiv \left( \frac{1}{2A}\log2 - \frac{1}{2A} \log{\frac{2T}{A+T}} 
    + \frac{1}{2T}\log2 - \frac{1}{2T} \log{\frac{2A}{A+T}} 
    \right) > 0 \quad  \because  F_{2} - F_{1} > 0 \\
    &\equiv \frac{1}{2A}\log{\frac{A + T}{T}} + \frac{1}{2T}\log{\frac{A + T}{A}} > 0
\end{align*}
The inequality holds since $\log(x)>0$ holds for any $x>1$. Hence, we show $\mathcal{L}(\bar{p}_{i} , \bar{p}^{*}_{i}) > \mathcal{L}(\bar{p}_{j}, \bar{p}^{*}_{j})$ and thus the inequality~\eqref{eq:axiom2asummary} hold for Case $(i)$.

% \red{Thus,} if $\mathcal{L}(\bar{p}_{i} , \bar{p}^{*}_{i}) > \mathcal{L}(\bar{p}_{j}, \bar{p}^{*}_{j})$ holds, then \red{Eq.~\ref{eq:axiom2asummary} of case $(i)$} can be satisfied. 

\smallsection{Case $(ii)$}
For Case $(ii)$, 
we first show that the inequality~\eqref{eq:axiom2asummary} is equivalent to $\mathcal{L}(\bar{p}_{i} , \bar{p}^{*}_{i}) \geq \mathcal{L}(\bar{p}_{j} , \bar{p}^{*}_{j})$.
%can be identical, while the intersection of the target arc's tail set and the reciprocal arc's head set is greater in $e_{j}$ than in $e_{i}$. 
The inequality can be rewritten as
% \begin{comment}
% \begin{align*}
% &\lvert H_{i} \setminus T'_{i} \vert\times\log{2} + \lvert H_{i} \cap T'_{i} \vert\times\mathcal{L}(p_{h, i} , p^{*}_{h, i})  \geq \lvert H_{j} \setminus T'_{j} \vert\times\log{2} + \lvert H_{j} \cap T'_{j} \vert\times\mathcal{L}(p_{h, i} , p^{*}_{h, i})  \\
% & > \lvert H_{j} \setminus T'_{j} \vert\times\log{2} + \lvert H_{j} \cap T'_{j} \vert\times\mathcal{L}(p_{h, j} , p^{*}_{h, j})
% \end{align*}
% \end{comment}
\begin{align*}
    &(\lvert H_{i}\vert - \lvert H_{i} \cap T'_{i}\vert ) \times \log2 +  \lvert H_{i} \cap T'_{i}\vert \times \mathcal{L}(\bar{p}_{i} , \bar{p}^{*}_{i}) \\
    &> (\lvert H_{j}\vert - \lvert H_{j} \cap T'_{j}\vert ) \times \log2 +  \lvert H_{j} \cap T'_{j}\vert \times \mathcal{L}(\bar{p}_{j} , \bar{p}^{*}_{j}) \\ 
    &\equiv \lvert H_{i} \cap T'_{i}\vert \times (\mathcal{L}(\bar{p}_{i} , \bar{p}^{*}_{i}) - \log2) > 
    \lvert H_{j} \cap T'_{j}\vert \times (\mathcal{L}(\bar{p}_{j} , \bar{p}^{*}_{j}) - \log2) \\
    %&\equiv \lvert H_{j} \cap T'_{j}\vert \times (\log2 - \mathcal{L}(p_{h, j} , p^{*}_{h, j})) > 
    %\lvert H_{i} \cap T'_{i}\vert \times (\log2 - \mathcal{L}(p_{h, i} , p^{*}_{h, i})) \\ 
    &\equiv \frac{\lvert H_{j} \cap T'_{j}\vert}{\lvert H_{i} \cap T'_{i} \vert} > 
    \frac{\log2 - \mathcal{L}(\bar{p}_{i} , \bar{p}^{*}_{i})}{\log2 - \mathcal{L}(\bar{p}_{j} , \bar{p}^{*}_{j})} \\
    &\because \lvert H_{j} \cap T'_{j}\vert, \lvert H_{j} \cap T'_{j}\vert > 0 \text{ and } \log2 > \mathcal{L}(\bar{p}_{i} , \bar{p}^{*}_{i}), \mathcal{L}(\bar{p}_{j} , \bar{p}^{*}_{j}).
\end{align*}
By the condition of the axiom, the intersection of the target arc's head set and the reciprocal arc's tail set is larger than for $e_{j}$ than for $e_{i}$, and thus $\frac{\lvert H_{j} \cap T'_{j}\vert}{\lvert H_{i} \cap T'_{i} \vert} > 1$ holds. 
Thus, Eq.~\eqref{eq:axiom2asummary} is implied by $\frac{\log2 - \mathcal{L}(\bar{p}_{i} , \bar{p}^{*}_{i})}{\log2 - \mathcal{L}(\bar{p}_{j} , \bar{p}^{*}_{j})} \leq 1$, which is equivalent to $\mathcal{L}(\bar{p}_{i} , \bar{p}^{*}_{i}) \geq \mathcal{L}(\bar{p}_{j} , \bar{p}^{*}_{j})$, holds.
Now we show that $\mathcal{L}(\bar{p}_{i} , \bar{p}^{*}_{i}) \geq \mathcal{L}(\bar{p}_{j}, \bar{p}^{*}_{j})$ holds, which is rewritten as
\begin{align*}
    &\mathcal{L}(p_{h, i} , p^{*}_{h, i}) - \mathcal{L}(p_{h, j} , p^{*}_{h, j}) \geq 0 \\ 
    &\equiv ( F_{1} - F_{2})\times\ell\left(\frac{1}{T}, \frac{1}{A}\right) + 
    ( F_{2} - F_{1} )\times\left(\frac{1}{2A} + \frac{1}{2T}\right)\log{2} \geq 0 \\
    &\equiv ( F_{2} - F_{1} ) \times \left(\frac{1}{2A}\log2 + \frac{1}{2T}\log2 - \ell\left(\frac{1}{T}, \frac{1}{A}\right) \right) \geq 0 
\end{align*}
Note that, unlike Case $(i)$, where $F_{1} < F_{2}$ holds, $F_{1} \leq F_{2}$ holds for Case $(ii)$.
If $F_{2} = F_{1}$, then ths LHS above becomes 0, and thus above inequality holds.
If $F_{2} > F_{1}$, %then the following inequality holds:
\begin{align*}
    &( F_{2} - F_{1} ) \times \left(\frac{1}{2A}\log2 + \frac{1}{2T}\log2 - \ell\left(\frac{1}{T}, \frac{1}{A}\right) \right) \geq 0 \\
    &\equiv \left( \frac{1}{2A}\log2 - \frac{1}{2A} \log{\frac{2T}{A+T}} 
    + \frac{1}{2T}\log2 - \frac{1}{2T} \log{\frac{2A}{A+T}} 
    \right) > 0 \quad  \because  F_{2} - F_{1} > 0 \\
    &\equiv \frac{1}{2A}\log{\frac{A + T}{T}} + \frac{1}{2T}\log{\frac{A + T}{A}} > 0.
\end{align*}
The inequality holds since $\log(x)>0$ holds for any $x>1$.
Hence, we show $\mathcal{L}(\bar{p}_{i} , \bar{p}^{*}_{i}) \geq \mathcal{L}(\bar{p}_{j}, \bar{p}^{*}_{j})$ and thus the inequality~\eqref{eq:axiom2asummary} hold for Case $(ii)$.
\end{proof}
\begin{comment}
\begin{subaxiom2}[Degree of Inverse Overlap: Small difference]
\label{subaxiom2b}{
Consider two arcs  $e_{i}$ and $e_{j}$ of the same size $($i.e., ${(\lvert H_{i} \vert  = \lvert H_{j} \vert )} \wedge {(\lvert T_{i} \vert = \lvert T_{j} \vert)})$.
If $R_i = \{e'_i\}$ and $R_j = \{e'_j\}$ satisfy
\begin{gather*}
\lvert H'_i \vert > \lvert H'_j \vert, \quad \lvert T'_i \vert = \lvert T'_j \vert, \quad 0 < \lvert H'_i \cap T_i \vert = \lvert H'_j \cap T_j \vert, \ \text{and} \ 0 < \lvert T'_i \cap H_i \vert = \lvert T'_j \cap H_j \vert,
\end{gather*}
then the following inequality should hold:
$$
r(e_i , R_i) < r(e_j , R_j).
$$
}
\end{subaxiom2}
\end{comment}

\begin{subprop2}{\label{proposition:2b}
\measure \normalfont{(}i.e., defining $r(e_i , R_i)$ as in Eq.~\eqref{eq:proposal}\normalfont{)}
satisfies  Generalized Axiom~\ref{subaxiom2b}.}
\end{subprop2}

\begin{proof}{
Since $\lvert R_{i} \vert  = \lvert R_{j} \vert  = 1$, the cardinality penalty terms can be ignored. 
The overall inequality is re-written as
\begin{align*}
    & r(e_{i} , R_{i}) < r(e_{j} , R_{j}) \\ & \equiv 
    1 - \frac{\sum_{v_{i} \in H_{i}} \mathcal{L}(p_{h} , p^{*}_{h})}
    {\lvert H_{i} \vert \cdot \mathcal{L}_{max}}  < 
    1 - \frac{\sum_{v_{j} \in H_{j}} \mathcal{L}(p_{h} , p^{*}_{h})}
    {\lvert H_{j} \vert \cdot \mathcal{L}_{max}} \\ 
    & \equiv \frac{\sum_{v_{j} \in H_{j}} \mathcal{L}(p_{h} , p^{*}_{h})}
    {\lvert H_{j} \vert \cdot \mathcal{L}_{max}} <  \frac{\sum_{v_{i} \in H_{i}} \mathcal{L}(p_{h} , p^{*}_{h})}
    {\lvert H_{i} \vert \cdot \mathcal{L}_{max}}
\end{align*}
As in the previous proof, $\lvert H_{i} \vert  = \lvert H_{j} \vert $, and $\mathcal{L}(p_{h} , p^{*}_{h})$ is identical for every $v_{h} \in H_{k} \cap T'_{k}$. 
Let {$\bar{p}_{k}, \forall k = i,j$ be the transition probability distribution regarding the target arc $e_{k}$ and its reciprocal set $R_{k}$.}
Here, $\lvert H_{i} \cap T'_{i} \vert = \lvert H_{j} \cap T'_{j} \vert$,  $\lvert T'_{i} \vert = \lvert T'_{j} \vert$, and the number of target arc's head set nodes $v_{h}$ that satisfy $\mathcal{L}(p_{h} , p^{*}_{h}) < \log{2}$ is identical for both cases. Thus, the above inequality is re-written as
\begin{align*}
    & \frac{\sum_{v_{j} \in H_{j}} \mathcal{L}(p_{h} , p^{*}_{h})}
    {\lvert H_{j} \vert \cdot \mathcal{L}_{max}} <  \frac{\sum_{v_{i} \in H_{i}} \mathcal{L}(p_{h} , p^{*}_{h})}
    {\lvert H_{i} \vert \cdot \mathcal{L}_{max}} \\ 
    & \equiv {\lvert H_{j} \cap T'_{j} \vert\times\mathcal{L}(\bar{p}_{j}, \bar{p}^{*}_{j})} < 
    {\lvert H_{i} \cap T'_{i} \vert\times\mathcal{L}(\bar{p}_{i}, \bar{p}^{*}_{i})} \\
    & \equiv \mathcal{L}(\bar{p}_{j}, \bar{p}^{*}_{j}) < \mathcal{L}(\bar{p}_{i}, \bar{p}^{*}_{i}).
\end{align*}

Now, we only need to show that the probabilistic distance between transition probability and the optimal transition probability is greater in $e_{i}$ than in $e_{j}$. 
Let $A = \lvert H'_{i} \vert \ > \ B = \lvert H'_{j} \vert$. We can decompose the domain of $v \in V$ into four parts as
$$
T_{k} \cap H'_{k}, \quad T_{k} \setminus H'_{k}, \quad 
H'_{k} \setminus T_{k}, \text{  and  } V \setminus \{H'_{k} \cup T_{k}\}, \quad \forall k = i,j
$$
Here, $\jsdpq{P}{Q}$ in the second and fourth parts is identical for both cases. 
That is, we only need to compare the probabilistic distances that are related to the first and third parts of the above four domains. That is,
\begin{align*}
&\mathcal{L}(\bar{p}_{j}, \bar{p}^{*}_{j}) < \mathcal{L}(\bar{p}_{i}, \bar{p}^{*}_{i}) \\
& \equiv F \times\ell \left(\frac{1}{B}, \frac{1}{T}\right) + \frac{B - F}{2B}\log{2} < 
F \times \ell \left(\frac{1}{A}, \frac{1}{T}\right) + \frac{A - F}{2A}\log{2}, 
\end{align*}
where $F = \lvert H'_{i} \cap T_{i} \vert = \lvert H'_{j} \cap T_{j} \vert$ and $T = \lvert T_{i} \vert=\lvert T_{j} \vert$. 
Note that $A > B$.  
Overall inequality is rewritten as
\begin{align*}
     &\mathcal{L}(\bar{p}_{j}, \bar{p}^{*}_{j}) < \mathcal{L}(\bar{p}_{i}, \bar{p}^{*}_{i}) \\
     & \equiv \frac{F}{2}\left(\frac{1}{B} - \frac{1}{A}\right)\log{2} > 
     F \times \left(\ell \left(\frac{1}{B}, \frac{1}{T}\right) - \ell \left(\frac{1}{A}, \frac{1}{T}\right) \right)
\end{align*}
To simplify the equation, we unfold $\ell(p,q)$ as 
\begin{align*}
    &\frac{F}{2}\left(\frac{1}{B} - \frac{1}{A}\right)\log{2} > 
     F \times \left(\ell \left(\frac{1}{B}, \frac{1}{T}\right) - \ell \left(\frac{1}{A}, \frac{1}{T}\right)\right) \\
     & \equiv \frac{1}{2}\left(\frac{1}{B} - \frac{1}{A}\right)\log{2} > 
     \frac{1}{2T} \log{\frac{2B}{B+T}} + \frac{1}{2B} \log{\frac{2T}{B+T}} 
     - \frac{1}{2T} \log{\frac{2A}{A+T}} - \frac{1}{2A} \log{\frac{2T}{A+T}} \\ 
     & \equiv \left(\frac{1}{B} - \frac{1}{A}\right)\log{2} > 
     \frac{1}{T} \log{\frac{2B}{B+T}} + \frac{1}{B} \log{\frac{2T}{B+T}} 
     - \frac{1}{T} \log{\frac{2A}{A+T}} - \frac{1}{A} \log{\frac{2T}{A+T}} \\ 
     & \equiv 0 > \frac{1}{T} \log{\frac{B}{B+T}} + \frac{1}{B} \log{\frac{T}{B+T}} 
     - \frac{1}{T} \log{\frac{A}{A+T}} - \frac{1}{A} \log{\frac{T}{A+T}} \\ 
     &\because \text{pull 2 inside each log term out} \\ 
     & \equiv 0 > \frac{1}{T} \log{\frac{B(A+T)}{A(B+T)}} - \frac{1}{B} \log{\frac{B + T}{T}} 
      + \frac{1}{A} \log{\frac{A + T}{T}}  \\
      & \equiv 0 > \log{\frac{AB + BT}{AB + AT}} 
      + \frac{T}{A} \log{(1 + \frac{A}{T})} - \frac{T}{B} \log{( 1 +\frac{B}{T})}  \quad \because \text{multiply both sides by $T$}
\end{align*}
We show that the last inequality holds by splitting it into two parts: $\log{\frac{AB + BT}{AB + AT}} < 0$ and  $\frac{T}{A} \log{(1 + \frac{A}{T})} - \frac{T}{B} \log{( 1 +\frac{B}{T})} < 0$. 
The first part is trivial since $B < A$ implies 
$$\frac{AB + BT}{AB + AT} < 1.$$
In the second part, 
%we prove that the inequality holds by using its functional form.
%\begin{gather}\label{eq:axiomb2}
%    \frac{T}{A} \log{(1 + \frac{A}{T})} - \frac{T}{B} \log{( 1 +\frac{B}{T})} < 0 
%\end{gather}
%Here, terms regarding $A$ and $B$ have 
each term is in the form of $f(x) = \frac{1}{x}\log{(1 + x)}$.} 
Since $f(x)$ is decreasing at $x>0$,\footnote{Note that $(1+\frac{1}{x})^{x}$ is a well-known increasing function whose limit as $x\rightarrow \infty$ is $e$. Thus, $\log(1+\frac{1}{x})^{x}=x\log(1+\frac{1}{x})$ is also an increasing function, and since $x'=1/x$ is decreasing at $x>0$, $x'\log(1+\frac{1}{x'})=\frac{1}{x}\log(1+x)$ is decreasing at $x>0$.}
%\red{Note that $\log$ is an increasing function. Thus, its input and output of function is proportional i.e.,  $\log {(1+x)^{\frac{1}{x}}} \propto (1+x)^{\frac{1}{x}}$.
% Here, let $\frac{1}{x} = x'$. As $x$ and $x'$ are in an inverse-proportional relation, to show $(1+x)^{\frac{1}{x}}$ is a decreasing function, it is enough to show that $(1+\frac{1}{x'})^{x'}$ is a increasing function. 
% Notice that this is a common form of $e$, which increases until its convergence at positive infinite for x' > 0.
% Thus, as $(1+\frac{1}{x'})^{x'}$ is an increasing function at $x'>0$, we can conclude that $(1+x)^{\frac{1}{x}}$ is a decreasing function at $x > 0$, which result in $\frac{1}{x}\log(1+x)$ is a decreasing function either.}
$A > B$ implies
\begin{gather*}\label{eq:axiomb2}
    \frac{T}{A} \log{(1 + \frac{A}{T})} - \frac{T}{B} \log{( 1 +\frac{B}{T})} < 0. %\qedhere
\end{gather*} 
\end{proof}
%For this, we show $f'(x) < 0, \  \forall x > 0$. 
    %$f'(x) = -\frac{1}{x^{2}}\log{(x+1)} + \frac{1}{x(x+1)}$, and 
    %if we multiply both terms by $x^{2}(x+1)$, then
    %$f'(x) = -(x+1)\log{(x+1)} + x$, where $f'(0) = 0$. 
    %$f''(x) = -1 -\log(x+1) + 1 < 0, \forall x > 0$.
%Since $f''(x) < 0$, and $f'(x) = 0$ we can derive  $f'(x) < 0 $ for $x > 0$. This implies \blue{that} $f(x)$ is a decreasing function.

\subsubsection{Proof of the Fact that \measure Satisfies Axiom 3} \label{subsec:axiom3}

Through an example, \axiomnum{3} states that
{when two arcs inversely overlap equally with their reciprocal sets, an arc with a single reciprocal arc is more reciprocal than one with multiple reciprocal arcs.}
%arc requiring fewer candidates to inversely overlap to the same extent should be more reciprocal.
\axiomnum{3} is split into two cases and each case is formalized in \textbf{\textsc{Generalized Axioms}~\ref{subaxiom3a}} and {\textbf{\ref{subaxiom3b}}} respectively, where an arc with a single reciprocal arc and an arc with exactly two reciprocal arcs are compared.
{In \textsc{\textbf{Remark~\ref{remark:canbegeneral}}}, we further generalize them to encompass a comparison of the former and an arc with two or more reciprocal arcs and provide a proof sketch to show that these extended statements hold true for our proposed measure.}
%Below, we generalize \axiomnum{3} and prove it by dividing it into two cases.
% \begin{itemize}
%     \item \textbf{\textsc{Axiom 3A:}} Tail sets of two arcs in a less reciprocal arc's reciprocal set are identical, while their head sets do not overlap.
%     \item \textbf{\textsc{Axiom 3B:}} Head sets of two arcs in a less reciprocal arc's reciprocal set are identical, while their tail sets do not overlap.
% \end{itemize}}

% \red{\axiomnum{3} states that when two different arcs' respective reciprocal sets cover same extent of pairwise relation, an arc with a reciprocal set consists of a single arc is more reciprocal than an arc with a reciprocal set consist of two arcs.
% 
\begin{comment}
\begin{subaxiom3}[Number of Reciprocal Arcs Differs: Identical Tail Sets]\label{subaxiom3a}
Let $e'_{k} \subseteq_{(R)} e_{k}$ indicate $H'_{k} \subseteq T_{k}$ and $T'_{k} \subseteq H_{k}$.
Consider two arcs  $e_{i}$ and $e_{j}$ of the same size $($i.e., ${(\lvert H_{i} \vert  = \lvert H_{j} \vert )} \wedge {(\lvert T_{i} \vert = \lvert T_{j} \vert)})$.
If $R_i = \{e'_{i1}, e'_{i2}\}$ and $R_j = \{e'_j\}$ satisfy 
\begin{gather*}
e'_{i1} \subseteq_{(R)} e_{i}, \quad e'_{i2} \subseteq_{(R)} e_{i}, \quad e'_{j} \subseteq_{(R)} e_{j}, \quad 
T'_{i1} = T'_{i2}, \quad \lvert T'_{i1} \vert = \lvert T_{j} \vert, \\ 
H'_{i1} \cap H'_{i2} = \emptyset, \quad \text{and} \quad 
\lvert (H'_{i1} \cup H'_{i2}) \cap T_{i} \vert = \lvert H'_{j}\cap T_{j} \vert,
\end{gather*}
then the following inequality should hold: 
$$
r(e_i , R_i) < r(e_j , R_j).
$$
\end{subaxiom3}
\end{comment}
\begin{subprop3}\label{prop3a}{
\measure \normalfont{(}i.e., defining $r(e_i , R_i)$ as in Eq.~\eqref{eq:proposal}\normalfont{)}
satisfies  Generalized Axiom~\ref{subaxiom3a}.}
\end{subprop3}

\begin{proof} 
% Unlike previous cases where there exists only a single arc in a reciprocal set, there are two arcs in the former case. 
Since the sizes of the reciprocal sets differ, the cardinality penalty term should be considered. 
Here, $r(e_{i} , R_{i})$ and $r(e_{j} , R_{j})$ is rewritten as
\begin{align*}
    r(e_{i} , R_{i}) &= \left(\frac{1}{2}\right)^{\alpha}  \left(1 - 
    \frac{\sum_{v_{h} \in H_{i}} \mathcal{L}(p_{h}, p_{h}^{*})}{\lvert H_{i} \vert  \cdot \mathcal{L}_{max}}\right), \\
    r(e_{j} , R_{j}) &= \left(1 - 
    \frac{\sum_{v_{h} \in H_{j}} \mathcal{L}(p_{h}, p_{h}^{*})}{\lvert H_{j} \vert \cdot \mathcal{L}_{max}}\right).
\end{align*}
Since $\alpha > 0$, $r(e_{i} , R_{i}) < r(e_{j} , R_{j})$ is implied by
\begin{equation*}
\left(1 - \frac{\sum_{v_{h} \in H_{i}} \mathcal{L}(p_{h}, p_{h}^{*})}{\lvert H_{i} \vert  \cdot \mathcal{L}_{max}}\right) \leq
    \left(1 - \frac{\sum_{v_{h} \in H_{j}} \mathcal{L}(p_{h}, p_{h}^{*})}{\lvert H_{j} \vert  \cdot \mathcal{L}_{max}}\right).
\end{equation*}
% From the inequality $r(e_{i} , R_{i}) < r(e_{j} , R_{j})$, we derive that
% \begin{align*}
%     &r(e_{i} , R_{i}) <  r(e_{j} , R_{j}) \\
%     &= \left(\frac{1}{2}\right)^{\alpha}  \left(1 - 
%     \frac{\sum_{v_{h} \in H_{i}} \mathcal{L}(p_{h}, p_{h}^{*})}{\lvert H_{i} \vert  \cdot \mathcal{L}_{max}}\right) < \left(1 - 
%     \frac{\sum_{v_{h} \in H_{j}} \mathcal{L}(p_{h}, p_{h}^{*})}{\lvert H_{j} \vert  \cdot  \mathcal{L}_{max}}\right) \\
%     &\equiv \left(\frac{1}{2}\right)^{\alpha}\left(1 - 
%     \frac{\sum_{v_{h} \in H_{i}} \mathcal{L}(p_{h}, p_{h}^{*})}{\lvert H_{i} \vert  \cdot \mathcal{L}_{max}}\right) < 
%     \left(1 - \frac{\sum_{v_{h} \in H_{i}} \mathcal{L}(p_{h}, p_{h}^{*})}{\lvert H_{i} \vert  \cdot \mathcal{L}_{max}}\right) \leq
%     \left(1 - \frac{\sum_{v_{h} \in H_{j}} \mathcal{L}(p_{h}, p_{h}^{*})}{\lvert H_{j} \vert  \cdot \mathcal{L}_{max}}\right) 
% \end{align*}
% The last less or equal relation can be induced because $\alpha > 0$. 
%In sum, proof is rewritten as
\text{Since $\lvert H_{i} \vert  = \lvert H_{j} \vert $, the inequality is rewritten as}
\begin{align}
& \left(1 - \frac{\sum_{v_{h} \in H_{i}} \mathcal{L}(p_{h}, p_{h}^{*})}{\lvert H_{i} \vert  \cdot \mathcal{L}_{max}}\right) \leq
    \left(1 - \frac{\sum_{v_{h} \in H_{j}} \mathcal{L}(p_{h}, p_{h}^{*})}{\lvert H_{j} \vert  \cdot \mathcal{L}_{max}}\right) \nonumber \\
&  \equiv   \frac{\sum_{v_{h} \in H_{j}} \mathcal{L}(p_{h}, p_{h}^{*})}{\lvert H_{j} \vert  \cdot \mathcal{L}_{max}} \leq
 \frac{\sum_{v_{h} \in H_{i}} \mathcal{L}(p_{h}, p_{h}^{*})}{\lvert H_{i} \vert  \cdot \mathcal{L}_{max}} \nonumber \\
 & \equiv \sum_{v_{h} \in H_{j}} \mathcal{L}(p_{h}, p_{h}^{*}) \leq \sum_{v_{h} \in H_{i}} \mathcal{L}(p_{h}, p_{h}^{*}). \label{eq:axiom3final}
\end{align}
% \begin{gather*}
%  \quad \text{Here, as $\lvert H_{i} \vert  = \lvert H_{j} \vert $, left less or equal relation is rewritten as}
% \end{gather*}
% \begin{equation}
% \end{equation}
For the target arc $e_{i}$, since $T'_{i1} = T'_{i2}$, $p_{h}$ for every $v_{h} \in H_{i}$ has the same distribution. 
For the target arc $e_{j}$, since there is only one single reciprocal arc, $p_{h}$ for every $v_{h} \in H_{i}$ has the same distribution. 
Let {$\bar{p}_{k}, \forall k = i,j$} be the transition probability distribution regarding the target arc $e_{k}$ and its reciprocal set $R_{k}$.
Here, the inequality~\eqref{eq:axiom3final} is rewritten as 
\begin{align*}
    &\sum_{v_{h} \in H_{j}} \mathcal{L}(p_{h}, p_{h}^{*}) \leq \sum_{v_{h} \in H_{i}} \mathcal{L}(p_{h}, p_{h}^{*}) \\
    & \equiv 
    \lvert H_{j} \cap T'_{j} \vert \times \mathcal{L}(\bar{p}_{j}, \bar{p}^{*}_{j}) \leq \lvert H_{i} \cap T'_{i1} \vert \times \mathcal{L}(\bar{p}_{i}, \bar{p}^{*}_{i})
\end{align*}
Since $e'_{i1} \subseteq_{(R)} e_{i}, \ e'_{i2} \subseteq_{(R)} e_{i}, \ e'_{j} \subseteq_{(R)} e_{j}$, and $\lvert T'_{j} \vert = \lvert T'_{i1} \vert$, the last inequality is rewritten as
\begin{align*}
    &\lvert H_{j} \cap T'_{j} \vert \times \mathcal{L}(\bar{p}_{j}, \bar{p}^{*}_{j}) \leq \lvert H_{i} \cap T'_{i1} \vert \times \mathcal{L}(\bar{p}_{i}, \bar{p}^{*}_{i})  \\
    &\equiv \lvert T'_{j} \vert \times \mathcal{L}(\bar{p}_{j}, \bar{p}^{*}_{j}) \leq \lvert T'_{i1} \vert \times \mathcal{L}(\bar{p}_{i}, \bar{p}^{*}_{i}) \\
    &\equiv \mathcal{L}(\bar{p}_{j}, \bar{p}^{*}_{j}) \leq \mathcal{L}(\bar{p}_{i}, \bar{p}^{*}_{i}).
\end{align*}
The proof can be done by showing the last inequality, $\mathcal{L}(\bar{p}_{j}, \bar{p}^{*}_{j}) \leq \mathcal{L}(\bar{p}_{i}, \bar{p}^{*}_{i})$.

In order to show $\mathcal{L}(\bar{p}_{j}, \bar{p}^{*}_{j}) \leq \mathcal{L}(\bar{p}_{i}, \bar{p}^{*}_{i})$, 
we should take a close look at the transition probability in $e_{i}$. 
%since it is different from previous arcs who are given a single reciprocal arc. 
Since the head sets of the two reciprocal arcs do not overlap, the transition probability is
\begin{equation*}
p_{h, i}(v) = \begin{cases}
    \frac{1}{2\lvert H'_{i1} \vert} & \text{if $v \in H'_{i1}$} \\
    \frac{1}{2\lvert H'_{i2} \vert} & \text{if $v \in H'_{i2}$} \\
    0 & \text{otherwise.} \\
\end{cases}
\end{equation*}
Since $e'_{i1} \subseteq_{(R)} e_{i}, \ e'_{i2} \subseteq_{(R)} e_{i}$, and $e'_{j} \subseteq_{(R)} e_{j}$, the domain of $v\in V$ can be divided into
\begin{align*}
&H'_{i1}, \quad H'_{i2}, \quad T_{i}\setminus \{H'_{i1} \cup H'_{i2}\}, \text{  and  }  V \setminus T_{i}, \quad \text{for $e_{i}$,} \\
& H'_{j}, \quad T_{j} \setminus H'_{j}, \text{  and  } V \setminus T_{j}, \quad \text{for $e_{j}$.}
\end{align*}
Since $\lvert T_{i} \vert = \lvert T_{j} \vert$, the probability mass for the last part is identical for both cases. 
Let $A = \lvert H_{j} \vert $, $B = \lvert H_{i1} \vert$, and $T = \lvert T_{i} \vert = \lvert T_{j} \vert$. {Since $H'_{i1}, H'_{i2} \subseteq T_{i}$, $H'_{j} \subseteq T_{j}$, $H'_{i1} \cap H'_{i2} = \emptyset$, and $\lvert (H'_{i1} \cup H'_{i2}) \cap T_{i} \vert = \lvert H'_{j}\cap T_{j} \vert$, $\lvert H_{i1} \vert + \lvert H_{i2} \vert = \lvert H_{j} \vert$ holds.}
Then, based on the above fact, we rewrite $\mathcal{L}(\bar{p}_{j}, \bar{p}_{j}^{*}) \leq \mathcal{L}(\bar{p}_{i}, \bar{p}_{i}^{*})$ as
\begin{align}
    &\mathcal{L}(\bar{p}_{j}, \bar{p}_{j}^{*}) \leq \mathcal{L}(\bar{p}_{i}, \bar{p}_{i}^{*}) \nonumber \\ 
    &\equiv  A\times \ell\left(\frac{1}{T}, \frac{1}{A}\right) + \frac{T - A}{2T}\log{2} \nonumber \\
    &\leq B\times \ell\left(\frac{1}{T}, \frac{1}{2B}\right)
    + (A-B)\times \ell\left(\frac{1}{T}, \frac{1}{2(A-B)}\right) + \frac{T - B - (A - B)}{2T} \log{2} \nonumber \\ 
    &\equiv A\times \ell\left(\frac{1}{T}, \frac{1}{A}\right) \leq 
    B\times \ell\left(\frac{1}{T}, \frac{1}{2B}\right) + (A - B)\times \ell\left(\frac{1}{T}, \frac{1}{2(A-B)}\right) \label{eq:canbegeneral}\\ 
    &\equiv \frac{A}{2T}\log{\frac{2A}{A + T}} + \frac{A}{2A}\log{\frac{2T}{A + T}}  \nonumber \\ % First Term
    &\leq 
    \frac{B}{2T}\log{\frac{4B}{2B + T}} + \frac{B}{4B}\log{\frac{2T}{2B + T}}  \nonumber\\
    &+\frac{(A-B)}{2T}\log{\frac{4(A-B)}{2(A-B) + T}} + \frac{(A-B)}{4(A-B)}\log{\frac{2T}{2(A-B) + T}}, \quad \because  \text{unfold $\ell(p,q)$} \nonumber \\
    &\equiv \frac{A}{2T}\log{\frac{A}{A + T}} + \frac{A}{2A}\log{\frac{T}{A + T}} \nonumber \\ % First Term
    &\leq 
    \frac{B}{2T}\log{\frac{2B}{2B + T}} + \frac{B}{4B}\log{\frac{T}{2B + T}} \nonumber \\ 
    &+ \frac{(A-B)}{2T}\log{\frac{2(A-B)}{2(A-B) + T}} + \frac{(A-B)}{4(A-B)}\log{\frac{T}{2(A-B) + T}}, \nonumber
\end{align}
where, for the last equivalence, we subtract $\left(\frac{A}{2T} + \frac{1}{2} \right)\log{2} = \left( \frac{A}{2T} + \frac{A}{2A} \right)\log{2}  = 
\left(\frac{B}{2T} + \frac{B}{4B} + \frac{A-B}{2T} + \frac{B-A}{4(B-A)}\right)\log{2}$ from both sides.
% Since $\lvert H_{i1} \cup H_{i2} \vert = \lvert H_{j} \vert $ and $\lvert H_{i1} \cap H_{i2} \vert = 0$, we derive . By using it, we simplify the size of each set as follows: .
% If we pull out $\log{2}$ terms from both sides
% $$
% \left( \frac{A}{2T} + \frac{A}{2A} \right)\log{2}  \quad \text{and} \quad 
% \left(\frac{B}{2T} + \frac{B}{4B} + \frac{A-B}{2T} + \frac{B-A}{4(B-A)}\right)\log{2}
% $$
% LHS and RHS have identical terms. By erasing them, original inequality can be re-written as 
% \begin{align*}
% \end{align*}
We show the last inequality by dividing it into two and proving each. 
If the following two inequality holds, the proof is done.
\begin{align}
    \label{eq:axiomcomp3a1} \frac{A}{2T}\log{\frac{A}{A + T}} &\leq 
    \frac{B}{2T}\log{\frac{2B}{2B + T}} + \frac{(A-B)}{2T}\log{\frac{2(A-B)}{2(A-B) + T}}, \\
    \label{eq:axiomcomp3a2} \frac{1}{2}\log{\frac{T}{A + T}} &\leq 
    \frac{1}{4}\log{\frac{T}{2B + T}} + \frac{1}{4}\log{\frac{T}{2(A-B) + T}}.
\end{align}

We first show the inequality~\eqref{eq:axiomcomp3a1}. By multiplying by $2T$ both sides, we get
\begin{align}
    \frac{A}{2T}\log{\frac{A}{A + T}} &\leq 
    \frac{B}{2T}\log{\frac{2B}{2B + T}} + \frac{(A-B)}{2T}\log{\frac{2(A-B)}{2(A-B) + T}} \nonumber \\
    \equiv A\log{\frac{A}{A + T}} &\leq 
    B\log{\frac{2B}{2B + T}} + (A-B)\log{\frac{2(A-B)}{2(A-B) + T}} \label{eq:axiomcomp3a1modified}
\end{align}
Here, we prove this inequality by using the functional form of $f(B) = B\log{\frac{2B}{2B + T}} + (A-B)\log{\frac{2(A-B)}{2(A-B) + T}}$ where $0 < B < A$. 
%Since $f(B) = B[\log{(2B)} - \log{(2B + T)}]  + (A-B)[\log{(2(A-B))} - \log{(2(A-B) + T)}]$,
Its derivative is
\begin{align*}
    \frac{\partial f(B)}{\partial B} &= \log{(2B)} - \log{(2B + T)} + B\left(\frac{2}{2B} - \frac{2}{2B + T}\right) \\ 
    &- \log{(2(A-B))} + \log{(2(A-B) + T)} - (A-B)\left(\frac{2}{2(A-B)} - \frac{2}{2(A-B) + T}\right). \\
    &= \log{(2B)} - \log{(2B + T)} - \frac{2B}{2B + T} \\
    &- \log{(2(A-B))} + \log{(2(A-B) + T)} +  \frac{2(A-B)}{2(A-B) + T} \\ 
    &= \log{\frac{2B}{2B+T}} - \frac{2B}{2B + T} - \log{\frac{2(A-B)}{2(A-B) + T}} + \frac{2(A-B)}{2(A-B) + T}.
\end{align*}
Thus, $\frac{\partial f(B)}{\partial B} = \log{x} - x - (\log{y} - y)$ 
for $x = \frac{2B}{2B+T}$ and $y = \frac{2(A-B)}{2(A-B) + T}$, which satisfy $0<x,y<1$, and it has the following  properties:
%Here, we derive the functional from of $f(B)$ by using following facts. 
\begin{itemize}
    \item If we plug in $B = \frac{A}{2}$, $f'(B) = 0$ holds,
    \item $\log{x} - x$ is an increasing function at $0 < x < 1$,
    \item If $0 < B < \frac{A}{2}$, then $(\log{y}) - y > (\log{x}) - x$, which implies $f'(B) < 0$,
    \item If $\frac{A}{2} < B < A $, then $(\log{x}) - x > (\log{y}) - y$, which implies $f'(B) > 0$. 
\end{itemize}
From these properties, we can derive 
\begin{equation*}
    f'(B) \begin{cases}
    < 0 & \text{if $0 < B < \frac{A}{2}$} \\
    = 0 & \text{if $B = \frac{A}{2}$} \\
    > 0 & \text{if $\frac{A}{2} < B < A$}.
    \end{cases}
\end{equation*}
Hence, when $0<B<A$, $f(B)$ has its minimum value at $B=A/2$, and therefore the inequality \eqref{eq:axiomcomp3a1modified}, which is equivalent to the inequality \eqref{eq:axiomcomp3a1}, holds. 
%Also note that the equality holds when $B = A/2$. 
% \begin{figure*}
%     \centering
%     \includegraphics[width=0.6\textwidth]{FIG_ONLINE_APPENDIX/FIG_ONLINE_AXIOM3.pdf}
%     \caption{Visualization of the inequality~\eqref{eq:axiomcomp3a1}.}
%     \label{fig:axiom3a}
% \end{figure*}
% Thus, we infer that $f(B)$ is a convex function at $0 < B < A$.
% Value of $f(B = \frac{A}{2}) = \frac{A}{2}\log{\frac{A}{A+T}} + \frac{A}{2}\log{\frac{A}{A+T}} = A\log{\frac{A}{A+T}}$. Rewind that our original goal is to show the below inequality.
% $$
% A\log{\frac{A}{A + T}} \leq B\log{\frac{2B}{2B + T}} + (A-B)\log{\frac{2(A-B)}{2(A-B) + T}}
% $$
% That is, the RHS term of inequality is a convex function that has its minimum value at $B=A/2$, and corresponding minimum value is equal to the LHS term. 
% Thus, we can guarantee the inequality~\eqref{eq:axiomcomp3a1} is satisfied. 
% The overview of this result is illustrated in Figure~\ref{fig:axiom3a}.
Now we show the inequality~\eqref{eq:axiomcomp3a2}, which is rewritten as 
\begin{align*}
    &\frac{1}{2}\log{\frac{T}{A + T}} \leq 
    \frac{1}{4}\log{\frac{T}{2B + T}} + \frac{1}{4}\log{\frac{T}{2(A-B) + T}} \\
    %&\equiv \frac{1}{4}\log{\frac{T}{A + T}} + \frac{1}{4}\log{\frac{T}{A + T}} \leq 
    %\frac{1}{4}\log{\frac{T}{2B + T}} + \frac{1}{4}\log{\frac{T}{2(A-B) + T}} \\
    &\equiv \frac{1}{4}\log{\frac{T^{2}}{(A + T)(A+T)}} \leq 
    \frac{1}{4}\log{\frac{T^{2}}{(2B + T)(2(A-B) + T)}}, \\
    &\equiv (2B + T)(2(A-B) + T) \leq (A + T)(A + T)  \\
    &= 4B(A-B) + 2AT + T^{2} \leq A^{2} + 2AT + T^{2} \\
    &\equiv A^{2} - 4AB + 4B^{2} = (A-2B)^{2} \geq 0.
\end{align*}
The last inequality trivially holds.%, and the equality holds when $B = A/2$. 
%Since both inequalities~\eqref{eq:axiomcomp3a1} and~\eqref{eq:axiomcomp3a2} hold, %, where equality occurs at $B = A/2$ in both terms, 
%$\mathcal{L}(p_{h, j}, p_{h, j}^{*}) \leq \mathcal{L}(p_{h, i}, p_{h, i}^{*})$ also holds.
\end{proof}

\color{black}
\begin{comment}
\begin{subaxiom3}[Number of Reciprocal Arcs: Identical Head Sets.]\label{subaxiom3b}{
Let $e'_{k} \subseteq_{(R)} e_{k}$ indicate $H'_{k} \subseteq T_{k}$ and $T'_{k} \subseteq H_{k}$.
Consider two arcs  $e_{i}$ and $e_{j}$ of the same size $($i.e., ${(\lvert H_{i} \vert  = \lvert H_{j} \vert )} \wedge {(\lvert T_{i} \vert = \lvert T_{j} \vert)})$.
If $R_i = \{e'_{i1}, e'_{i2}\}$ and $R_j = \{e'_j\}$ satisfy
\begin{gather*}
e'_{i1} \subseteq_{(R)} e_{i}, \quad e'_{i2} \subseteq_{(R)} e_{i}, \quad e'_{j} \subseteq_{(R)} e_{j}, \quad 
H'_{i1} = H'_{i2}, \quad \lvert H'_{i1} \vert = \lvert H_{j} \vert , \\ 
T'_{i1} \cap T'_{i2} = \emptyset, \quad and \quad \lvert (T'_{i1} \cup T'_{i2}) \cap H_{i} \vert = \lvert T'_{j}\cap H_{j} \vert,
\end{gather*}
then the following inequality should hold: 
$$
r(e_i , R_i) < r(e_j , R_j).
$$
}
\end{subaxiom3}
\end{comment}
\begin{subprop3}\label{subprop3}{
\measure \normalfont{(}i.e., defining $r(e_i , R_i)$ as in Eq.~\eqref{eq:proposal}\normalfont{)}
satisfies  Generalized Axiom~\ref{subaxiom3b}.}
\end{subprop3}
\begin{proof}
%Setting is all the same as the proof of Proposition~\ref{prop3a} until the inequality~\eqref{eq:axiom3final}. 

By following the proof of Proposition~\ref{prop3a}, we can show that $r(e_i \, R_i) < r(e_j \, R_j)$ is implied by
$$ \sum_{v_{h} \in H_{i}} \mathcal{L}(p_{h}, p_{h}^{*}) \leq \sum_{v_{h} \in H_{j}} \mathcal{L}(p_{h}, p_{h}^{*}).
$$
%Since $H'_{i1} = H'_{i2}$, the same transition probability is assigned to $e_i$'s tail nodes that belong to $H'_{i1} = H'_{i2}$.} 

%Since $T'_{i1} \cap T'_{i2} = \emptyset$, the transition probability for every $v_{h} \in H_{i} \cap \{T'_{i1} \cup T'_{i2}\}$ is identical for $e_{i}$. Thus, the transition probability f in this case can be %written as
%$$
%p_{i}(v) = p_{h, i} (v), \quad  \forall v_{h} \in H_{i} \cap \{T'_{i1} \cup T'_{i2}\}. 
%$$
%This is also true for $e_{j}$, whose reciprocal set has only one arc.
%Since $\lvert H_{i} \cap \{T'_{i1} \cup T'_{i2}\} \vert = \lvert H_{j} \cap T'_{j} \vert$, the above inequality is rewritten as
%$$
%\mathcal{L}(p_{h, i}, p_{h, i}^{*}) \leq \mathcal{L}(p_{h, j}, p_{h, j}^{*}).
%$$
%Since $e'_{i1} \subseteq_{(R)} e_{i}$, $e'_{i2} \subseteq_{(R)} e_{i}$, $e'_{j} \subseteq_{(R)} e_{j}$, and $\lvert H'_{i1} \vert = \lvert H'_{j} \vert$, the probabilistic distances for $e_{i}$ and $e_{j}$ are identical. Thus,
%$$
%\mathcal{L}(p_{h, i}, p_{h, i}^{*}) = \mathcal{L}(p_{h, j}, p_{h, j}^{*}), 
%$$
%and thus $\mathcal{L}(p_{h, i}, p_{h, i}^{*}) \leq \mathcal{L}(p_{h, j}, p_{h, j}^{*})$ also holds.
\noindent Since $T'_{i1} \cap T'_{i2} = \emptyset$ and {$H'_{i1} = H'_{i2}$}, the transition probability for every $v_{h} \in H_{i} \cap \{T'_{i1} \cup T'_{i2}\}$ is identical for $e_{i}$. 
%Thus, the transition probability $p_{h,i}(v)$ in this case is rewritten as
%$$
%p_{i}(v) = p_{h, i} (v), \quad  \forall v_{h} \in H_{i} \cap \{T'_{i1} \cup T'_{i2}\}. 
%$$
This is also true for $e_{j}$, whose reciprocal set has only one arc.
{Let {$\bar{p}_{k}, \forall k = i,j$} be the transition probability distribution regarding the target arc $e_{k}$ and its reciprocal set $R_{k}$.}
Since $\lvert H_{i} \cap \{T'_{i1} \cup T'_{i2}\} \vert = \lvert H_{j} \cap T'_{j} \vert$, the above inequality is rewritten as
$$
\mathcal{L}(\bar{p}_{i}, \bar{p}_{i}^{*}) \leq \mathcal{L}(\bar{p}_{j}, \bar{p}_{j}^{*}).
$$
Since $e'_{i1} \subseteq_{(R)} e_{i}$, $e'_{i2} \subseteq_{(R)} e_{i}$, $e'_{j} \subseteq_{(R)} e_{j}$, and $\lvert H'_{i1} \vert = \lvert H'_{j} \vert$, the probabilistic distances for $e_{i}$ and $e_{j}$ are identical. 
That is,
\begin{equation}\label{eq:newgeneral4}
    \mathcal{L}(\bar{p}_{i}, \bar{p}_{i}^{*}) = \mathcal{L}(\bar{p}_{j}, \bar{p}_{j}^{*}),    
\end{equation}
and thus $\mathcal{L}(\bar{p}_{i}, \bar{p}_{i}^{*}) \leq \mathcal{L}(\bar{p}_{j}, \bar{p}_{j}^{*})$ also holds.
% By following the proof of Proposition~\ref{prop3a}, we can show that the two inequalities below are equivalent.
% $$
% r(e_i \, R_i) < r(e_j \, R_j) \equiv \sum_{v_{h} \in H_{i}} \mathcal{L}(p_{h}, p_{h}^{*}) \leq \sum_{v_{h} \in H_{j}} \mathcal{L}(p_{h}, p_{h}^{*}).
% $$

% \blue{
% Since $H'_{i1} = H'_{i2}$, the same transition probability is assigned to $e_i$'s tail nodes that belong to $H'_{i1} = H'_{i2}$.} Furthermore, since $T'_{i1} \cap T'_{i2} = \emptyset$, the transition probability for every $v_{h} \in H_{i} \cap \{T'_{i1} \cup T'_{i2}\}$ is identical. Thus, the transition probability of $e_{i}$ in this case can be written as
% $$
% p_{i}(v) = p_{h, i} (v), \quad  \forall v_{h} \in H_{i} \cap \{T'_{i1} \cup T'_{i2}\} 
% $$
% Since $\lvert H_{i} \cap \{T'_{i1} \cup T'_{i2}\} \vert = \lvert H_{j} \cap T'_{j} \vert$, the above inequality is rewritten as
% $$
% \mathcal{L}(p_{h, i}, p_{h, i}^{*}) \leq \mathcal{L}(p_{h, j}, p_{h, j}^{*})
% $$
% Note that $e'_{i1} \subseteq_{(R)} e_{i} \ e'_{i2} \subseteq_{(R)} e_{i} \ e'_{j} \subseteq_{(R)} e_{j}$ and $\lvert H'_{i1} \vert = \lvert H'_{j} \vert$, the probabilistic distances for $e_{i}$ and $e_{j}$ are identical. Thus,
% $$
% \mathcal{L}(p_{h, i}, p_{h, i}^{*}) = \mathcal{L}(p_{h, j}, p_{h, j}^{*})
% $$
% While their probabilistic distances are identical, the penalty term $\left(\frac{1}{\lvert R_{i} \vert }\right)^{\alpha} = 
% \left(\frac{1}{2}\right)^{\alpha} < 1, \ \forall \alpha > 0$ causes $r(e_i , R_i) < r(e_j , R_j)$ to hold. 
\end{proof}

\begin{remark}[Extension of \textsc{\textbf{Propositions~\ref{prop3a}}} and ~\textsc{\textbf{\ref{subprop3}}} to Multiple Hyperarc Cases]\label{remark:canbegeneral}
Although the statement of \textsc{\textbf{Proposition}~\ref{prop3a}} presents a case with a single arc in $R_{i}$ and two arcs in $R_{j}$, it can be further generalized: a single arc in $R_{i}$ and $K \geq 2$ hyperarcs in $R_{j}$ under the following conditions, which are equivalent to the current conditions for $K=2$: 
(1) the head sets of the arcs in $R_{i}$ are disjoint and their tail sets are identical, (2) all arcs in $R_{i}$ satisfy the condition of $\subseteq_{(R)}$, and {(3) the coverage of $T_{i}$ by the head sets of the arcs in $R_{i}$ is of the same size as the coverage of $T_{j}$ by the head set of the arc in $R_{j}$.}

We provide a proof sketch to demonstrate the validity of \measure in this generalized setting.
{As in the proof of \textsc{\textbf{Proposition}~\ref{subaxiom3a}}, it suffices to show that Eq.~\eqref{eq:newgeneral} holds,} which generalizes  Eq.~\eqref{eq:canbegeneral},
\begin{align}\label{eq:newgeneral}
    A\times \ell \left(\frac{1}{T}, \frac{1}{A}\right) \leq B_{1}\ell \left(\frac{1}{T}, \frac{1}{KB_{1}} \right) + \cdots + B_{K}\ell \left(\frac{1}{T}, \frac{1}{KB_{K}} \right), 
\end{align}
where $A = \sum_{i=1}^{K}B_{i}$. 
Considering that the Jensen-Shannon Divergence (JSD) is an average of two KL-divergence terms, which is a well-known convex function when one probability distribution is fixed ($1/T$ in our case), 
we can derive that each term $\ell$ in Eq.~\eqref{eq:newgeneral} is also a convex function.

From this fact, we can apply Jensen's Inequality, which states:
$ f(a_{1}x_{1} + \cdots + a_{K}x_{K}) \leq a_{1}f(x_{1}) + \cdots + a_{K}f(x_{K})$ holds for a convex function $f$ and non-negative coefficients $a_{1},\cdots, a_{K}$ where $\sum_{i=1}^{K} a_{i} = 1$. 
By considering the fact that $\ell(\frac{1}{T}, x)$ is a convex function with respect to $x$, we derive Eq.~\eqref{eq:newgeneral2} by setting $a_{i} = B_{i}/A$ and $x_{i} = 1/(KB_{i})$.
\begin{align}~\label{eq:newgeneral2}
    \ell \left(\frac{1}{T}, \frac{1}{A}\right) & = \ell \left(\frac{1}{T}, \frac{K}{AK}\right) = \ell \left(\frac{1}{T}, \frac{1}{KB_{1}} \times \frac{B_{1}}{A} + \cdots \frac{1}{KB_{K}} \times \frac{B_{K}}{A}\right)
      \nonumber \\
    &\leq \frac{B_{1}}{A} \ell \left(\frac{1}{T}, \frac{1}{KB_{1}}\right) + \cdots \frac{B_{K}}{A} \ell \left(\frac{1}{T}, \frac{1}{KB_{K}}\right). 
\end{align}
Multiplying both sides of Eq.~\eqref{eq:newgeneral2} by $A$ implies Eq.~\eqref{eq:newgeneral}, which is the result we aim to show.
% \begin{align}\label{eq:newgeneral3}
%     A\times \ell \left(\frac{1}{T}, \frac{1}{A}\right) \leq B_{1}\ell \left(\frac{1}{T}, \frac{1}{KB_{1}} \right) + \cdots + B_{K}\ell \left(\frac{1}{T}, \frac{1}{KB_{K}} \right). 
% \end{align}

Similarly, we can further generalize \textsc{\textbf{Proposition}~\ref{subprop3}} to case with a single arc in $R_{i}$ and $K \geq 2$ arcs in $R_{j}$ under the following conditions, which are equivalent to the current conditions when $K=2$:
(1) the tail sets of the arcs in $R_{i}$ are disjoint and their head sets are identical, (2) all arcs in $R_{i}$ satisfy the condition of $\subseteq_{(R)}$, and {(3) the coverage of $H_{i}$ by the tail sets of the arcs in $R_{i}$ is of the same size as the coverage of $H_{j}$ by the tail set of the arc in $R_{j}$.} The proof provided for \textbf{\textsc{Proposition}}~\ref{subprop3} can be directly applied in this generalized setting to show the validity of \measure.

\end{remark}

\subsubsection{Proof of the Fact that \measure Satisfies Axiom 4}\label{subsec:axiom4}

% \textbf{equally} with their reciprocal sets, an arc whose reciprocal arcs are equally reciprocal to all nodes in the arc is more reciprocal than one with reciprocal arcs \textbf{biased} towards a subset of nodes in the arc.

Through an example, \axiomnum{4} states that an arc whose reciprocal arcs are \textbf{equally} reciprocal to all nodes in the arc is more reciprocal than one with reciprocal arcs \textbf{biased} towards a subset of nodes in the arc.
\kijung{The generalized statement of \axiomnum{4} is formalized in \textbf{\textsc{Generalized Axiom~\ref{subaxiom4}}}.}
\begin{newpropfull}{
\measure \normalfont{(}i.e., defining $r(e_i , R_i)$ as in Eq.~\eqref{eq:proposal}\normalfont{)}
satisfies  Generalized Axiom~\ref{subaxiom4}.}
\end{newpropfull}

\begin{proof} 

By the definition, the inequality is rewritten as
\begin{align*}
&r(e_i , R_i) < r(e_j , R_j) \\
& =  \left( \frac{1}{\lvert R_{i} \vert } \right)^\alpha \left(1 - \frac{\sum_{v_h \in H_i}
        \mathcal{L}(p_{h}, p_{h}^{*})}{\lvert H_i \vert \cdot \mathcal{L}_{max} }\right) < 
        \left( \frac{1}{\lvert R_{j} \vert } \right)^\alpha \left(1 - \frac{\sum_{v_h \in H_j}
        \mathcal{L}(p_{h}, p_{h}^{*})}{\lvert H_j \vert  \cdot \mathcal{L}_{max} }\right).
\end{align*}
{Let {$\bar{p}_{k}, \forall k = i,j$} be the transition probability distribution regarding the target arc $e_{k}$ and its reciprocal set $R_{k}$ which does not rely on the starting node $v_{h}$.}
The above inequality is rewritten as 

\begin{align*}
    &r(e_i , R_i) < r(e_j , R_j)   \\
    & \equiv   \left(1 - \frac{\sum_{v_h \in H_i}
        \mathcal{L}(p_{h}, p_{h}^{*})}{\lvert H_i \vert \cdot \mathcal{L}_{max} }\right) < 
        \left(1 - \frac{\sum_{v_h \in H_j}
        \mathcal{L}(p_{h}, p_{h}^{*})}{\lvert H_j \vert  \cdot \mathcal{L}_{max} }\right) \quad 
        \because \lvert R_{i} \vert = \lvert R_{j} \vert   \\
    & \equiv  \sum_{v_h \in H_j}\mathcal{L}(p_{h}, p_{h}^{*}) < \sum_{v_h \in H_i}\mathcal{L}(p_{h}, p_{h}^{*})  \quad \because \lvert H_{i} \vert  = \lvert H_{j} \vert \\
    & \equiv \mathcal{L}(\bar{p}_{j}, \bar{p}_{j}^{*}) < \mathcal{L}(\bar{p}_{i}, \bar{p}_{i}^{*})
    \quad \because T'_{i} = H_{i} \text{ and } T'_{j} = H_{j}
\end{align*}

Here, we prove the last inequality by showing that, in the above setting, (a) $R_{j}$ minimizes the distance (i.e., $\mathcal{L}(p_{h, j}, p_{h, j}^{*}) \equiv \mathcal{L}(\bar{p}_{j}, \bar{p}_{j}^{*})$), and (b) the distance is inevitably larger in all other cases.
% we prove \axiomnum{4} by showing that given $e_{j}$ is the optimal reciprocal case from \axiomnum{4}'s setting (where probabilistic distance is being minimized) and other cases are inevitably less reciprocal than it. 
By the assumptions,
for the target arc $e_j$, the corresponding reciprocal arcs have a head set of size $2$, and the number of reciprocal arcs equals the number of tail nodes (i.e., $\lvert R_{j} \vert  = \lvert T_{j} \vert$).
In addition, the head set of every reciprocal arc is a subset of the tail set $T_{j}$ of $e_j$ and $T'_{j} = H_{j}$.
Thus, Eq.~\eqref{eq:axiom4b}, which is about $e_{j}$, implies that every node $v \in T_{j}$ in the tail set is included in two of the head sets of the reciprocal arcs. 
% In summary, 
% \begin{itemize}
%     \item There are $\lvert T_{j} \vert$ reciprocal arcs, whose head set sizes are all $2$. 
%     \item Every node in the tail set of $T_{j}$ the target arc $e_j$ is included in two of the head sets of the reciprocal arcs.
%     \item The head set of every reciprocal arc is a subset of the tail set of the target arc.
% \end{itemize}
Because of these facts, the transition probability can be written as
\begin{equation*}
    p_{h, j}(v) = \begin{cases}
    \frac{1}{2\lvert T_{j} \vert} + \frac{1}{2\lvert T_{j} \vert} = \frac{1}{\lvert T_{j} \vert} & \text{if $v \in T_{j}$,} \\
    0 & \text{otherwise,}
    \end{cases}
\end{equation*}
Note that this is identical to the optimal transition probability.

Now, consider the case of $e_{i}$. Here, %since there exists at least one inequality between the number of inclusion of each target arc's tail set node to reciprocal arcs' head sets,
due to Eq.~\eqref{eq:axiom4a},
the transition probability cannot be uniform as in the case of $e_{j}$. 
Assume a node $v'_{i1} \in T_{i}$ belongs to reciprocal arcs' head sets $K \neq 2$ times. 
Then, the transition probability assigned to $v'_{i1}$ is $p(v'_{i1}) = \frac{1}{2\lvert T_{i} \vert} \times K \neq \frac{1}{\lvert T_{i} \vert}$. 
This result indicates that the transition probability of $e_{i}$ is not the only optimal one. Thus, the following inequality holds:
\begin{equation*}
\mathcal{L}(\bar{p}_{j}, \bar{p}^{*}_{j}) < \mathcal{L}(\bar{p}_{i}, \bar{p}^{*}_{i})\hfill %\qedhere
\end{equation*}
%\begin{equation*}
%\mathcal{L}(p_{h, j}, p^{*}_{h, j}) < \mathcal{L}(p_{h, i}, p^{*}_{h, i}), \quad \forall v_{h, i} \in %T'_{i}, \ \forall v_{h, j} \in T'_{j}. \hfill %\qedhere
%\end{equation*}
\end{proof}

\subsubsection{Proof of the Fact that \measure Satisfies Axioms 5-8}  
\begin{newpropfull}{
\measure \normalfont{(}i.e., defining $r(e_i , R_i)$ as in Eq.~\eqref{eq:proposal}\normalfont{)}
satisfies \axiom~\textbf{\ref{axiom5}}.}
\end{newpropfull}
\begin{proof} This can be shown by using the known range of the probabilistic distance  $\mathcal{L}(p,q)$ as follows: 
\begin{align*}
    & 0 \leq \sum_{v_{h} \in H_{i}}\frac{\mathcal{L}(p_{h},p^{*}_{h})}{\mathcal{L}_{max}}  \leq \lvert H_{i} \vert   \quad (\because 0 \leq \mathcal{L}(p,q) \leq \mathcal{L}_{max}, \forall p,q) \\
    &\equiv  0 \leq \frac{\sum_{v_{h} \in H_{i}} \mathcal{L}(p_{h},p^{*}_{h})}{\lvert H_{i} \vert  \cdot \mathcal{L}_{max}}  \leq 1 \\ 
    &\equiv 0 \leq 1 - \frac{\sum_{v_{h} \in H_{i}} \mathcal{L}(p_{h},p^{*}_{h})}{\lvert H_{i} \vert  \cdot \mathcal{L}_{max}}  \leq 1 \\ 
    &\equiv 0 \leq \left( \frac{1}{\lvert R_{i} \vert } \right)^\alpha \left(1 - \frac{\sum_{v_h \in H_i}
        \mathcal{L}(p_{h}, p_{h}^{*})}{\lvert H_i \vert \cdot \mathcal{L}_{max} }\right) \leq 1 \quad (\because \alpha > 0) %\qedhere
\end{align*} 
\end{proof}
%\subsubsection{Proof of How \textsc{HyperRec} Satisfies \textsc{Axiom 6}}
\begin{newpropfull}{
\measure \normalfont{(}i.e., defining $r(G)$ as in Eq.~\eqref{eq:hypergraphreciprocity}\normalfont{)}
satisfies \axiom~\textbf{\ref{axiom6}}.}
\end{newpropfull}
\begin{proof}
% Note that this reciprocity measure is identical to the normal digraph reciprocity \cite{newman2002email, garlaschelli2004fitness}. 
Recall that the hypergraph level reciprocity of \measure is defined as 
$$
r(G) = \frac{1}{\lvert E \vert} \sum_{i=1}^{\lvert E \vert}r(e_{i} ,  R_{i})
$$
By the assumption, the size of every arc's head set is $1$, and thus each $r(e_{i} , R_{i})$ is rewritten as 
$$
r(e_{i}) = \max_{R_{i} \subseteq E, R_i\neq \emptyset} \left(\frac{1}{\lvert R_{i} \vert }\right)^{\alpha} 
\left(1 - \frac{\mathcal{L}{(p_{h} , p^{*}_{h})}}{\mathcal{L}_{max}}\right),
$$
where $\{v_{h}\} = H_{i}$.
Here, the optimal transition probability is 
\begin{equation*}
    p^{*}_{h}(v) = \begin{cases}
    1 & \text{if $\{v\} = T_{i}$,} \\
    0 & \text{otherwise.}
    \end{cases}
\end{equation*}

For a case where the perfectly reciprocal opponent $e'_i= \langle H'_{i} = T_{i} , T'_{i} = H_{i} \rangle$ of $e_i$ exists (i.e., $e'_{i} \in E$), $R_i=\{e'_i\}$ maximizes  $\left(\frac{1}{\lvert R_{i} \vert }\right)^{\alpha} 
\left(1 - \frac{\mathcal{L}{(p_{h} , p^{*}_{h})}}{\mathcal{L}_{max}}\right)$
since it minimizes both $\lvert R_{i} \vert $ (to $1$) and $\mathcal{L}{(p_{h} , p^{*}_{h})}$ (to $0$).
Thus, $r(e_{i})$ becomes $1$.

For a case where the perfectly reciprocal opponent $e'_i= \langle H'_{i} = T_{i} , T'_{i} = H_{i} \rangle$ of $e_i$ does not exist (i.e., $e'_{i} \notin E$).
Then, for each arc $e_k$ in the reciprocal set $R_i$, 
since $\lvert H_{i} \vert =\lvert T_{i} \vert=\lvert H_{k} \vert=\lvert T_{k} \vert=1$, $H_{i}\cap T_{k}=\emptyset$ or $T_{i}\cap H_{k}=\emptyset$ should hold.
Thus, there is no transition possibility from any node in $\in H_{i}$ to any node in $T_{i}$, and as a result, $p_h(v)=0$, $\forall v_h\in H_{i}$, $\forall v\in T_{i}$.
Hence, for every $R_{i}\subseteq E$, by \itemzero, $\mathcal{L}{(p_{h} , p^{*}_{h})}=\mathcal{L}_{max}$, and thus $r(e_i)=0$.

% It is trivial that only one arc is required for the reciprocal set since this is a common digraph case. 
% Thus, we only need to consider whether there exists an arc inversely-overlap, as $max$ term will filter out arcs which are not necessary.
% For $R_{i} = \{e_{k}\}$ where $T_{k} \neq H_{i}$, by \itemzeroone, reciprocity for such an arc becomes zero.
% For $R_{i} = \{e_{k}\}$ where $T_{k} = H_{i}$, transition probability for the target arc $e_{i}$ is defined as
% \begin{equation*}
%     p^{*}_{h}(v) = \begin{cases}
%     1 & \text{if $\{v\} = H_{k}$} \\
%     0 & \text{otherwise}
%     \end{cases}
% \end{equation*}
% Here, the transition probability and the optimal transition probability become identical if and only if $H_{k} = T_{i}$. 
% In this case, probabilistic distance becomes 0 as they have identical distribution (i.e., $\mathcal{L}{(p_{h} , p^{*}_{h})}=0$).
% Otherwise, as their non-zero probability domains do not overlap, their distance is maximized (i.e., $\mathcal{L}{(p_{h} , p^{*}_{h})}=\log{2}=\mathcal{L}_{max}$). 
% In sum, $r(e_{i} , R_{i} = \{e_{k}\})$ is formally re-written as
% \begin{equation*}
%     r(e_{i} , R_{i} = \{e_{k}\}) = \begin{cases}
%     1 & \text{if $H_{k} = T_{i}$ and $T_{k} = H_{i}$} \\
%     0 & \text{otherwise}
%     \end{cases}
% \end{equation*}
If we consider both cases together, $r(e_{i})$ becomes an indicator function that gives $1$, if there exists the perfectly reciprocal opponent, and $0$, otherwise. Formally, 
\begin{align*}
r(G) &= \frac{1}{\lvert E \vert} \sum_{i = 1}^{\lvert E \vert} r(e_{i}) \\
     &= \frac{1}{\lvert E \vert} \sum_{i = 1}^{\lvert E \vert} \mathbbm{1}
    (\exists e'_{i}\in E \text{ such that } H'_{i} = T_{i} \text{ and } T'_{i} = H_{i}),% \\
    %&= \frac{\lvert E^{\leftrightarrow} \vert}{\lvert E \vert},
\end{align*}
where $\mathbbm{1}(\text{TRUE}) = 1$ and $\mathbbm{1}(\text{FALSE}) = 0$; and this is identical to the digraph reciprocity measure \citep{newman2002email, garlaschelli2004fitness}, i.e., ${\lvert E^{\leftrightarrow} \vert}/{\lvert E \vert}$.
\end{proof}

%\subsubsection{Proof of How \textsc{HyperRec} Satisfies \textsc{Axiom 7}}

\begin{newpropfull}{
\measure \normalfont{(}i.e., defining $r(G)$ as in Eq.~\eqref{eq:hypergraphreciprocity}\normalfont{)}
satisfies \axiom~\textbf{\ref{axiom7}}.}
\end{newpropfull}
\begin{proof}
Recall that the hypergraph-level reciprocity of \measure is defined as 
$$
r(G) = \frac{1}{\lvert E \vert} \sum_{i=1}^{\lvert E \vert} r(e_{i})= \frac{1}{\lvert E \vert} \sum_{i=1}^{\lvert E \vert} \max_{R_{i}\subseteq E,R_{i} \neq\emptyset } r(e_{i} , R_{i})
$$
By \textsc{\textbf{Axiom 5}}, $0\leq r(e_{i} \, R_{i})\leq 1$ for any $e_{i}$ and $R_{i}$. 
This implies $0 \leq \sum_{i=1}^{\lvert E \vert} r(e_{i}) \leq \lvert E \vert$, which is equivalent to $0 \leq r(G)=\frac{1}{\lvert E \vert} \sum_{i=1}^{\lvert E \vert} r(e_{i}) \leq 1$. 

\end{proof}

%\subsubsection{Proof of How \textsc{HyperRec} Satisfies \textsc{Axiom 8}}

\begin{newpropfull}{
\measure \normalfont{(}i.e., defining $r(G)$ as in Eq.~\eqref{eq:hypergraphreciprocity}\normalfont{)}
satisfies \axiom~\textbf{\ref{axiom8}}.}
\end{newpropfull}
\begin{proof}
\kijung{We first show that the maximum value of \measure is attainable under the given condition of \axiom~\textbf{\ref{axiom8}}.}
From an arbitrary hypergraph $G$, let $E'=\{e_{i} \in E : \langle T_{i} , H_{i}\rangle \notin E \}$ be the set of arcs whose perfectly reciprocal opponents do not exist in $G$. 
Let $E^{add} = \bigcup_{e_{ki} \in E^{'}} {\langle T_{i} , H_{i} \rangle}$ be the set of perfectly reciprocal opponents of the arcs in $E'$. 
If we add $E^{add}$ to $G$, which gives $G^{+}=(V,E^{+}=E \cup E^{add})$,
then for each arc $e_i\in E^{+}$, 
the perfectly reciprocal opponent $e'_i= \langle H'_{i} = T_{i} , T'_{i} = H_{i} \rangle$ of $e_i$ exists (i.e., $e'_{i} \in E^{+}$), and thus
\begin{align*}
 r(e_{i})=r(e_{i},\{e'_i\}) & = \left(1 - \frac{\sum_{v_h \in H_i}
        \mathcal{L}(p_{h}, p_{h}^{*})}{\lvert H_i \vert \cdot \mathcal{L}_{max} }\right)  \\
        &= \left( 1 - \frac{0 + \cdots + 0}{\lvert H_i \vert \cdot \mathcal{L}_{max}}\right) \quad (\because p_{h} = p^{*}_{h}) \\
        & = 1,
\end{align*}
which implies that $r(G^{+})=\frac{1}{\lvert E^+ \vert } \sum_{i = 1}^{\lvert E^+ \vert } r(e_{i})=1$.

\kijung{We now show that the minimum value of \measure is attainable under the given condition of \axiom~\textbf{\ref{axiom8}}.
From an arbitrary hypergraph $G=(V,E)$, let $E^{-} = \{e_{i}\}$ be a hyperarc set that contains any single hyperarc $e_{i} \in E$.
For a hypergraph $G^{-} = (V, E^{-})$, the only possible choice of $R_{i}$ is $R_{i} = \{e_{i}\}$ since $R_{i}$ should be a non-empty set and there exists only a single hyperarc $e_{i}$ in $G^{-}$, and thus
\begin{align*}
    r(e_{i})=r(e_{i},\{e_i\}) & = \left(1 - \frac{\sum_{v_h \in H_i}
        \mathcal{L}(p_{h}, p_{h}^{*})}{\lvert H_i \vert \cdot \mathcal{L}_{max} }\right)  \\
        & = \left(1 - \frac{\vert H_{i}\vert \cdot \mathcal{L}_{max}}{\vert H_{i}\vert \cdot \mathcal{L}_{max}}\right) \quad (\because \text{\itemzeroone and \itemzerotwo}) \\
        & = 0,
\end{align*}
which implies that $r(G^{-}) = r(e_{i}) = 0.$
}
% Here, let . For $E^{*}$, there always exists a perfectly reciprocal opponent for each arc.
% \begin{align*}
% \text{Let } R_{i} = \{\langle T_{i} , H_{i} \rangle \} \quad \text{Then}\quad r(e_{i} , R_{i}) &= \left(1 - \frac{\sum_{v_h \in H_i}
%         \mathcal{L}(p_{h}, p_{h}^{*})}{\lvert H_i \vert \cdot \mathcal{L}_{max} }\right) \\
%         &= \left( 1 - \frac{0 + \cdots + 0}{\lvert H_i \vert \cdot \mathcal{L}_{max}}\right)  = 1 \quad \because p_{h} = p^{*}_{h}
% \end{align*}
% As $R_{i} = \{\langle T_{i} , H_{i} \rangle \}$, the target arc $e_{i}$ can always achieves maximum arc reciprocity. 
% In addition, the corresponding reciprocal set can always be found from $E^{*}$ due to the $max$ searching. As every arc reciprocity $r(e_{i}, R_{i}), \ \forall e_{i} \in E$ is equal to 1, its hypergraph reciprocity $r(G)$ also becomes 1.
\end{proof}

\subsection{Proof of Theorem \ref{thm:algo:basic}}\label{subsec:thm2}

\begin{proof}
{Refer to Section~\ref{sec:ferret} for the definition of $\Phi_{i}(\langle H'_{i}, T'_{i}\rangle)$.}
Given a target arc $e_{i}$, let $e_{a}$ and $e_{b}$ be two arcs in a set $\Phi_{i}(\langle H'_{i}, T'_{i}\rangle)$ where $A=\lvert H_a \vert \leq B=\lvert H_b\vert$.
{Consider an arbitrary reciprocal set $R_{i}\subseteq E$.
We use $p_{(i, a, h)}$ to denote the probability distribution at each node $v_{h} \in H'_{i}$ when the reciprocal set is $R_{i} \cup \{e_{a}\}$. 
Then, probabilistic distance between $p_{(i, a, h)}$ and $p^{*}_{h}$ is rewritten as
\begin{multline}
\mathcal{L}(p_{(i, a, h)}, p^{*}_{h}) = \sum_{v \in T'_{i}} \ell (\frac{1}{\lvert T_{i} \vert }, \frac{Kq_{(i, h)}(v)}{K+1} + \frac{1}{A(K+1)}) 
+ \sum_{v \in T_{i} \setminus T'_{i}} \ell (\frac{1}{ \vert T_{i} \vert }, \frac{Kq_{(i, h)}(v)}{K+1})  \\  + \sum_{v \in (\bigcup_{e_{k} \in E'_{(i,h)}} H_{k}) \setminus T_{i}} \frac{Kq_{(i,h)}(v)}{2(K+1)}\log2 
+ \sum_{v \in H_{a} \setminus T_{i}} \frac{1}{2A(K+1)}\log2 \label{eq:overall:a},
\end{multline} 
% \label{eq:overall:a}
where $E'_{(i, h)} = \{e_{k} \in R_{i} :v_{h} \in T_{k}\}$, $K = \lvert E'_{i,h} \vert $, and $q_{(i, h)}$ is the probability distribution at each node $v_{h} \in H'_{i}$ when the reciprocal set is $R_{i}$.
It should be noticed that the definition of $H'_{i}$ and $T'_{i}$, $H'_{i} = T_{a} \cap H_{i} = T_{b} \cap H_{i}$ and $T'_{i} = H_{a} \cap T_{i} = H_{b} \cap T_{i}$ hold. 
%Hence, $\mathcal{L}(p_{(i, a, h)}, p^{*}_{h})$ can be expressed as Eq.~\eqref{eq:overallfirst}-\eqref{eq:overallfourth}.
In the same way, we define $p_{(i,b,h)}$ is as the probabilistic distribution at each node $v_{h} \in H'_{i}$ when $R_{i} \cup \{e_{b}\}$ is the reciprocal set.
Then, $\mathcal{L}(p_{(i, b, h)}, p^{*}_{h})$ can be rewritten as in Eq.~\eqref{eq:overall:a}.
}

We prove the theorem by showing that $\mathcal{L}(p_{(i, a, h)}, p^{*}_{h}) \leq \mathcal{L}(p_{(i, b, h)}, p^{*}_{h})$ holds for every $v_h\in H'_{i}=H_{i} \cap T_{a} = H_{i} \cap T_{b}$.
%Since $$ holds, the remaining step is to show that 
Note that the second and third terms of the RHS do not depend on $e_{a}$ and  $e_{b}$, and they are identical in $\mathcal{L}(p_{(i, a, h)}, p^{*}_{h})$ and $\mathcal{L}(p_{(i, b, h)}, p^{*}_{h})$. 
Thus, we rewrite $\mathcal{L}(p_{(i, a, h)}, p^{*}_{h}) \leq \mathcal{L}(p_{(i, b, h)}, p^{*}_{h})$ as
\begin{align*}
&\mathcal{L}(p_{(i, a, h)}, p^{*}_{h}) \leq \mathcal{L}(p_{(i, b, h)}, p^{*}_{h}) \\ 
&\equiv \sum_{v \in T'_{i}} \ell (\frac{1}{\lvert T_{i} \vert }, \frac{Kq_{(i, h)}(v)}{K+1} + \frac{1}{A(K+1)}) + \sum_{v \in H_{a} \setminus T_{i}} \frac{1}{2A(K+1)}\log2 \\
&\leq \sum_{v \in T'_{i}} \ell (\frac{1}{\lvert T_{i} \vert }, \frac{Kq_{(i, h)}(v)}{K+1} + \frac{1}{B(K+1)}) + \sum_{v \in H_{b} \setminus T_{i}} \frac{1}{2B(K+1)}\log2. 
\end{align*}
%where $\lvert H_{b} \vert = B$.
For simplicity, let $\lvert T_{i} \vert = T$ abd $\lvert T'_{i} \vert = F$. Then, the above inequality is rewritten as
\begin{align}
&\sum_{v \in T'_{i}} \ell (\frac{1}{T}, \frac{Kq_{(i, h)}(v)}{K+1} + \frac{1}{A(K+1)})
- \ell (\frac{1}{T}, \frac{Kq_{(i, h)}(v)}{K+1} + \frac{1}{B(K+1)}) \nonumber \\
&\leq \frac{B - F}{2B(K+1)}\log2 - \frac{A - F}{2A(K+1)}\log2~\label{eq:fulleq}.
\end{align}
Let $v' = \argmax_{v \in T'_{i}} \ell (\frac{1}{T}, \frac{Kq_{(i, h)}(v)}{K+1} + \frac{1}{A(K+1)})
- \ell (\frac{1}{T}, \frac{Kq_{(i, h)}(v)}{K+1} + \frac{1}{B(K+1)})$,
and let $p'=q_{(i,h)}(v')$. 
Then, the following inequality holds:
\begin{align*}
&\sum_{v \in T'_{i}} \ell (\frac{1}{T}, \frac{Kq_{(i, h)}(v)}{K+1} + \frac{1}{A(K+1)})
- \ell (\frac{1}{T}, \frac{Kq_{(i, h)}(v)}{K+1} + \frac{1}{B(K+1)}) \\
&\leq F \times (\ell (\frac{1}{T}, \frac{Kp'}{K+1} + \frac{1}{A(K+1)})
- \ell (\frac{1}{T}, \frac{Kp'}{K+1} + \frac{1}{B(K+1)}))
\end{align*}
Thus, the inequality \eqref{eq:fulleq} is implied by 
\begin{align}
&F \times (\ell (\frac{1}{T}, \frac{Kp'}{K+1} + \frac{1}{A(K+1)})
- \ell (\frac{1}{T}, \frac{Kp'}{K+1} + \frac{1}{B(K+1)})) \nonumber \\ 
&\leq \frac{F}{2A(K+1)}\log2 - \frac{F}{2B(K+1)}\log2 ~\label{eq:fulleq2}.
\end{align}
By unfolding $\ell$ in the LHS and dividing both sides by $F$, the inequality~\eqref{eq:fulleq2} is rewritten as
\begin{align*}
& \frac{1}{2(K+1)}(\frac{1}{A} - \frac{1}{B}) \log 2 \geq \\
&  \frac{1}{2T}\log(\frac{\frac{2}{T}}{\frac{1}{T} + \frac{Kp'}{K+1} + \frac{1}{A(K+1)}}) + \frac{1}{2}(\frac{Kp'}{K+1} + \frac{1}{A(K+1)}) 
    \log (\frac{2(\frac{Kp'}{K+1} + \frac{1}{A(K+1)})}{\frac{1}{T} + \frac{Kp'}{K+1} + \frac{1}{A(K+1)}})  \\ 
    &  -\frac{1}{2T}\log(\frac{\frac{2}{T}}{\frac{1}{T} + \frac{Kp'}{K+1} + \frac{1}{B(K+1)}})
    - \frac{1}{2}(\frac{Kp'}{K+1} + \frac{1}{B(K+1)}) 
    \log (\frac{2(\frac{Kp'}{K+1} + \frac{1}{B(K+1)})}
    {\frac{1}{T} + \frac{Kp'}{K+1} + \frac{1}{B(K+1)}}) \\
&  \equiv 
\frac{1}{(K+1)}(\frac{1}{A} - \frac{1}{B}) \log 2 - \frac{1}{(K+1)}(\frac{1}{A} - \frac{1}{B})\log 2 \geq  \\
& \frac{1}{T}\log(\frac{\frac{1}{T}}{\frac{1}{T} + P' + \frac{1}{A(K+1)}}) + (P' + \frac{1}{A(K+1)}) 
    \log (\frac{P' + \frac{1}{A(K+1)}}{\frac{1}{T} + P' + \frac{1}{A(K+1)}}) \\ 
&  -\frac{1}{T}\log(\frac{\frac{1}{T}}{\frac{1}{T} + P' + \frac{1}{B(K+1)}})
    - (P' + \frac{1}{B(K+1)}) 
    \log (\frac{P' + \frac{1}{B(K+1)}}{\frac{1}{T} + P' + \frac{1}{B(K+1)}}),
\end{align*}
where $P = \frac{Kp'}{K+1}$.
Let $P_{A} = P' + \frac{1}{A(K+1)}$ and $P_{B} = P' + \frac{1}{B(K+1)}$, where $P_{A} \geq P_{B}$.
Then, by cancelling out all identical terms, the above inequality is simplified as
\begin{align*}
    % \resizebox{0.3\textwidth}{!}{$P_{A} = (P' + \frac{1}{A(K+1)}) \quad \text{and} \quad P_{B} = (P' + \frac{1}{B(K+1)}) $} \\
    & 0 \geq \frac{1}{T}\log(\frac{\frac{1}{T}}{\frac{1}{T} + P_{A}}) 
    -  \frac{1}{T}\log(\frac{\frac{1}{T}}{\frac{1}{T} + P_{B}}) \\
    & + P_{A} \log (\frac{P_{A}}{\frac{1}{T} + P_{A}}) - P_{B} 
    \log (\frac{P_{B}}{\frac{1}{T} + P_{B}}) \\
    & \equiv 0 \geq \frac{1}{T}\log(\frac{\frac{1}{T} + P_{B}}{\frac{1}{T} + P_{A}}) + P_{B} \log (\frac{\frac{1}{T} + P_{B}}{P_{B}}) - P_{A} \log (\frac{\frac{1}{T} + P_{A}}{P_{A}}) 
\end{align*}
If we multiply by $T$ both sides,
this inequality is implied by the two following inequalities:
\begin{equation}
    \log(\frac{\frac{1}{T} + P_{B}}{\frac{1}{T} + P_{A}}) \leq 0, \label{eq:finalineq1}
\end{equation}
\begin{equation}
    \log\left( \frac{(1 + \frac{1}{TP_{B}})^{TP_{B}}}
    {(1 + \frac{1}{TP_{A}})^{TP_{A}}}\right) \leq 0. \label{eq:finalineq2}
\end{equation}
The inequality~\eqref{eq:finalineq1} is trivial since $P_{A} \geq P_{B}$.
For the inequality~\eqref{eq:finalineq2}, the numerator and the denominator are in the form of $f(n) = (1+\frac{1}{n})^{n}$, which is a non-decreasing function.
Thus, the denominator is always greater than or equal to the numerator, thus satisfying the inequality.
% As original inequality holds, proof is done.
\end{proof}

\subsection{Proof of Corollary~\ref{corr:algo}}\label{subsec:col1}
% By Theorem~2, we can choose the one arc from each $\langle H'_{i}, T'_{i}\rangle$.
% Tail sets of arcs in $\Psi_{i}$ are disjoint since $\lvert T_{j} \vert = 1 ~ \forall e_{j} \in E$.
% If any arc $e_{k} \in \Psi_{i}$ is excluded from $R_{i}$, $v' = H_{i} \cap T_{k}$ does not have transition probability, which increases the probabilistic distance. 
% Thus, the inclusion of $e_{k} \in \Psi_{i}$ to $R_{i}$ is always better for the maximization of the reciprocity, as the size penalty is safely discarded with our assumption of $\alpha \rightarrow 0$. Finally, we induce $\argmax_{R_{i} \subseteq E} r(e_{i} , R_{i}) = \Psi_{i}$.

\begin{proof}
Since \measure measures a weighted average of the probabilistic distances that are defined for each head set node of the target arc, we consider each head set node $v_h\in H_i$ of the target arc $e_i$.
Since the tail set size of every arc is identical to 1 (i.e., $\lvert T_{i} \vert = 1$), there exists at most one arc in $\Psi_{i}$ that covers a specific node $v_{h}$ in the head set $H_i$ of the target arc, i.e., $\forall v_h \in H_i$,
\begin{equation}
    %\lvert \Psi'_{i,h} \vert = 
    \lvert \{e_{k} \in \Psi_{i} : v_{h} \in T_{k}\} \vert \leq 1,  \label{eq:atmostone}
\end{equation}
%Example is illustrated in Figure~\ref{fig:corr1} that when a target arc is $e_{2}$, $v \in H_{2}$ can be covered by at most one node.
%Intuitively, $\Gamma_{i, k}$ consists of $k$ number of arcs which are picked from $\Psi_{i}$ in an ascending order of their head set size.
Let $v'$ be the target arc's tail set node, i.e., $T_{i}=\{v'\}$.
Then, by the definition of $\Psi_{i}$, $v'$ is included in the head set of every $e_{k} \in \Psi_{i}$, i.e., $\forall e_{k} \in \Psi_{i}$,
\begin{equation}
    v'\in H_{k}. \label{eq:mustinclude}
\end{equation}
Eq.~\eqref{eq:atmostone} and Eq.~\eqref{eq:mustinclude} imply that, $\forall e_{k} \in R_{i} \subseteq \Psi_{i}$, Eq.~\eqref{eq:decreasing} holds.
\begin{equation}
   \jsdpq{p_{h}}{p^{*}_{h}}=
   \ell(1,\frac{1}{\lvert H_{k} \vert})
   +\sum_{v\in H_{k}\setminus \{v'\}}\ell(0,\frac{1}{\lvert H_{k}\vert }).
   \label{eq:decreasing}
\end{equation}

Let $A = \lvert H_{k1} \vert$ and $B = \lvert H_{k2} \vert$.
Then, $\forall e'_{h1},e'_{h2} \in R_{i} \subseteq \Psi_{i} \text{ s.t. }  \lvert H'_{h1} \vert \leq \lvert H'_{h2} \vert$, 
\begin{align}
    & \jsdpq{p_{h1}}{p^{*}_{h1}}
    \leq \jsdpq{p_{h2}}{p^{*}_{h2}} \label{eq:decreasing:jsd}  \\
    & \equiv
    \ell(1,\frac{1}{\lvert H_{k1} \vert})
   +\sum_{v\in H_{k1}\setminus \{v'\}}\ell(0,\frac{1}{\lvert H_{k1}\vert })
   \leq 
\ell(1,\frac{1}{\lvert H_{k2} \vert})
   +\sum_{v\in H_{k2}\setminus \{v'\}}\ell(0,\frac{1}{\lvert H_{k2}\vert }) \nonumber \\
   &\equiv \frac{1}{2A}\log (\frac{2}{1 + A}) + \frac{1}{2}\log (\frac{2A}{1 + A})+ \frac{A-1}{2A} \log2 \nonumber \\
& \leq \frac{1}{2B}\log (\frac{2}{1 + B}) + \frac{1}{2}\log (\frac{2B}{1 + B}) + \frac{B-1}{2B} \log2 \nonumber  \\
&\equiv \log(\frac{A(B+1)}{B(A+1)}) + (\frac{1}{A} - \frac{1}{B}) \log2 + \log((1 + B)^{\frac{1}{B}}) - \log((1 + A)^{\frac{1}{A}})
\leq (\frac{1}{A} - \frac{1}{B}) \log2 \nonumber  \\
&\equiv \log(\frac{A(B+1)}{B(A+1)}) + \log(\frac{(1 + B)^{\frac{1}{B}}}{(1 + A)^{\frac{1}{A}}})
\leq 0, ~\label{eq:whydecreasing2}
\end{align}
We prove the inequality \eqref{eq:decreasing:jsd} by showing that
both first and second terms of the LHS of the inequality \eqref{eq:whydecreasing2} are smaller than or equal to $0$. Since $A \leq B$, it is trivial that the first term is smaller than or equal to 0. The second term has a functional form of $\log(f(x)/f(x'))$ where $f(x)$ is a non-increasing function and $x\leq x'$, and thus the second term is also smaller than or equal to $0$.

%and this is a non-decreasing function of $\vert H_k \vert$.
%The reason is that Eq.~\eqref{eq:whydecreasing} holds for  since 
%the first and second term of LHS of Eq.~\eqref{eq:whydecreasing2} are both smaller than or equal to 0.
% \begin{equation}

%   ~\label{eq:whydecreasing}
% \end{equation}
% \begin{align}
% &\frac{1}{2A}\log (\frac{2}{1 + A}) + \frac{1}{2}\log (\frac{2A}{1 + A})+ \frac{A-1}{2A} \log2 \nonumber \\
% &\leq \frac{1}{2B}\log (\frac{2}{1 + B}) + \frac{1}{2}\log (\frac{2B}{1 + B}) + \frac{B-1}{2B} \log2 \nonumber  \\
% &\equiv \log(\frac{A(B+1)}{B(A+1)}) + (\frac{1}{A} - \frac{1}{B}) \log2 + \log((1 + B)^{\frac{1}{B}}) - \log((1 + A)^{\frac{1}{A}})
% \leq (\frac{1}{A} - \frac{1}{B}) \log2 \nonumber  \\
% &\equiv \log(\frac{A(B+1)}{B(A+1)}) + \log(\frac{(1 + B)^{\frac{1}{B}}}{(1 + A)^{\frac{1}{A}}})
% \leq 0. ~\label{eq:whydecreasing2}
% \end{align}

In addition, let $e'_{h} \in \Psi_{i}$ be the only arc where $T'_{h}=\{v_{h}\}$.
%As explained in the previous paragraph, at most one such an arc $e'_{h}$ exists.
Then, the probabilistic distance at a each head set node $v_{h}$ depends on whether $e'_{h}$ is in any reciprocal set $R_i\subseteq \Psi_{i}$ as follows:
\begin{equation}
    \mathcal{L}(p_{h}, p_{h}^{*}) = \begin{cases}
    {\jsdpq{p'_{h}}{p_{h}^{*}}} & \text{if $e'_{h} \in R_{i}$} \\
    \mathcal{L}_{max} & \text{otherwise,} \\
\end{cases} \label{eq:onlyinclusion}
\end{equation}
where $p'_{h}$ is the probability distribution  
at $v_h$ 
when the reciprocal set $R_i=\{e'_h\}$, $p^{*}_{h}$ is the optimal probability distribution at a node $v_{h} \in H_{i}$.

%\red{That is, the tail sets of arcs in $\Psi_{i}$  are singleton sets, they consist of different head set nodes of the target arc.
%Their head set includes the only tail set node of the target arc.} 

% an arc with smaller head-set size has smaller probabilistic distance. 
% In summary,
% \begin{itemize}
%     \item Inclusion of a specific arc $e'_{h} \in \Psi_{i}$ in $R_{i}$ only affects $p_{h}$ where $v_{h} \in H_{i} \cap T'_{h} \neq \emptyset$.
%     \item There exists at most one $e'_{h}$ for each $v_{h} \in H_{i}$.
%     \item 
%      \red{Consider two hyperarcs $e'_{h1},e'_{h2} \in R_{i}$.}
%     If $\lvert H'_{h1} \vert \leq \lvert H'_{h2} \vert$ holds, then $\jsdpq{p'_{h1}}{p^{*}_{h1}} \leq \jsdpq{p'_{h2}}{p^{*}_{h2}}$ holds. 
% \end{itemize}
By Eq.~\eqref{eq:atmostone} and Eq.~\eqref{eq:onlyinclusion}, for any $R'_{i} = \{e'_{h1}, \cdots , e'_{hk}\}\subseteq \Psi_{i}$, Eq.~\eqref{eq:fullequation} holds.
% (assuming $v_{h1}, \cdots v_{hk}$ are all distinct nodes).
\begin{equation}
    r(e_{i}, R'_{i}) = \left( \frac{1}{k}\right)^{\alpha} \left(1 - \frac{\jsdpq{p'_{h1}}{p^{*}_{h1}} + \cdots + \jsdpq{p'_{hk}}{p^{*}_{hk}} + (\lvert H_{i} \vert  - k) \times \mathcal{L}_{max}}{\lvert H_{i} \vert  \times \mathcal{L}_{max}} \right), \label{eq:fullequation}
\end{equation}
where $T'_{hj}=\{v_{hj}\}$ for every $j\in \{1,\cdots, k\}$.
In addition, Eq.~\eqref{eq:fullequation} and the inequality \eqref{eq:decreasing:jsd} imply that drawing $k$ reciprocal arcs from $\Psi_{i}$ in ascending order of their head set size achieves the maximum reciprocity for fixed $k$.
Let $\Gamma_{i,k}$ be such a reciprocal set.
Formally, if we let $\Gamma_{i, k}$ be a subset of $\Psi_{i}$ such that $ \lvert \Gamma_{i,k} \vert = k \text{ and }\lvert H_{s} \vert \leq \lvert H_{t} \vert,  \ \forall e_{s} \in \Gamma_{i, k}, \ \forall e_{t} \in \{\Psi_{i} \setminus \Gamma_{i, k}\}$, 
then %Thus, when $k$ is fixed as a constant, 
$\argmax_{R_{i}  \subseteq \Psi_{i}  s.t. \lvert R_{i} \vert  = k} r(e_i,R_i) = \Gamma_{i, k}$ holds.
%As there exists the size penalty term (i.e., $(1/\lvert R_{i} \vert )^{\alpha}$), one still need to compare $\Gamma_{i, 1}, \Gamma_{i, 2}, \cdots, \Gamma_{i, |\Psi_{i} \vert}$ in order to find $\argmax_{R_{i} \subseteq \Psi_{i}, R_{i}\neq \emptyset} r(e_i,R_i)$.
% Thus, overall algorithm's framework is formalized as $\argmax_{R_{i} \subseteq \Psi_{i}, R_{i}\neq \emptyset} r(e_i,R_i)$ = $\argmax_{R_{i} \in \{\Gamma_{i, 1}, \cdots \Gamma_{i, |\Psi_{i} \vert}\}} r(e_i,R_i)$, resulting in the complexity of $O(|\Psi_{i} \vert)$.
\end{proof}

}

\section{Appendix: Limitations of Baseline Measures}{
    \label{section:baselinefail}
    
In this section, we show why several baseline measures fail in satisfying some of
\axiomnum{1-8}.
Below, We use $G_{i}=(V_{i},E_{i})$ and $G_{j}=(V_{j},E_{j})$ to denote the hypergraphs on the left side and the right side, respectively, of each subfigure of Figure~\ref{fig:AXIOMs}. 

%\begin{figure}[t]
%  \centering
%  \includegraphics[width=0.4\textwidth]{FIG_ONLINE_APPENDIX/FIG_Proof3.pdf}
%  \caption{An example regarding Corollary~\ref{corr:algo}}\label{fig:corr1}.
%\end{figure}

%\begin{figure}[t]
    %\centering
    %\subfigure[Axiom 3A]{\includegraphics[width=0.336\textwidth]{FIG_ONLINE_APPENDIX/FIG_ONLINE_B1a.pdf}}
    %\hspace{10mm}
    %\subfigure[Axiom 3B]{\includegraphics[width=0.336\textwidth]{FIG_ONLINE_APPENDIX/FIG_ONLINE_B1b.pdf}}
    %\caption{\label{fig:axiom3false} Counterexamples for baseline measures. The left hypergraph of each subfigure is $G_{1}$, and the right hypergraph of each subfigure is $G_{2}$.}
%\end{figure}

\begin{table}[t]
    \caption{Violation of \textsc{\textbf{Axiom 3}}.
    We use $G_{i}$ and $G_{j}$ to denote the hypergraphs on the left side and the right side, respectively, of each subfigure of Figure~\ref{fig:AXIOMs}. 
    This table reports the computed reciprocity values, $r(G_{i})$ and $r(G_{j})$, of \citep{pearcy2014hypergraph} (B1), and the computed reciprocity values, $r(e_{i}, R_{i})$ and $r(e_{j}, R_{j})$, of the ratio of covered pairs (B2) and \measure without size penalty (B5), for Figures~\ref{fig:AXIOMs}(d)-(e). In order to satisfy \axiomnum{3}, $r(e_{i}, R_{i}) < $  $r(e_{j}, R_{j})$ (or $r(G_{i}) < $  $r(G_{j})$) should hold in both subfigures. Note that, B1, B2, and B5 cannot satisfy the inequality in at least one subfigure. } % title of Table
    \label{tab:violation3}
    \vspace{3mm}
    \centering % used for centering table
    %\scalebox{0.73}{
    \resizebox{\columnwidth}{!}{
    \renewcommand{\arraystretch}{1.2}{
    \begin{tabular}{l  c  c  c  c } % centered columns (4 columns)
        \toprule %inserts double horizontal lines
        \multirow{2}{*}{} &  \multicolumn{2}{c}{Figure~\ref{fig:AXIOMs}(d) (\axiomnum{3A})} & \multicolumn{2}{c}{Figure~\ref{fig:AXIOMs}(e) (\axiomnum{3B})} \\
        & $r(e_{i}, R_{i})$ or $r(G_{i})$ & 
        $r(e_{j}, R_{j})$ or $r(G_{j})$& 
        $r(e_{i}, R_{i})$ or $r(G_{i})$&
        $r(e_{j}, R_{j})$ or $r(G_{j})$\\
        \midrule
        B1 (\citet{pearcy2014hypergraph}) & 0.2093 & 0.2093 & 0.2093 & 0.2093 \\
        B2 (Ratio of Covered Pairs) & 0.5625 & 0.5625 & 0.5625 & 0.5625 \\
        B5 (\measure w/o Size Penalty) & 0.6333 & 0.6446 & 0.6446 & 0.6446 \\
        \bottomrule
    \end{tabular}
    }
    }
\end{table} 

% \begin{figure}[t]
%     \centering
%     \includegraphics[width=0.5\textwidth]{FIG/AXIOMs/AXIOM2_LABEL.pdf}
%     \caption{\label{fig:axiom3false} For simplicity, we denote each node of Figure~\ref{fig:AXIOMs} as above.}
%     \label{fig:nodenamesofaxiom3}
% \end{figure}

\subsection{Violations of Axiom 3}

We show how several baseline measures violate \axiomnum{3}. In Figures~\ref{fig:AXIOMs}(d)-(e), $r(e_{i}, R_{i}) < r(e_{j}, R_{j})$ should hold in order to satisfy \axiomnum{3}. However, we numerically verify that $r(e_{i}, R_{i}) = r(e_{j}, R_{j})$ hold for some baseline measures, which violates \axiomnum{3}.

%% Sunwoo 

\smallsection{B1. \citep{pearcy2014hypergraph}} They use clique expansion, which transforms every hyperedge of a hypergraph to a clique of a pairwise graph (e.g., $\langle \{v_{1} \}, \{v_{2}, v_{3} \} \rangle  \rightarrow \{\langle \{v_{1}\} , \{v_{2}\}\rangle ,  \langle \{v_{1}\} , \{v_{3}\}\rangle\}$).
Through this process, the original hypergraph is transformed into a weighted digraph (see Section~\ref{sec:relatedwork}). 
Since \citet{pearcy2014hypergraph} do not propose any arc-level reciprocity, we compare its hypergraph-level reciprocity for counterexamples regarding \axiomnum{3}.
That is, we compare $r(G_{i})$ and $r(G_{j})$ in Figures~\ref{fig:AXIOMs}(d)-(e).
% \red{Specifically, 
% we use $G_{i}$ to denote the hypergraph (with three arcs) on the left side of Figure~\ref{fig:AXIOMs}(d) and use $G_{j}$ to denote the hypergraph (with two arcs) on the right side of the same figure.}
%we compare $r(G_{i})$ and $r(G_{j})$ in Figure~\ref{fig:AXIOMs}(d), where $G_{i} = (V_{i} = \{v_{k} : k \in \{1, \cdots 8\}\} , E_{i} = \{e'_{i1}, e'_{i2} \})$ and $G_{j} = (V_{j} = \{v_{t} : t \in \{9, \cdots 16\}\} , E_{j} = \{e'_{j}\})$ (if we denote each node of Figure~\ref{fig:AXIOMs} as Figure~\ref{fig:nodenamesofaxiom3})}.
As reported in Table~\ref{tab:violation3}, $r(G_{i}) = r(G_{j}) = 0.2093$ and $r(G_{i}) = r(G_{j}) = 0.2093$ hold in  Figure~\ref{fig:AXIOMs}(d) and Figure~\ref{fig:AXIOMs}(e), respectively, violating \axiomnum{3}.
\smallsection{B2. Ratio of Covered Pairs} %Unlike the previous case, 
We compare the ratio of covered pairs (B2), which is arc-level reciprocity, in Figures~\ref{fig:AXIOMs}(d)-(e).
%Here we show that this metric results in $r(e_{i} , R_{i} = \{e'_{i1},e'_{i2}\}) = r(e_{j} , R_{j} = \{e'_{j}\})$, which is again a violation of \axiomnum{3}. 
As reported in Table~\ref{tab:violation3}, $r(e_{i}, R_{i}=\{e'_{i1},e'_{i2}\}) = r(e_{j}, R_{j} = \{e'_{j}\}) = 0.5625$ and $r(e_{i}, R_{i}) = r(e_{j}, R_{j}) = 0.5625$ hold in Figure~\ref{fig:AXIOMs}(d) in Figure~\ref{fig:AXIOMs}(e), respectively, which violates \axiomnum{3}.

\smallsection{B5. \measure w/o Size Penalty} 
As reported in Table~\ref{tab:violation3}, $r(e_{i} , R_{i}=\{e'_{i1},e'_{i2}) = r(e_{j} , R_{j} = \{e'_{j}\}) = 0.6446$ holds in Figure~\ref{fig:AXIOMs}(e), which violates \axiomnum{3}.

\begin{comment}
\smallsection{B5. \measure w/o Size Penalty} While proving general \measure, original inequality $r(e_{i} , R_{i}) < r(e_{j} , R_{j})$ has been relaxed to $\mathcal{L}(p_{h, j}, p^{*}_{h, j}) \leq \mathcal{L}(p_{h, i}, p^{*}_{h, i})$ 
with the help of the size penalty term $\left(1/|R_{i}|\right)^{\alpha}$. 
On the other hand, if we remove this size penalty term, an equality term in $\mathcal{L}(p_{h, j}, p^{*}_{h, j}) \leq \mathcal{L}(p_{h, i}, p^{*}_{h, i})$ cannot guarantees the original reciprocity inequality.
However, we have shown several cases where $\mathcal{L}(p_{h, j}, p^{*}_{h, j}) = \mathcal{L}(p_{h, i}, p^{*}_{h, i})$ occur, which result in the violation of \axiomnum{3}.
In summary, \measure without the size penalty term $(1/|R_{i}|)^{\alpha}$ cannot satisfies \axiomnum{3}.
\end{comment}

\begin{table}[t]
    \caption{Violation of \axiomnum{4}. The computed reciprocity values, $r(e_{i}, R_{i})$ and $r(e_{j}, R_{j})$, of the ratio of covered pairs (B2) and the penalized ratio of covered pairs (B3) with $\alpha=1$ in Figure~\ref{fig:AXIOMs}(f). While $r(e_{i}, R_{i}) < $  $r(e_{j}, R_{j})$ should hold to satisfy \axiomnum{4}, the inequality does not hold for B2 and B3.}
    \label{tab:violation4}
    \vspace{3mm}
    \centering % used for centering table
    \scalebox{1.0}{
    \renewcommand{\arraystretch}{1.2}{
    \begin{tabular}{l  c  c} % centered columns (4 columns)
        \toprule %inserts double horizontal lines
         &  \multicolumn{2}{c}{Figure~\ref{fig:AXIOMs}(f) (\axiomnum{4})} \\
        & $r(e_{i}, R_{i})$ & $r(e_{j}, R_{j})$\\
        \midrule
        B2 (Ratio of Covered Pairs) & 1.00 & 1.00 \\
        B2 (Penalized Ratio of Covered Pairs) & 0.25 & 0.25\\
        \bottomrule
    \end{tabular}
    }
    }
\end{table} 

\subsection{Violations of Axiom 4}

We show how several baseline measures violate \axiomnum{4}. In Figure~\ref{fig:AXIOMs}(f), $r(e_{i}, R_{i}) < r(e_{j}, R_{j})$ should hold in order to satisfy \axiomnum{4}. However, we numerically verify that $r(e_{i}, R_{i}) = r(e_{j}, R_{j})$ hold for some baseline measures, which violates \axiomnum{4}.

\smallsection{B2. Ratio of Covered Pairs}  %\red{We show $r(e_{i} , R_{i} = \{e'_{i}\}) = r(e_{j} , R_{j} = \{e'_{j}\})$, which is a violation of \axiomnum{4}. 
As reported in Table~\ref{tab:violation4}, $r(e_{i} , R_{i}=(E_{i}\setminus \{e_{i}\}) = r(e_{j} , R_{j}=E_{j}\setminus \{e_{j}\}) = 1.00$ holds in Figure~\ref{fig:AXIOMs}(f), which violates \axiomnum{4}.

\smallsection{B3. Penalized Ratio of Covered Pairs} %\red{Similar to the B2's violation, we verify that $r(e_{i} , R_{i} = \{e'_{i}\}) = r(e_{j} , R_{j} = \{e'_{j}\})$ hold. 
%As shown in in Table~\ref{tab:violation4}, $r(e_{i} , R_{i}) = r(e_{j} , R_{j}) = 0.25$ hold in \axiomnum{4}'s case, which violates \axiomnum{4}}.
As reported in Table~\ref{tab:violation4}, $r(e_{i} , R_{i}=(E_{i}\setminus \{e_{i}\}) = r(e_{j} , R_{j}=E_{j}\setminus \{e_{j}\}) = 0.25$ holds in Figure~\ref{fig:AXIOMs}(f), which violates \axiomnum{4}.

\begin{comment}
\subsection{Violations of Axiom 4}
\smallsection{B2. Ratio of Covered Pairs} If every pair of a target arc has been covered, then reciprocity for such a case is always equal to 1. 
That is, how many times each node has been pointed is indistinguishable from this metric. 
Here, reciprocity for both cases become identical (i.e., $r(e_{i} , R_{i}) = r(e_{j} , R_{j})$), which violates \axiomnum{4}. \\

\smallsection{B3. Penalized Ratio of Covered Pairs} Although size penalty  has been added to \textbf{B2}, cardinality of reciprocal sets are identical for both cases. 
Thus, this also has the same issue of identical reciprocity, which violates \axiomnum{4}. 
\end{comment}

\subsection{Violations of Axiom 5}

\smallsection{B4. \measure w/o Normalization} 
In order to satisfy \axiomnum{5}, reciprocity should always lie in a fixed finite range.
%In order to satisfy \axiomnum{5}, reciprocity should always lie between $0$ and $1$.
%Here, we demonstrate that a certain baseline measure violates \axiomnum{5} by showing that its reciprocity value can lie outside $[0, 1]$.
%Here, we demonstrate that a certain baseline measure violates \axiomnum{5} \modify{by showing that its reciprocity value can become infinite.
Here, we demonstrate that (B4) violates \axiomnum{5} by showing that its reciprocity value can become infinite.
Recall that (B4) is defined as
\begin{equation}\label{eq:b4}
r(e_{i}, R_{i}) = \left( \frac{1}{\lvert R_{i} \vert } \right)^\alpha \left(\lvert H_{i} \vert  - \frac{\sum_{v_h \in H_i}
        \mathcal{L}(p_{h}, p_{h}^{*})}{\mathcal{L}_{max} }\right).    
\end{equation}
Consider a case where $R_{i}=\{e'_{i}=\langle T_{i}, H_{i}\rangle \}$.
Then, for each $v_{h} \in H_{i}$, $\mathcal{L}(p_{h}, p^{*}_{h})=0$ holds.
In turn, Eq~\eqref{eq:b4} becomes $r(e_{i},R_{i})=\vert H_{i}\vert$.
In this case, as $\vert H_{i}\vert$ approaches infinity, $r(e_{i},R_{i})$ also becomes infinite.
Since the value of (B4) does not lie in a fixed finite range, (B4) violates \axiomnum{5}.

\color{black}
\subsection{Violations of Axiom 6}

\begin{figure*}[h]
    \centering
    \includegraphics[width=0.3\textwidth]{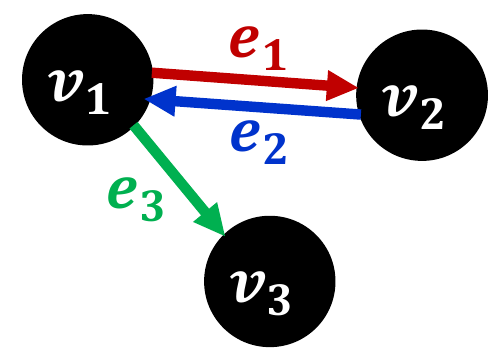}
    \caption{A counterexample that shows that some baseline measures fail to satisfy \axiomnum{6}.}
    \label{fig:axiom6counter}
\end{figure*}

\smallsection{B1. \citep{pearcy2014hypergraph}} Consider the digraph in Figure~\ref{fig:axiom6counter}
The digraph reciprocity of the digraph is $r(G) = \frac{2}{3}$ since 
$E = \{ e_{1},e_{2},e_{3} \}$, and $E^{\leftrightarrow} = \{e_{1}, e_{2}\}$.
In this case, however, the clique-expanded adjacency matrices of the digraph and the perfectly reciprocal hypergraph are
\begin{equation*}
    \bar{A} = 
    \begin{pmatrix}
        0 & 1 & 1\\
        1 & 0 & 0\\
        0 & 0 & 0
        \end{pmatrix}   
    \text{ and } \bar{A'} = 
    \begin{pmatrix}
        0 & 1 & 1\\
        1 & 0 & 0\\
        1 & 0 & 0
        \end{pmatrix}.    
\end{equation*} Thus, according to the definition in \citep{pearcy2014hypergraph}, the reciprocity becomes $\frac{2}{4} = 0.5$ because $tr(\bar{A}^{2}) = 2$ and $tr(\bar{A'}^{2}) = 4$. Since $\frac{2}{4} \neq \frac{2}{3}$, \citep{pearcy2014hypergraph} (B1) violates \axiomnum{6}. 

%\subsection{Violations of Axiom 7}

\smallsection{B6. \measure with All Arcs as Reciprocal Set} 
According to \axiomnum{6}, a hypergraph reciprocity value should equal 2/3 $\approx$ 0.6667 in Figure~\ref{fig:axiom6counter}.
%For each target arc $e_i$, since all arcs are included in $R_{i}$, this measure cannot achieve the arc-level reciprocity of 1 even if there exists the perfectly reciprocal opponent of $e_i$. 
%In Figure~\ref{fig:axiom6counter}, although $e_{2}$ has its perfectly reciprocal arc (i.e., $e_{1}$), it $e_{3}$ is also included in $R_{2}$, which results in the transition probability of $p_{1} = [v_{1}, v_{2}, v_{3}, v_{sunk}] = [0, 0.5, 0.5, 0]$. 
%Here, the arc reciprocity becomes $r(e_{2}, R_{2} = \{e_{1}, e_{3}\}) = 0.7842 \neq 1$. 
If we let $\alpha = 1$,
then the overall hypergraph-level reciprocity ($r(G)$) based on \measure with all arcs as the reciprocal set (B6) is  
\begin{gather*}
r(e_{1}, \{e_{2}, e_{3}\}) = 0.5, \quad r(e_{2}, \{e_{1}, e_{3}\}) = 0.3444, \ \ \text{and} \ \ r(e_{3}, \{e_{1}, e_{2}\}) = 0 \\ 
r(G) = \frac{r(e_{1}, \{e_{2}, e_{3}\}) + r(e_{2}, \{e_{1}, e_{3}\}) + r(e_{3}, \{e_{1}, e_{2}\})}{3} \\
= \frac{0.5 + 0.3444 + 0}{3} = 0.2815 \neq 0.6667
\end{gather*}
which violates \axiomnum{6}.

%which implies this metric cannot works as a proper indicator function. 
%Thus, \axiomnum{6} cannot be achieved.

\subsection{Violations of Axiom 8}

\smallsection{B6. \measure with All the Arcs as Reciprocal Set} Even when there exists the perfect reciprocal opponent of a specific arc, the transition probability cannot be identical to the optimal transition probability if there exists another inversely overlapping arc (see the target arc $e_{2}$'s case in Figure~\ref{fig:axiom6counter}). 

\smallsection{B7. \measure with Inversely Overlapping Arcs as Reciprocal Set} As in the previous case, if there exist multiple inversely overlapping arcs, all of them are included in the reciprocal set. 
As a result, for such arcs, the cardinality penalty term gets smaller than $1$ (i.e., $(1/\lvert R_{i} \vert )^{\alpha} < 1$), resulting in $r(e_{i} , R_{i}) < 1$. 
Consequently, the overall hypergraph reciprocity becomes smaller than 1. %, which cannot be fully reciprocal.
}

\section{Appendix: Data Description}{
    \label{section:datadescription}
    
In this section, we provide the sources of the considered datasets and describe how we preprocess them.

\smallsection{Metabolic datasets} We use two metabolic hypergraphs, \textbf{iAF1260b} and \textbf{iJO1366}, which are provided by \citep{yadati2020nhp}. 
They are provided in the form of directed hypergraphs, and they do not require any pre-processing.
We remove one hyperarc from each dataset since their head set or tail set is abnormally large. Specifically, the size of their head sets is greater than $20$, while the second largest one is $8$. %(second largest hyperarc's head set size is 8, but abnormal hyperarc head set size is greater than 20).
Each node corresponds to a gene, and each hyperarc indicates a metabolic reaction among them. 
Specifically, a hyperarc $e_{i}$ indicates that a reaction among the genes in the tail set $T_{i}$ results in the genes in the head set $H_{i}$.

\smallsection{Email datasets} We use two email hypergraphs, \textbf{email-enron} and \textbf{email-eu}.
The \textbf{Email-enron} dataset is provided by \citep{chodrow2020annotated}. 
We consider each email as a single hyperarc. Specifically, the head set is composed of the receiver(s) and cc-ed user(s), and the tail set is composed of the sender.
The \textbf{Email-eu} dataset is from SNAP \citep{snapnets}.   
The original dataset is a dynamic graph where each temporal edge from a node $u$ to a node $v$ at time $t$ indicates that $u$ sent an email to $v$ at time $t$.
The edges with the same source node and timestamp are replaced by a hyperarc, where the tail set consists only of the source node and the head set is the set of destination nodes of the edges.
Note that every hyperarc in these datasets has a unit tail set, i.e., $\lvert T_{i} \vert = 1, \forall i = \{1, \cdots , \lvert E \vert \}$.

%We consider di-arcs which have identical source node and time stamp are one single group of email, and we transform these di-arcs into one single hyperarc. 
%Here, identical source node becomes a tail set, and each destination node of diarcs create a head set of corresponding hyperarc.
%

\smallsection{Citation datasets} We use two citation hypergraphs, \textbf{citation-data mining} and \textbf{citation-software}, which we create from pairwise citation networks, as suggested by \citep{yadati2021graph}. %, and we follow the proposed steps. 
Nodes are the authors of publications. 
Assume that a paper $A$, which is co-authored by $\{v_{1}, v_{2}, v_{3}\}$, cited another paper $B$, which is co-authored by $\{v_{4}, v_{5}\}$. 
%Here, we consider each author group of publications as a (head or tail) set. 
Then, this citation leads to a hyperarc where the head set is $\{v_{4}, v_{5}\}$ and the tail set is $\{v_{1}, v_{2}, v_{3}\}$.
%Assume that a paper $B$ cited another paper $C$, which is written by $\{v_{6}, v_{7}\}$. (let this hyperarc $e_{2}$). 
%Now, $\{v_{4}, v_{5}\}$ becomes a tail set $T_{2}$ of a new hyperarc $e_{2}$. 
As pairwise citation networks, we use subsets of a DBLP citation dataset \citep{sinha2015overview}.
The subsets consist of papers published in the venues of data mining and software engineering, respectively.\footnote{We use the venues listed at \citep{csconference}}
%Then, we adopt papers which are published in the selected venues. 
In addition, we filter out all papers co-authored by more than $10$ authors to minimize the impact of such outliers.

\smallsection{Question answering datasets} We use two question answering hypergraphs, \textbf{qna-math} and \textbf{qna-server}. 
We create directed hypergraphs from the log data of a question answering site, \textit{stack exchange}, provided at \citep{qnadataset}. 
Among various domains, we choose \textit{math-overflow}, which covers mathematical questions, and \textit{server-fault}, which treats server related issues.
The original log data contains the posts of the site, and one questioner and one or more answerers are involved with each post.
We ignore all posts without any answerer.
We treat each user as a node, and we treat each post as a hyperarc.
For each hyperarc, the questioner of the corresponding post composes the head set, and the answerer(s) compose the tail set. 
Note that every hyperarc in these datasets has a unit head set, i.e., 1 $\lvert H_{i} \vert  = 1, \forall i = \{1, \cdots , \lvert E \vert \}$.

%\red{[KJ: I am here]}

\smallsection{Bitcoin transaction dataset} We use three bitcoin transaction hypergraphs, \textbf{bitcoin-2014}, \textbf{bitcoin-2015}, and \textbf{bitcoin-2016}. 
The original datasets are provided by \citep{wu2020detecting}, and they contain first 1,500,000 transactions in 11/2014, 06/2015, and 01/2016, respectively.
We model each account as a node, and we model each transaction as a hyperarc. 
As multiple accounts can be involved in a single transaction, the accounts from which the coins are sent compose the tail set, and the accounts to which the coins sent compose the head set.
%As change-making is also treated as a transaction, 
We remove all transactions where the head set and the tail set are exactly the same.
}

%\section{introduction}
%The Introduction section, of referenced text expands on the background of the work (some overlap with the Abstract is acceptable). The introduction should not include subheadings.
%Springer Nature does not impose a strict layout as standard however authors are advised to check the individual requirements for the journal they are planning to submit to as there may be journal-level preferences. When preparing your text please also be aware that some stylistic choices are not supported in full text XML (publication version), including coloured font. These will not be replicated in the typeset article if it is accepted. 


%\bibliography{sn-style-ref}



\end{document}